\documentclass[11pt,a4paper]{article}
\pdfoutput=1
\usepackage{jheppub}

\usepackage{graphicx} 
\usepackage{epstopdf} 
\usepackage{dcolumn} 
\usepackage{xcolor} 
\usepackage{amsmath,amssymb} 
\usepackage{hyperref} 
\usepackage[utf8]{inputenc} 
\usepackage[english]{babel} 
\usepackage[displaymath, mathlines]{lineno} 
\usepackage{blindtext} 
\usepackage{ifthen} 
\usepackage{orcidlink} 

\usepackage{booktabs}
\usepackage{subfigure}
\usepackage{tabularx}
\usepackage[titletoc, title]{appendix}
\usepackage[title]{appendix}
\usepackage{placeins}

\graphicspath{{figures/}} 

\newboolean{articletitles}
\setboolean{articletitles}{true} 

\input{belle2-symbols}

\begin{document}

\vspace*{-3\baselineskip}
\resizebox{!}{2cm}{\includegraphics{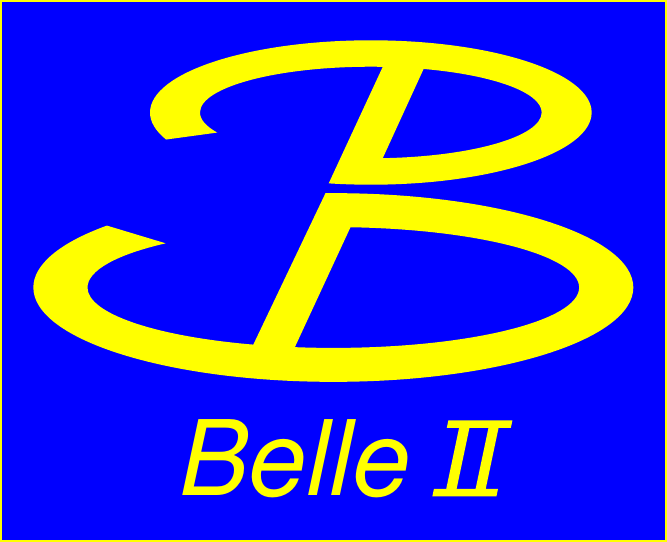}}
\title{Measurement of the branching fractions of \boldmath{$\overline B\to D^{(*)} K^- K^{(*)0}_{(S)}$} and \boldmath{$\overline B\to D^{(*)}D_s^{-}$} decays at Belle~II
}
\collaboration{The Belle II Collaboration}
  \author{I.~Adachi\,\orcidlink{0000-0003-2287-0173},} 
  \author{L.~Aggarwal\,\orcidlink{0000-0002-0909-7537},} 
  \author{H.~Aihara\,\orcidlink{0000-0002-1907-5964},} 
  \author{N.~Akopov\,\orcidlink{0000-0002-4425-2096},} 
  \author{A.~Aloisio\,\orcidlink{0000-0002-3883-6693},} 
  \author{N.~Althubiti\,\orcidlink{0000-0003-1513-0409},} 
  \author{N.~Anh~Ky\,\orcidlink{0000-0003-0471-197X},} 
  \author{D.~M.~Asner\,\orcidlink{0000-0002-1586-5790},} 
  \author{H.~Atmacan\,\orcidlink{0000-0003-2435-501X},} 
  \author{T.~Aushev\,\orcidlink{0000-0002-6347-7055},} 
% ...

\emailAdd{coll-publications@belle2.org}

\abstract{We present measurements of the branching fractions of eight {$\overline B{}^0\to D^{(*)+} K^- K^{(*)0}_{(S)}$,  $B^{-}\to D^{(*)0} K^- K^{(*)0}_{(S)}$} decay {channels}.
The results are based on data from SuperKEKB electron-positron collisions at the $\Upsilon(4S)$ resonance collected with the Belle II detector{,} corresponding to an integrated luminosity of ${362~\text{fb}^{-1}}$. 
The event yields are extracted from fits to the distributions of the difference between expected and observed $B$ meson energy, and are efficiency-corrected as a function of $m(K^-K^{(*)0}_{(S)})$ and $m(D^{(*)}K^{(*)0}_{(S)})$ in order to avoid dependence on the decay model.
These results include the first observation of $\overline B{}^0\to D^+K^-K_S^0$, $B^-\to D^{*0}K^-K_S^0$, and $\overline B{}^0\to D^{*+}K^-K_S^0$ decays and a significant improvement in the precision of the other {channels} compared to previous measurements. 
The helicity{-}angle distributions and the invariant mass distribution{s} of the $K^- K^{(*)0}_{(S)}$ systems are compatible with quasi-two-body decays via a resonant transition with spin-parity $J^P=1^-$ for the $K^-K_S^0$ systems and $J^P= 1^+$ for the $K^-K^{*0}$ systems. We also present measurements of the branching fraction{s} of four {$\overline B{}^0\to D^{(*)+} D_s^-$, $B^{-}\to D^{(*)0} D_s^- $} decay {channels} with a precision compatible to the current world averages.
}

\maketitle
\flushbottom

\section{Introduction}
Knowledge of $B$ meson hadronic decays is limited: about $40\%$ of the total $B$ width is not measured in terms of exclusive branching fractions ($\mathcal{B}$).
{Hence, unmeasured decays are usually simulated using the \texttt{PYTHIA} fragmentation model}~\cite{MC:pythia82},
which does not properly describe $B$ decays.  
The inclusive branching fraction for $B$ meson decays into a $D^{(*)}$ meson accompanied by a kaon pair and possibly pions could account for 6\% {of the $B$ width}, according to \texttt{PYTHIA},
However, only a small fraction of the exclusive components has been measured~\cite{Belle:DKK}. Improving the knowledge of {these decay channels} can be of {significant} benefit to understanding the background contamination of many analyses {of} hadronic $B$ decays. The unknown fraction of the total $B$ width is spread across many exclusive channels, therefore improvements are not expected from single branching fraction measurements, but require the systematic exploration of the large unmeasured or poorly measured channels. This paper is part of a program of measuring the missing components in the $B$ decay width.

Assuming factorization~\cite{Theory:factorization}, three-body ${B\to DK^- K}$  decays proceed via a tree-level amplitude with an external $W$ boson emission and the production of an $s\overline s$ pair. The symbol $B$ indicates {a} neutral or charged $B$ meson, the symbol $K$ indicates {a} $K_S^{0}$ or {a} $K^{*0}$ meson, while the symbol $D$ indicates {a} $D^{(*)+,0}$ meson, from now on.\footnote{We use natural units $\hbar=c=1$ and charge conjugation is implied throughout.} However, the quasi-two-body mechanism, $B\to DX^-(\to K^- K)$, with an intermediate resonance $X^-$ {that} decays strongly, is also possible. Several theoretical studies have been performed to interpret these decays in term of conventional~\cite{theory:rho_v0, theory:rho} or exotic~\cite{theory:molecular_a1,theory:exotic_a1} resonances. 
The factorization assumption forbids the production of resonance{s} with spin larger than one.
{Assuming factorization and exact isospin symmetry}, a spin-parity assignment $J^P=1^-$ is expected for the $K^- K^{0}_S$ system~\cite{Diehl:spinparity}. If isopin symmetry (or factorization) is not assumed $J^P=0^+$ states are also allowed. On the other hand, for the $K^-K^{*0}$ system the spin-parity assignments $J^P=0^-$, $1^-$, or $1^+$ are allowed.  

A search for $B\to DK^- K$ decays was previously performed by the Belle experiment,  on a sample with an integrated luminosity of $29~\text{fb}^{-1}$~\cite{Belle:DKK}.
Four $B\to DK^-K^{*0}$ {channels} and the $B^-\to D^0 K^- K^0_S$ mode were observed, while excesses in the remaining $K_S^0$ {channels} were seen with significances of about 2.5~standard deviations.\footnote{We use ``$K_S^0$ channels'' to refer to all the channels where a $K_S^0$ is present; similarly we use ``$K^{*0}$'', ``$D^0$'', ``$D^+$'', ``$D^{*0}$'' and ``$D^{*+}$ channels''.} 
A study of the $m(K^-K^{*0})$ invariant mass distribution was also reported{.} A dominant transition via a $J^P=1^+$ state was claimed for {${B \to D K^- K^{*0}}$} decays, and the intermediate state was interpreted as a $a_1(1260)^-$ resonance. 
Due to the small sample size, no conclusive claims were made for ${B\to DK^- K^{0}_S}$ decays. 
However, the $m(K^-K^{0}_S)$ invariant mass and angular distributions suggested the presence of a resonant component with $J^P=1^-$ in {${B\to D K^- K^{0}_S}$} decays. This was interpreted as a $\rho$-like resonance, but the $\rho(770)$ was disfavored given the limited available phase-space.
    
A potential application of an improvement in understanding the $B\to DK^-K$ sector is {an amelioration {of the}  $B$-tagging algorithms used at $B$-factories.  At an energy-asymmetric $e^+e^-$ collider such as SuperKEKB}, $B^0\overline B{}^0$ and $B^+B^-$ meson pairs ($B\overline B$ {pairs}) are produced at threshold from $\Upsilon(4S)$ meson decays. A large part of the Belle~II physics program~\cite{Belle2:PhysicsBook} relies on identifying the partner $B$ meson produced in association with the signal $B$ meson to infer the properties of the signal ($B$-tagging). 
In particular, in hadronic $B$-tagging the partner $B$ meson is fully reconstructed to infer the kinematic properties of the signal using initial-state constraints. 
The Belle~II $B$-tagging algorithm, Full Event Interpretation (FEI), is based on a set of multivariate classifiers trained on the Belle~II {Monte Carlo} (MC) simulation~\cite{Belle2:FEI}. A mismodeling in the simulation may introduce biases and uncertainties in the FEI efficiencies, and {leads} to suboptimal FEI performance, degrading Belle~II's physics reach. 
Three-body $B\to DK^-K$ decay channels are currently not used by the FEI algorithm. However, the high purity of these decays~\cite{Belle:DKK} makes them ideal candidates to improve the $B$-tagging efficiency.
Moreover, information on the branching fractions alone is insufficient, and knowledge of the final-state kinematic properties and the intermediate states is essential for an accurate description of these decays during FEI training.  {In conclusion, additional measurements of $B\to DK^-K$ decays and their dynamics} can lead to an improvement of the FEI efficiency and its background rejection. 

A more precise measurement of eight $B\to D K^- K$ absolute branching fractions including observation of the three unobserved ${B\to D K^- K^{0}_S}$ {channels} is reported in this work, using the 362~fb$^{-1}$ Belle~II data sample collected at the $\Upsilon(4S)$ resonance. This analysis aims to better understand the intermediate states {in} $B\to D K^- K$ decays by investigating their Dalitz plots~\cite{Dalitz}. In particular, the $m(K^-K^{0}_S)$ and $m(K^-K^{*0})$ distributions and the angular distributions of these systems are studied.

Measurements of the branching fractions of the four $B\to D D_s^- $ decays, {which are reconstructed in the same final states as the $B\to DK^- K$ channels,} are also included in this work. These channels are often adopted as control or normalization channels in several measurements~\cite{LHCb:DsD_use1,LHCb:DsD_use2,LHCb:DsD_use3,LHCb:DsD_use4} and improvements in $\mathcal B(B\to DD_s^-)$ precision will reduce their uncertainties.

The branching fractions are extracted independently for each channel. 
After the event reconstruction and selection, {a fit to the distribution of the difference between expected and observed $B$-meson energy} is performed to separate the signal from the backgrounds. 
In $B\to DK^-K$ channels, the $s$Plot procedure~\cite{splot} is used to obtain a background-subtracted $\bigl(m(K^-K), m(DK)\bigr)$ two-dimensional invariant mass distribution. The signal efficiency is evaluated using a simulated signal sample, as a function of $\bigl(m(K^-K), m(DK)\bigr)$.
The branching fraction is then obtained from the efficiency-corrected integral of the $\bigl(m(K^-K), m(DK)\bigr)$ distribution. 
The ${B\to DD_s^-(\to K^-K)}$ branching fractions are extracted using the same strategy, but the efficiency correction is applied directly to the yield obtained from the $\Delta E$ fit, since {there is no model dependence in these channels.}
We use simulation and data control samples to optimize the analysis.  In particular, the ${B\to DD_s^-(\to K^-K)}$ channels, along with the ${B^-\to D^{*0}\pi^-}$ control channel, are also used to validate the $B\to DK^-K$ analysis and assess some of the systematic uncertainties. 
To reduce experimenter's bias, data containing signal candidates are only analyzed after {the entire analysis procedure is finalized}. 

The paper is organized as follows. Section~\ref{sec:Belle2} gives a brief description of {the} Belle~II detector {and analysis sample}. Section~\ref{sec:reco} {discusses} the reconstruction and event selection. Section~\ref{sec:yield} and {Section}~\ref{sec:efficiency} describe the {signal} yield extraction and efficiency estimation, respectively. Section~\ref{sec:BR} and {Section}~\ref{sec:syst} present the measured branching fractions and the related systematic uncertainties, respectively. In {Section}~\ref{sec:mKK} {discusses} the $m(K^-K)$ invariant mass and angular distributions. 
\vspace{-0.2cm}
\section{The Belle~II detector and samples}\label{sec:Belle2}
\vspace{-0.2cm}
The Belle~II experiment~\cite{Belle2:TDR} is located at SuperKEKB, which collides electrons and positrons at and near the $\Upsilon(4S)$ resonance~\cite{SuperKEKB:TDR}. The Belle~II detector has a cylindrical geometry {with the symmetry axis, defined as the $z$ axis, almost coincident with the direction of the electron beam.} {The detector} includes a two-layer silicon-pixel detector surrounded by a four-layer double-sided silicon-strip detector~\cite{Belle2:SVD} and a 56-layer central drift chamber~(CDC). These detectors reconstruct {trajectories of charged particles} {(tracks)}.  Only one sixth of the second layer of the {pixel detector} was installed for the data analyzed here.   Surrounding the CDC, which also provides $dE/dx$ energy-loss measurements, is a time-of-propagation counter~\cite{Belle2:TOP} in the central region and an aerogel-based ring-imaging Cherenkov counter in the forward region.  These detectors provide charged-particle identification.  Surrounding {these subsystems} is an electromagnetic calorimeter based on CsI(Tl) crystals that primarily provides energy and timing measurements for photons and electrons. Outside of the calorimeter is a superconducting solenoid magnet. The solenoid magnet provides a 1.5~T magnetic field that is parallel to the $z$ axis. Its flux return is instrumented with resistive-plate chambers and plastic scintillator modules to detect muons, $K^0_L$ mesons, and neutrons. 

Belle~II integrates a hardware {trigger and an online software event selection} to suppress {background processes} and to select $b\overline b$, $c\overline c$, and $\tau^+\tau^-$ events with high efficiency. 

The data sample of Belle~II collected from 2019 to 2022 at {the} energy of the $\Upsilon(4S)$ is used in this analysis. The total integrated luminosity of the sample is $\mathcal L_\text{int}=(362\pm 2)\,\text{fb}^{-1}$, which corresponds to $(387\pm 6)\times 10^6$ $B\overline B$ pairs.

Simulated samples {of exclusive decay channels} are used to determine signal efficiencies, to define fit models, and to evaluate the systematic uncertainties. These samples have an integrated luminosity equivalent to at least 50 times the data. 
A simulated sample that reproduces the composition of Belle~II events, including $B\overline B$ and $e^+e^-\to q\overline q$ continuum (where $q$ indicates an $u,d,s$ or $c$ quark) backgrounds, and equivalent to an integrated luminosity of $1~\text{ab}^{-1}$, is used to investigate the sample composition and validate the analysis before examining the signal region in data. 

Simulated events are generated using \texttt{KKMC} generator for quark-antiquark production from $e^+e^-$ collisions~\cite{MC:KKMC}, \texttt{PYTHIA8} generator
for hadronization~\cite{MC:pythia82}, \texttt{EvtGen} software package and \texttt{PYTHIA8} generator for the decay of the generated hadrons~\cite{MC:evtgen, MC:evtgen_old}, and \texttt{Geant4} software package for the detector response~\cite{MC:geant4}. The simulation includes simulated beam-induced backgrounds~\cite{Belle2:bkg}. 
The data and the MC simulations are processed using the Belle~II analysis software~\cite{Belle2:basf2, Belle2:basf2_repo}.

\section{Reconstruction and event selection}\label{sec:reco}
Events are selected by the trigger based on the charged-particle multiplicity and total energy, to suppress low-multiplicity events. 

Candidate charged pions and kaons are identified from their tracks, which are required to have transverse and longitudinal impact parameter {calculated with respect to} the interaction point, (IP)  ${d_r<2~\text{cm}}$ and ${|d_z|<4~\text{cm}}$, respectively, polar angle $\theta$ within the acceptance of the CDC ($17^\circ\leq\theta\leq 150^\circ$), and at least 20 {measurement points (hits)} in the CDC. The candidates are identified as pions or kaons based on the ratio between the particle-identification (PID) likelihood for the test hypothesis and the sum of the likelihoods for all other hypotheses. The PID selection efficiency is between 70\% and 80\% {on average}, with misidentification probability {varying} between {3\% and 10\%}, depending on the transverse momentum $p_T$ and $\cos\theta$ of the charged particle, both for kaon- and pion-enriched samples. 

Photon candidates are selected by requiring calorimeter energy deposits (clusters) {contain} more than one crystal, {have} $\theta$ {within} the CDC acceptance, and a polar-angle-dependent energy threshold of 80~MeV, 30~MeV, and 60~MeV in the forward endcap, barrel, and backward endcap, respectively.
Clusters are required to be detected within 200~ns of the beam crossing to suppress energy deposits from beam background. The misreconstructed photons {from} hadronic clusters without a matched track, are suppressed with a dedicated multivariate classifier, which uses cluster-shape variables, calorimeter crystal pulse-shape, and track-to-cluster distance.

Neutral pions are reconstructed from photon candidate pairs, requiring the diphoton mass to be between $120~\text{MeV}$ and $145~\text{MeV}$ and the convergence of a mass-constrained fit.  
Candidate $K_S^0$ mesons are reconstructed from pairs of oppositely charged pions. We require the dipion mass to be within  $10~\text{MeV}$ of the known value, a successful vertex fit, the transverse distance between the interaction point and the dipion vertex to exceed 0.4~cm, and the cosine of the angle between the $K_S^0$ flight direction and its reconstructed momentum evaluated in the laboratory frame to exceed 0.8. Candidate $K^{*0}$ mesons are reconstructed in the $K^{*0}\to K^+\pi^-$ decay channel, requiring their invariant mass to be within 50 MeV of the known $K^{*0}$ mass value.  

Candidate $D^{(*)}$ mesons are reconstructed using the $D^0\to K^-\pi^+$, $D^+\to K^-\pi^+\pi^+$, $D^{*0}\to D^0\pi^0$, and $D^{*+}\to D^0\pi^+$ decay channels. The invariant masses of the $D^0$ and $D^+$ candidates {are} required to be within $15~\text{MeV}$, {corresponding to three units of resolution ($\sigma$)}, of the known values, and a mass- and vertex-constrained fit is performed. Similarly, the invariant mass of the $D^{*0}$ and $D^{*+}$ {candidates} is required to be within $3~\text{MeV}$ and $1.5~\text{MeV}$ of the known value  ($3\sigma$). 

Candidate $B$ mesons are reconstructed in eight ${B\to DK^- K}$ {channels} {and four $B\to DD_s^-$ channels}. 
We require the beam-constrained mass $M_{\rm bc}=\sqrt{E^{*2}_\text{beam}-p^{*2}_B}>5.272~\text{GeV}$ {and $-0.12~\text{GeV}<\Delta E = E_B^*-E_\text{beam}^*<0.3~\text{GeV}$, where $p_B$ and $E_B$ are the momentum and the energy of the $B$ meson, respectively, and {$E_\text{beam}^*=\sqrt{s}/2$} is the calibrated value of the beam energy. The symbol $^*$ indicates that the variable is evaluated in the $\Upsilon(4S)$ center-of-mass frame}. The asymmetric requirement {on $\Delta E$} suppresses cross-feed from other channels in which a pion is not reconstructed. 
To reduce continuum background, we require that the ratio of the second{- to zeroth Fox-Wolfram moments} satisfies $R_2<0.5$~\cite{theory:FWM}; the absolute value of the cosine of the angle between the thrust axis of the $B$ and the thrust axis of the rest of the event be smaller than 0.85, where the thrust axis is the direction that maximizes the total projection of the momenta of all particles in the event~\cite{Theory:thrust}; and the absolute value of the cosine of the angle between the $B$ momentum and the beam direction satisfies ${|\cos\theta_{p_B^*p_\text{beam}^*}|<0.9}$. 
The rest of the event includes all the remaining tracks and clusters selected with the following requirements. The tracks are required to be within the CDC polar {angle} acceptance, to have transverse momenta greater than $100~\text{MeV}$, and to have $d_r<0.5~\text{cm}$, and $|d_z|<3~\text{cm}$. The clusters are required to be in the CDC polar acceptance, to have energies above $50~\text{MeV}$, and to be detected within 200~ns of the beam crossing. 

In the $B\to DK^-K$ channels, we also require $m(K^-K)$ to differ by more than 20~MeV  ($4\sigma$) from the known $D_s^-$ mass, to suppress resonant $B\to DD_s^-(\to K^-K)$ {decays}, which has the same final state as the signal. To {identify} $B\to DD_s^-(\to K^-K)$ {decays}, the selection procedure for $B\to DK^-K$ channels is used; however, the {$D_s^-$ mass} veto is inverted. 

The average $B$ candidate multiplicity per event as determined from signal MC simulation is between $1.05$ and $1.08$ for the $D^0$, $D^+$, and $D^{*+}$ channels,  while it is about $1.4$ for the $D^{*0}$ channels. 
For each event, the $B$ candidate with $M_{\rm bc}$ closest to the known $B$ mass is selected. 
The efficiency of this candidate selection, defined as the ratio of the number of correct candidates chosen in multiple candidate events to the total number of candidates for events {with multiple candidates} in which the correct candidate is reconstructed, {is 67\% for {the} $\overline B{}^0\to D^{*0}K^-K_S^0$ channel, while it is between 80\% and 92\% for all the other channels}, based on simulation. 
After the candidate selection, the purity of the sample (defined as the ratio of correctly reconstructed events to the total number of reconstructed events) is {between 92\% and 96\%} for all channels, except for the $D^{*0}$ channels where it is 71\%, according to the signal MC simulation.

The $B^-\to D^{*0}\pi^-$ control sample is reconstructed {using the $D^{*0}$, $\pi^-$ and $B^-$ selections used in $B^-\to D^{*0}K^-K$ channels}. In addition the $D^{*0}$ momentum is required to be lower than $2.5~\text{GeV}$,  to obtain a momentum range similar to {that of} the $B^-\to D^{*0}K^-K$ signal.

We apply corrections for the track-momentum scale and photon-energy scale on data, which are derived from large data control samples. These corrections range from {$10^{-4}$} {to the} $10^{-3}$ level. 

\section{{Signal} yield extraction}\label{sec:yield}

The composition of the selected samples is investigated using the $\Delta E$ distribution from MC simulation. A small, smooth $e^+e^-\to q\overline q$ continuum background is expected in all channels, while the signal $B\to DK^-K$ {peaks} at $\Delta E\approx 0$.
In the ${\overline B{}^0\to D^+K^- K}$ channels a cross-feed component from ${\overline B{}^0\to D^{*+}K^-  K}$ channels is expected, in which the $\pi^0$ from the $D^{*+}\to D^+\pi^0$ decay is not reconstructed. 
This component peaks at {$\Delta E\approx-0.15~\text{GeV}$}. 
{A similar background} is present in the $B^-\to D^0K^-K$ channels, {which has} a cross-feed component from the $B^-\to D^{*0}K^-K$ and $\overline B{}^0\to D^{*+}K^-  K$ channels, that both mimic the signal when the $\pi^0$ or the $\pi^+$ from the $D^*$ decay is not reconstructed.  
In the $B^-\to D^{*0}K^-K$ channels, a cross-feed component from the $B^-\to D^0K^-K$ channel is reconstructed if an incorrectly reconstructed $\pi^0$ is associated with the $D^0$. This background component peaks around {$\Delta E\approx 0.15~\text{GeV}$}. In the $B^-\to D^{*0}K^-K$ channels a cross-feed component from the $\overline B{}^0\to D^{*+} K^- K$ channels is also expected, when the $\pi^+$  from the $D^{*+}$ decay is not reconstructed and an incorrect $\pi^0$ is included. 
This background component peaks under the signal, and thus it requires special treatment. The $\overline B{}^0\to D^{*+} K^- K$ channels have no peaking backgrounds.  The four $B\to DK^-K^{*0}$ channels have backgrounds from $B\to DK^-K^+\pi^-$, i.e.\ the non-$K^{*0}$-resonant component.  The $B\to DD_s^-$ channels have the same composition as the $B\to DK^-K$ channels{; however,} they have an additional background from $B\to DK^-K$. 

The \texttt{RooFit} package~\cite{ROOT:roofit} is used to perform extended maximum likelihood fits to the unbinned $\Delta E $ distribution {to extract the signal yield}.  The fit is performed in the $\Delta E$ range  ${[-0.12~\text{GeV}, 0.3~\text{GeV}]}$ for all channels. The {upper boundary is chosen} to better constrain the fit for the $B^- \to D^0 K^- K$, $\overline B{}^0 \to D^{+} K^- K$,  and $\overline B{}^0 \to D^{*+} K^- K$ channels.  In the $B^- \to D^{*0} K^- K$ channel {an} extended range is used to constrain the background peaking under the signal. 

The $\Delta E $ distribution is described with the following model:
\begin{equation}
    p(\Delta E)= p_\text{sig}(\Delta E)+p_\text{bkg}(\Delta E)+p_\text{cross-feed}(\Delta E)+p_{DKK\pi\text{-bkg}}(\Delta E)+p_{DKK\text{-bkg}}(\Delta E),
\end{equation}
where the details of the components are described in Table~\ref{tab:Fit_model}, and each individual component {(signal, combinatorial background, cross-feed in $B\to D^{*0}K^-K$ channels, $B\to DK^+K^-\pi$ background, $B\to DK^-K$ background in $B\to DD_s$ channels, respectively)} is included only in the specific $B$ decay channel indicated in the last column of the table.

\begin{table}[!htb]
\begingroup
\small
\renewcommand{\arraystretch}{1.5}
\centering
\caption{Details of the model for the $\Delta E$ fit. The symbol $\mathcal G(x,\mu,\sigma)$ stands for a Gaussian {function} with mean $\mu$ and width $\sigma$; $\mathcal G_A(x,\mu,\sigma_L,\sigma_R)$ stands for an asymmetric Gaussian {function} with mean $\mu$, left width $\sigma_L$, and right width $\sigma_R$; {$\mathcal B({D^{*+}})/\mathcal B({D^0})$} is the ratio of branching fractions from Eq.~\eqref{eq:DstpBkg_yield}; $f$ stands for the fraction of the $DKK\pi$ component as described in the text; $C$, $B$, $K$, $N$, and $D$ are yield parameters; $t$ is the tail-to-core Gaussian fraction; $r$ is the resolution scale factor; $d$ is the exponential decay constant; and $\Delta \mu$ is the shift of the mean for the cross-feed component.}\label{tab:Fit_model}
\hspace{-0.5 cm}
\begin{tabular}{ccc}
\toprule
 Component      &  Description      & $B$ {signal} channels \\
 \midrule
 $p_\text{sig}$       & $C\left[(1-t)\mathcal G(\Delta E,\mu,r\cdot \sigma)+t\mathcal G_A(\Delta E,\mu,\sigma_L,\sigma_R)\right]$ &  All \\
$p_\text{bkg}$    & $Be^{d\Delta E}+K$  &  All \\
$p_\text{cross-feed}$    &  $N\left[\mathcal G_A(\Delta E,\mu+\Delta\mu,\sigma_{L0},\sigma_{R0})+ \frac{\mathcal B(D^{*+})}{\mathcal B(D^0)}\mathcal G_A(\Delta E,0,\sigma_{L+},\sigma_{R+})\right]$ & $D^{*0}$ channels \\
$p_{DKK\pi\text{-bkg}}$ & $  fC\mathcal G(\Delta E,\mu,r\cdot \sigma)$ & $B\to DK^-K^{*0}$ channels \\
$p_{DKK\text{-bkg}}$ & $D \mathcal G(\Delta E,\mu,r\cdot \sigma)$ & $B\to DD_s^-$ channels \\
\bottomrule
\end{tabular}
\endgroup
\end{table}
\vspace{-0.2cm}
\subsection{{$B\to DK^-K_S^0$ channels}}

In the $B^- \to D^0 K^- K^0_S$, $\overline B{}^0 \to D^{+} K^- K^{0}_S$,  and $\overline B{}^0 \to D^{*+} K^- K^0_S$ channels, the signal is described by the sum of one symmetric (\textit{core}) and one asymmetric (\textit{tail}) Gaussian distributions sharing the same mean parameter. 
The widths of the Gaussians are fixed to the values obtained in a fit to the simulated signal sample.  However, a scale factor $r$ is assigned as a multiplier to the core Gaussian width to compensate for any difference in resolution between simulation and data. The resolution scale factor is a free parameter in the $D^{0}$  channel fit to data. In the $D^{+}$ and $D^{*+}$ channels, the resolution scale factors are fixed to the value from the $D^0$ fit ($r=1.1\pm 0.1$), since the expected yields are too small to constrain an additional free parameter. The fractions of the core and the tail Gaussian are fixed to the values obtained in a fit to the signal simulation sample.  The background is described by the sum of a falling exponential distribution and a constant. The fit has two or three free shape parameters (mean, exponential constant, and resolution scale factor for $D^0$ channel), along the signal yield, and two background yields. 

The ${B^- \to D^{*0} K^- K^0_S}$ channel parametrization requires a different approach, because of the background from the ${\overline B{}^0 \to D^{*+} K^- K^0_S}$ channel peaking under the signal. 
The yield of this cross-feed component can be constrained directly from data using the cross-feed from the $B^- \to D^{0} K^- K^0_S$ channel, {which } is shifted in $\Delta E$ by ${0.15~\text{GeV}}$. 
This $\Delta E$ fitting region is almost free from continuum background, {allowing an accurate determination of the cross-feed despite the low yield}. The yield $N_{D^0}^\text{bkg}$ of cross-feed from the $B^- \to D^{0} K^- K^0_S$ channel can be determined from data as 
\begin{equation}
N_{D^0}^\text{bkg}\propto \mathcal{B}(B^- \to D^{0} K^- K^0_S)f_{\pi^0}\varepsilon_{D^{0}},
\end{equation} 
where $\varepsilon_{D^{0}}$ is the efficiency of $B^- \to D^{0} K^- K^0_S$ reconstruction and $f_{\pi^0}$ is the probability of incorrect $\pi^0$ association. For the $\overline B{}^0 \to D^{*+} K^- K^0_S$ cross-feed, assuming that the efficiency of a multiparticle final state factorizes into the products of single-particle efficiencies, we have 
\begin{equation}\label{eq:Dstp_bkg_eff}
N_{D^{*+}}^\text{bkg}\propto \mathcal{B}(\overline B{}^0 \to D^{*+} K^- K^0_S) f_{\pi^0}\varepsilon_{ D^{0} \text{-in-} D^{*+}},
\end{equation}
where $\varepsilon_{ D^{0} \text{-in-} D^{*+}}$ is the efficiency of $B^- \to D^{0} K^- K^0_S$ reconstruction from $B^- \to D^{*+} K^- K^0_S$ events. 
The $\overline B{}^0 \to D^{*+} K^- K^0_S$ decay chain is identical to $\overline B{}^- \to D^{0} K^- K^0_S${,} except for the additional low-momentum $\pi^+$, and {shows no relevant kinematic differences} in the simulation of the $D^{0} K^- K^0_S$ channel that could lead to a significant difference in the efficiency. Therefore we assume
\begin{equation}\label{eq:Dst0_hypothesis}
\varepsilon_{D^{0}\text{-in-} D^{*+}}=\varepsilon_{D^{0}},
\end{equation}
to obtain
\begin{equation}\label{eq:DstpBkg_yield}
   N_{D^{*+}}^\text{bkg} = \frac{\mathcal{B}(\overline B{}^0 \to D^{*+} K^- K^0_S)}{\mathcal{B}(B^- \to D^{0} K^- K^0_S)}  N_{D^0}^\text{bkg}.
\end{equation}
Thus, by measuring the yield of the $\Delta E$-shifted cross-feed from the $B^- \to D^{0} K^- K^0_S$ channel, and adding as inputs the branching fractions of the two cross-feed backgrounds measured in this analysis, we constrain the yield of the cross-feed from $\overline B{}^0 \to D^{*+} K^- K^0_S$ decays peaking under the signal. A systematic uncertainty is assigned due to the assumptions {in} this procedure {(as described in Sec.~\ref{sec:syst})}.

The ${B^- \to D^{*0} K^- K^0_S}$ channel  is fitted using  functional forms described in Table~\ref{tab:Fit_model}. 
The resolution scale factor is fixed to unity, as estimated from the $B^-\to D^{*0}\pi^-$ control channel. 
The cross-feed component from the $B^- \to D^{0} K^- K^0_S$ channel is parametrized as an asymmetric Gaussian distribution with widths fixed from the signal simulation and mean equal to the signal mean with a fixed shift evaluated from the signal simulation. The $B^- \to D^{0} K^- K^0_S$ cross-feed yield is free in the fit. 
The cross-feed component from the $\overline B{}^0 \to D^{*+} K^- K^0_S$ channel is fitted as an asymmetric Gaussian distribution with mean fixed to zero and width fixed from the signal simulation. The yield is fixed based on the relation in Eq.~\eqref{eq:DstpBkg_yield}, where $N_{D^0}^\text{bkg}$ is the yield of the $B^- \to D^{0} K^- K^0_S$ cross-feed, which is fitted simultaneously, and the $D^{*+}$ and $D^0$ channel branching fractions are fixed. The fits have two shape parameters (mean and exponential constant), the signal yield, the yields of two backgrounds, and the yield of the $B^- \to D^{0} K^- K^0_S$ cross-feed component as free parameters.

\begin{figure}[!t]
\centering
\subfigure{\includegraphics[width=0.45\columnwidth]{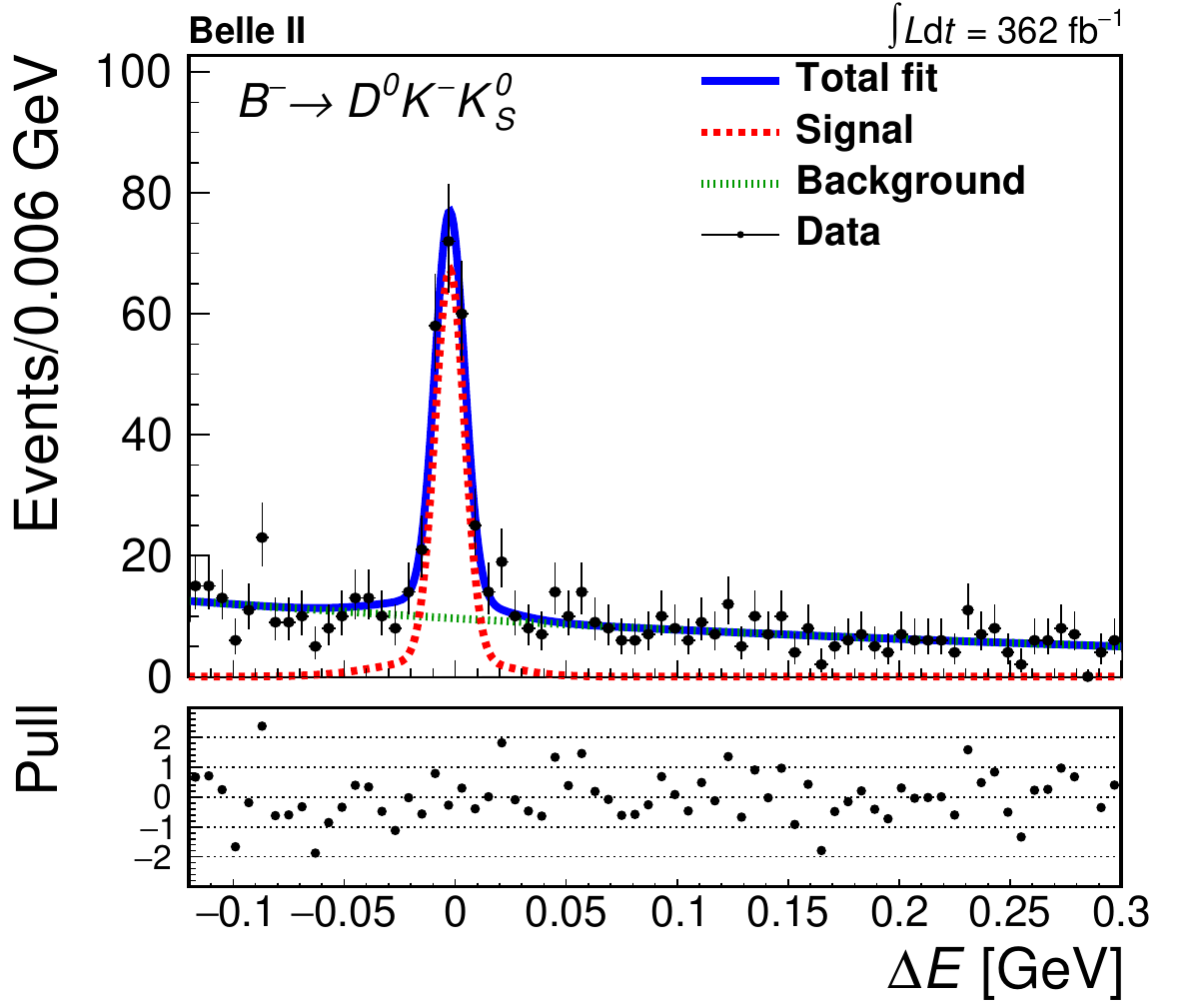}}
\subfigure{\includegraphics[width=0.45\columnwidth]{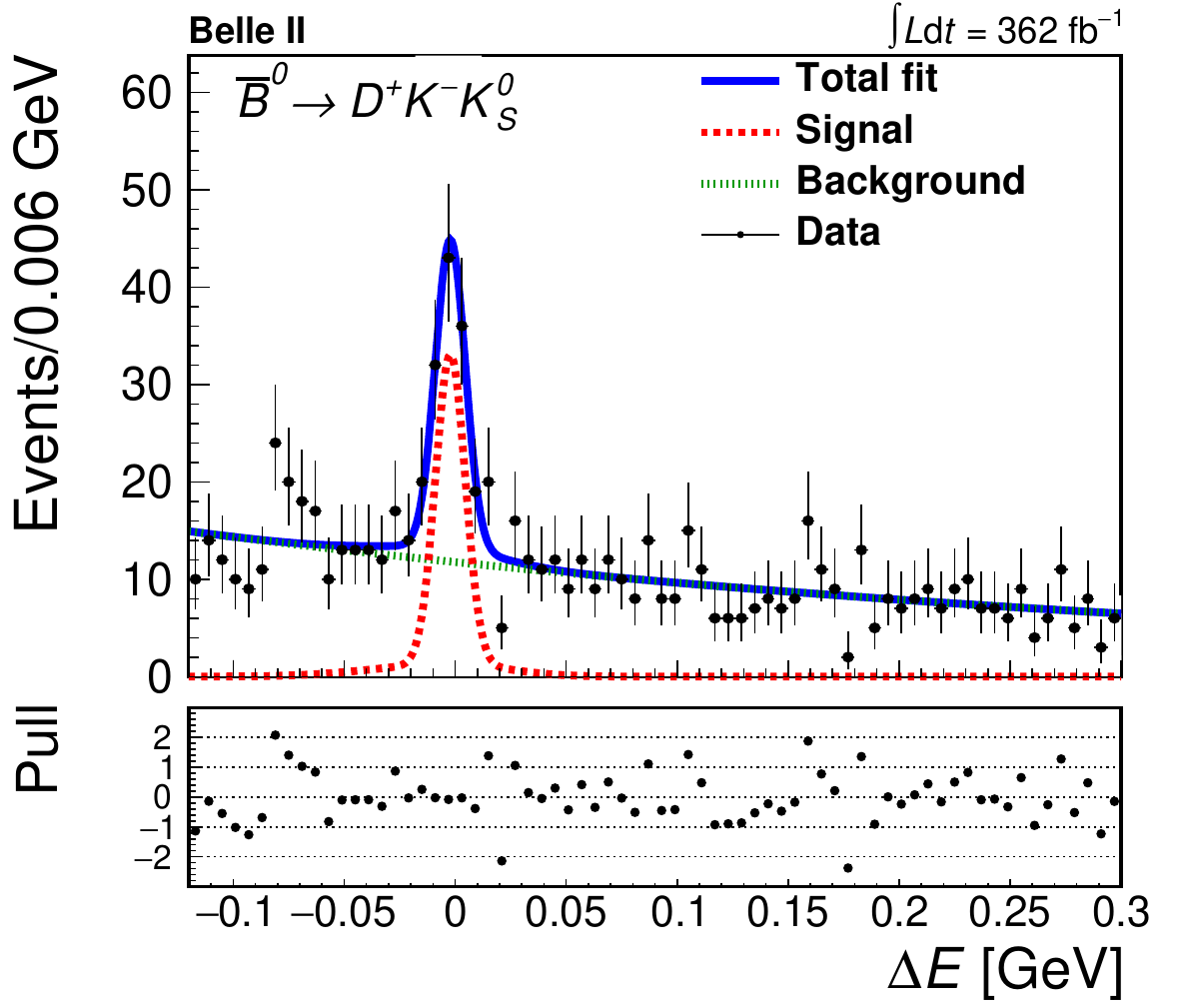}}
\subfigure{\includegraphics[width=0.45\columnwidth]{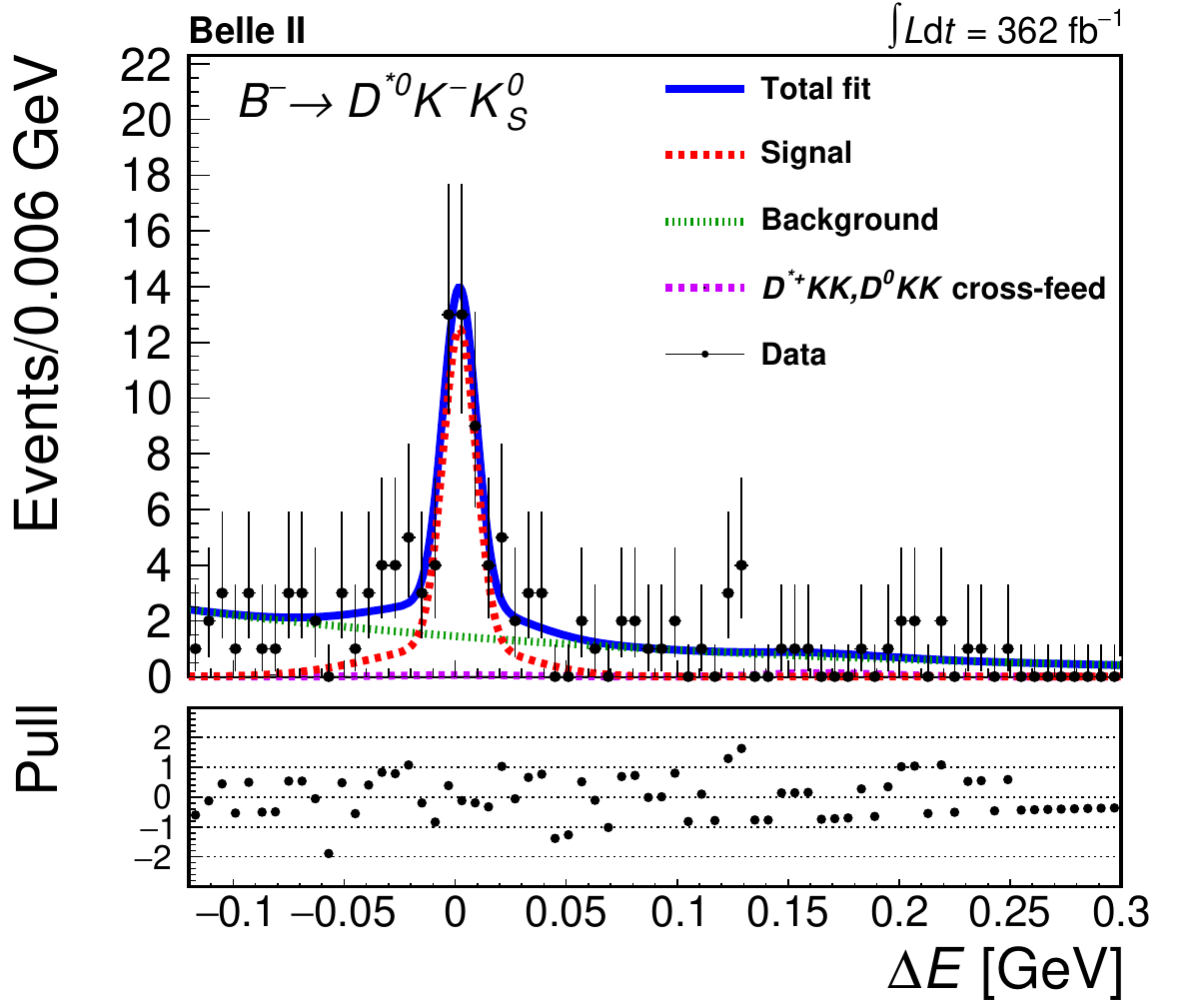}}
\subfigure{\includegraphics[width=0.45\columnwidth]{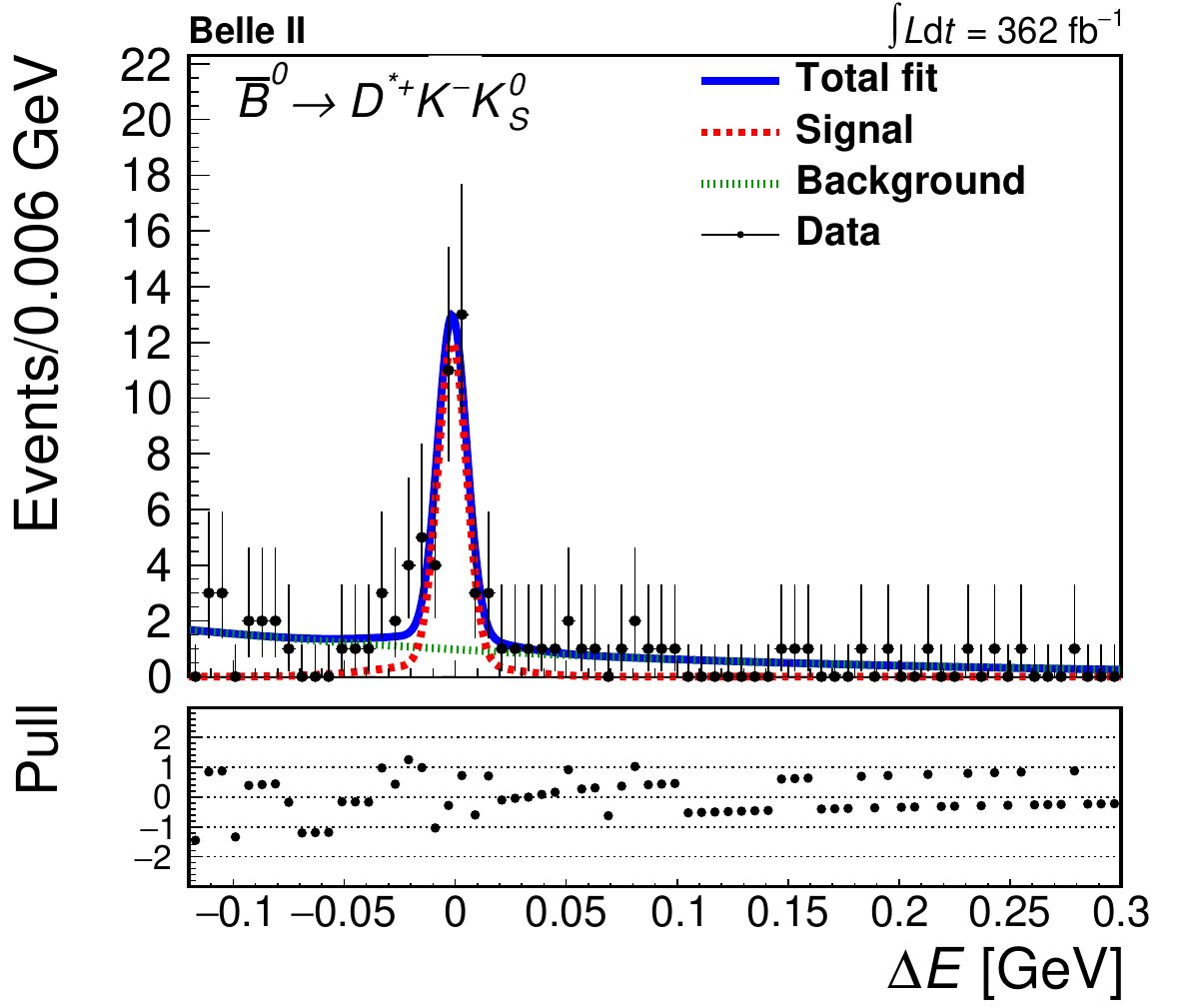}}
\caption{Distribution of $\Delta E$ for the $B^-\to D^0K^-K_S^0$ (top left), $\overline B{}^0\to D^+K^-K_S^0$ (top right), $B^-\to D^{*0}K^-K_S^0$ (bottom left),  and $\overline B{}^0\to D^{*+}K^-K_S^0$ (bottom right) channels. The projections of the fits are overlaid, the fit components are highlighted, and the pulls between the fit and the data are shown in the bottom panel of each plot.}  \label{fig:deltaE_fit_DKKS0}
\end{figure}

Data with fit projections overlaid are shown in Fig.~\ref{fig:deltaE_fit_DKKS0} {for the four $K_S^0$ channels}. 
The backgrounds are smooth and small as expected, with a signal-to-background ratio at $\Delta E\approx 0$ between 3 and 15. The cross-feed components in the $D^{*0}$ channel are well modeled. The four signals have statistical significances well above five standard deviations, calculated as $\sqrt{-2\ln(\mathcal L_0^\text{max}/\mathcal L^\text{max})}$, where $\mathcal L_0^\text{max}$ and $\mathcal L^\text{max}$ are the maximized likelihood {values} for {the} background-only and {the} signal-plus-background hypothesis, respectively. 
The pulls between the data distribution and the fit are also shown, defined as the difference between {the data yield and the fit value} divided by the data uncertainty. The agreement is {reasonable}. The yields are summarized in Table~\ref{tab:BR_data}.

\subsection{{$B\to DK^-K^{*0}$ channels}}

\begin{figure}[!h]
\centering
\subfigure{\includegraphics[width=0.45\columnwidth]{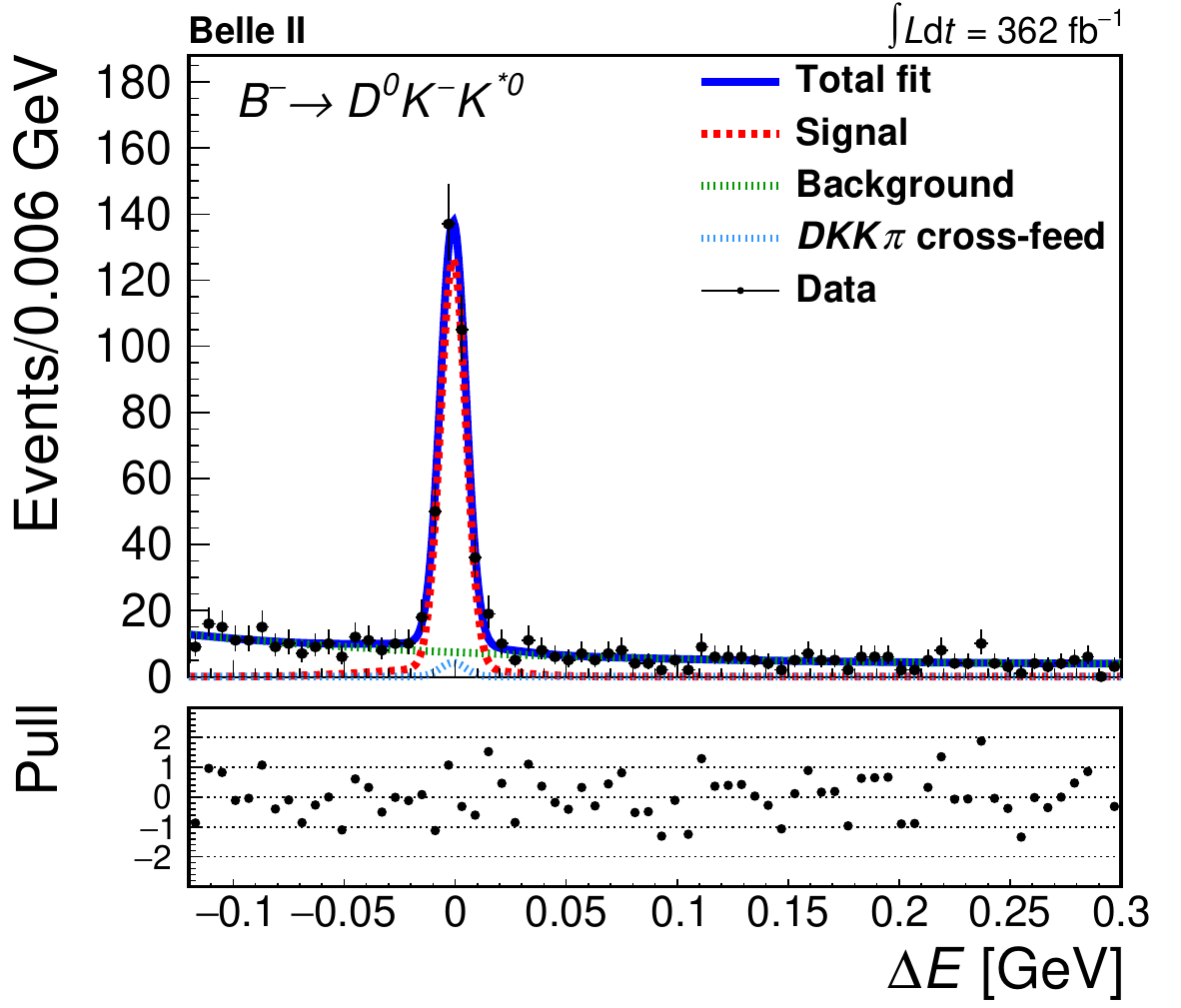}}
\subfigure{\includegraphics[width=0.45\columnwidth]{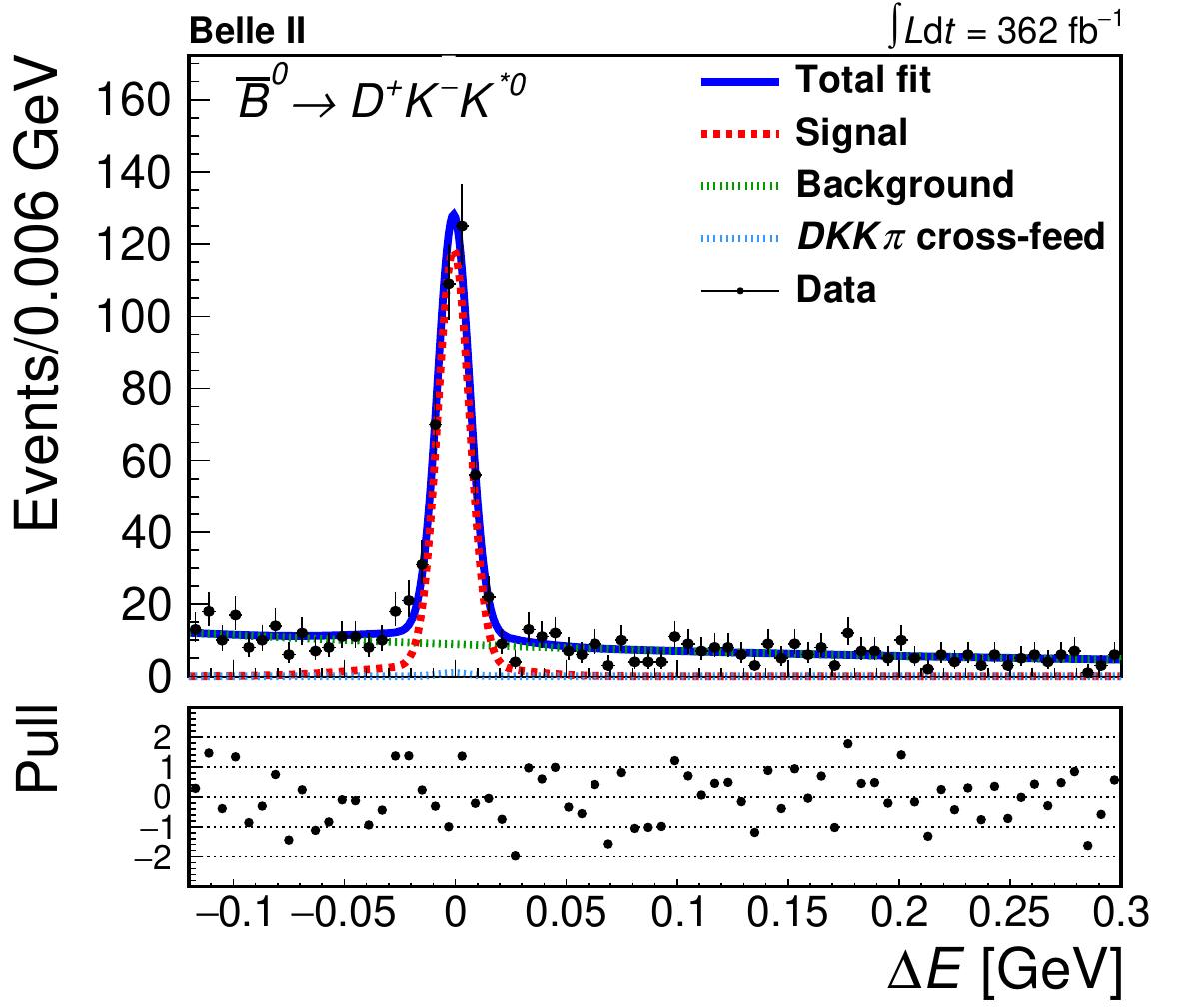}}
\subfigure{\includegraphics[width=0.45\columnwidth]{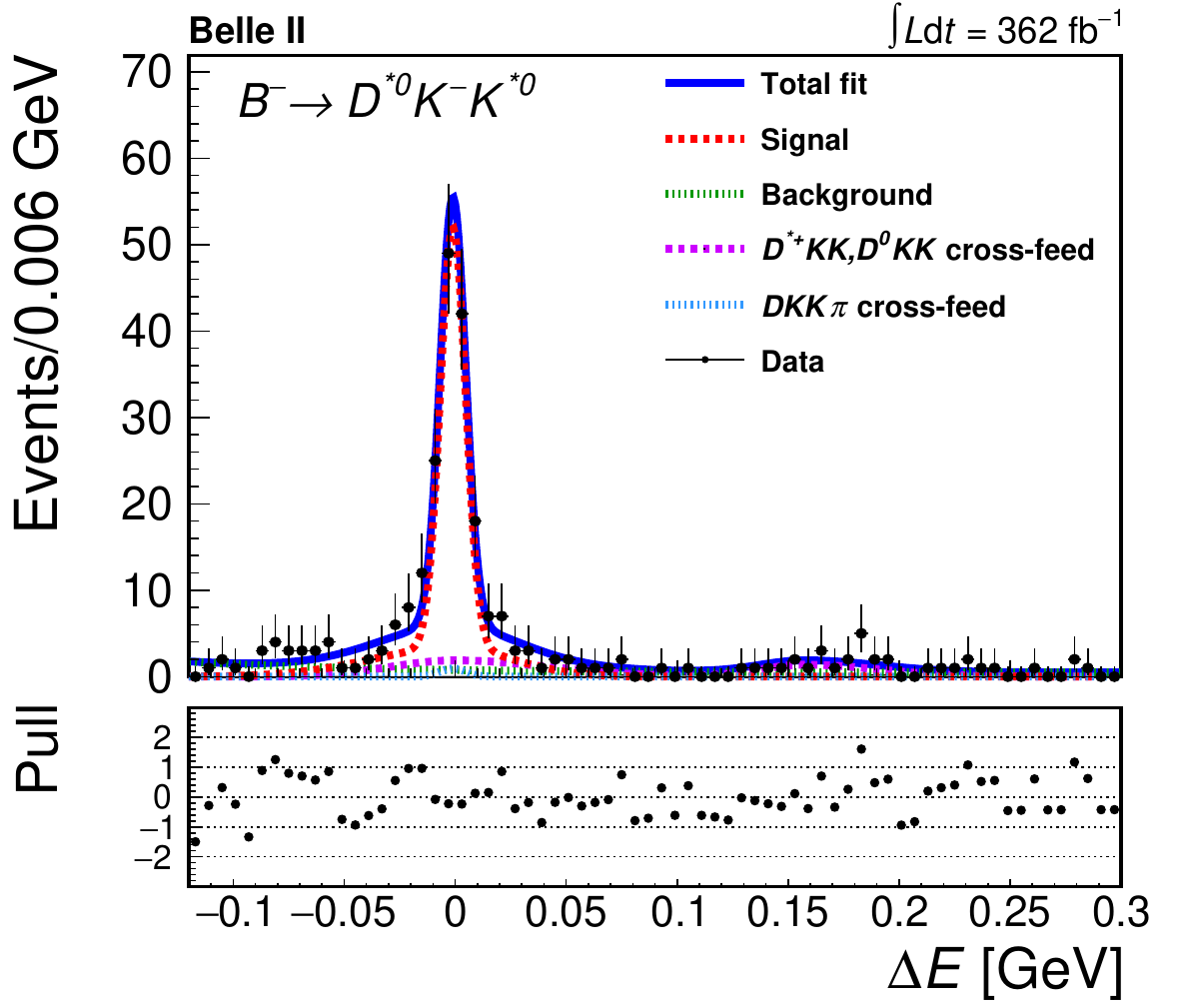}}
\subfigure{\includegraphics[width=0.45\columnwidth]{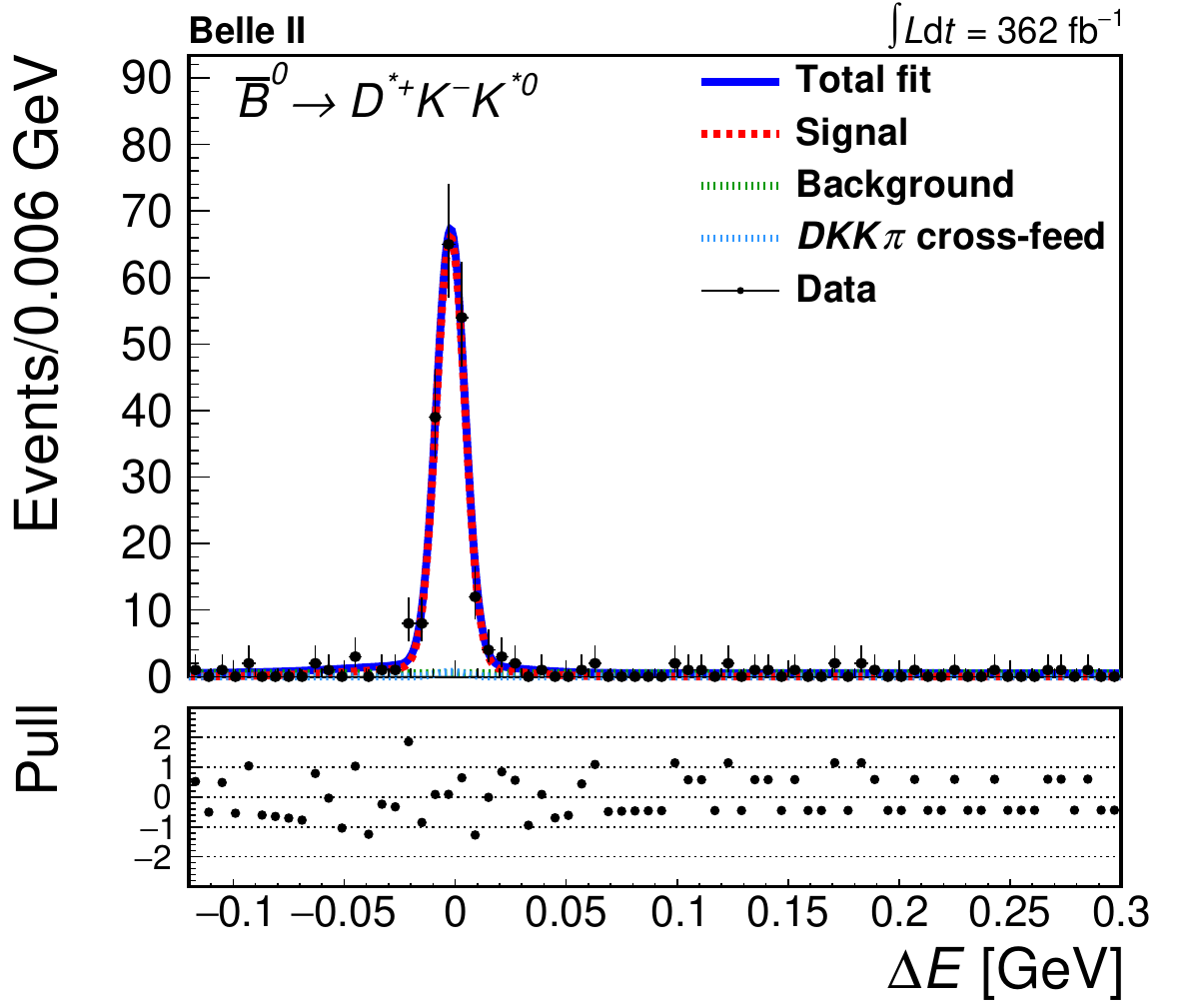}}
\caption{Distribution of $\Delta E$ for the $B^-\to D^0K^-K^{*0}$ (top left), $\overline B{}^0\to D^+K^-K^{*0}$ (top right), $B^-\to D^{*0}K^-K^{*0}$ (bottom left),  and $\overline B{}^0\to D^{*+}K^-K^{*0}$ (bottom right) channels. The projections of the fits are overlaid, the fit components are highlighted, and the pulls between the fit and the data are shown in the bottom panel of each plot.}  \label{fig:deltaE_fit_DKKst0}
\end{figure}

The {$B\to DK^-K^{*0}$} channels are fitted in $\Delta E$ with the same function as used for the $K_S^0$ channels, however the resolution scale factors are free parameters in the four channels.  An additional Gaussian function is added to each mode to describe the four body $B\to D K^- K^+\pi^-$ component. The mean and the width of the {additional} Gaussian {function} are fixed to that of the core {Gaussian function}. The {$B\to DK^-K^+\pi^-$ yield}  is constrained to be a fixed fraction of the signal yield. 
The fraction is estimated from a fit to the $m(K^+\pi^-)$ distribution. The latter is obtained by relaxing the $K^{*0}$ invariant mass selection, performing a fit to the resulting $\Delta E$ distribution with the same model {used to fit the $K_S^0$ channels}, and using the $s$Plot technique to obtain the continuum-background-subtracted $m(K^+\pi^-)$ distribution. The $m(K^+\pi^-)$ distribution is fitted in the range $0.65~\text{GeV}<m(K^+\pi^-)<2.5~\text{GeV}$, with the sum of a histogram template from the simulation for the signal component and a third-order Chebyshev polynomial for the $B\to DK^-K^+\pi^-$ component with parameters fixed from simulation. 
The [1.25 GeV,1.60 GeV] and [1.85 GeV,1.87 GeV] invariant mass regions are vetoed to exclude higher{-}mass kaon resonances and $B\to D^{(*)}D^{(*)}K$ decays, respectively. The  $m(K^+\pi^-)$ distributions with the projection of the fits overlaid are shown in Fig.~\ref{fig:mKpi} for the four channels. 
The non-$K^{*0}$-resonant fraction is obtained from the ratio between the $B\to DK^-K^+\pi^-$ and the $B\to DK^-K$ yields within the $K^{*0}$ signal region. The fractions are between 0.7\% and 3.1\% of the signal, depending on the channel. The fractions are {assumed to be} as uniform in the $\bigl(m(K^-K),m(DK)\bigr)$ plane, given their small values and large systematic uncertainties.  Data with fit projections overlaid {for $K^{*0}$ channels} and the pulls between {the} data distribution and the fit are shown in Fig.~\ref{fig:deltaE_fit_DKKst0}. {As} in the $K_S^0$ channels, the remaining backgrounds are small as expected, with signal-to-background ratios at $\Delta E\approx 0$ between 8 and 100. {All} the four signals have statistical significances well above five standard deviations. The yields are summarized in Table~\ref{tab:BR_data}.
\vspace{-0.2cm}
 
\begin{figure}[!b]
\centering
\subfigure{\includegraphics[width=0.45\columnwidth]{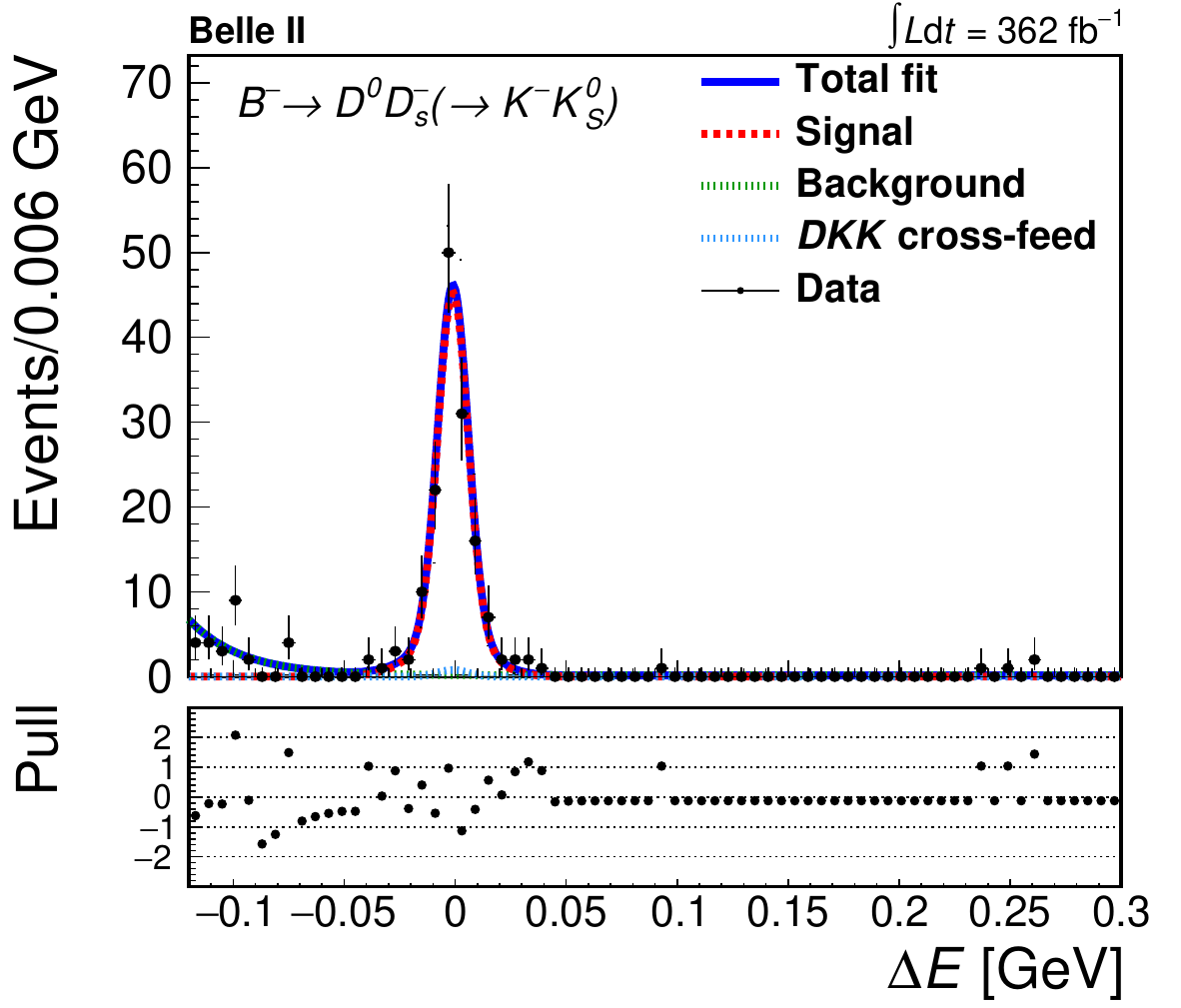}}
\subfigure{\includegraphics[width=0.45\columnwidth]{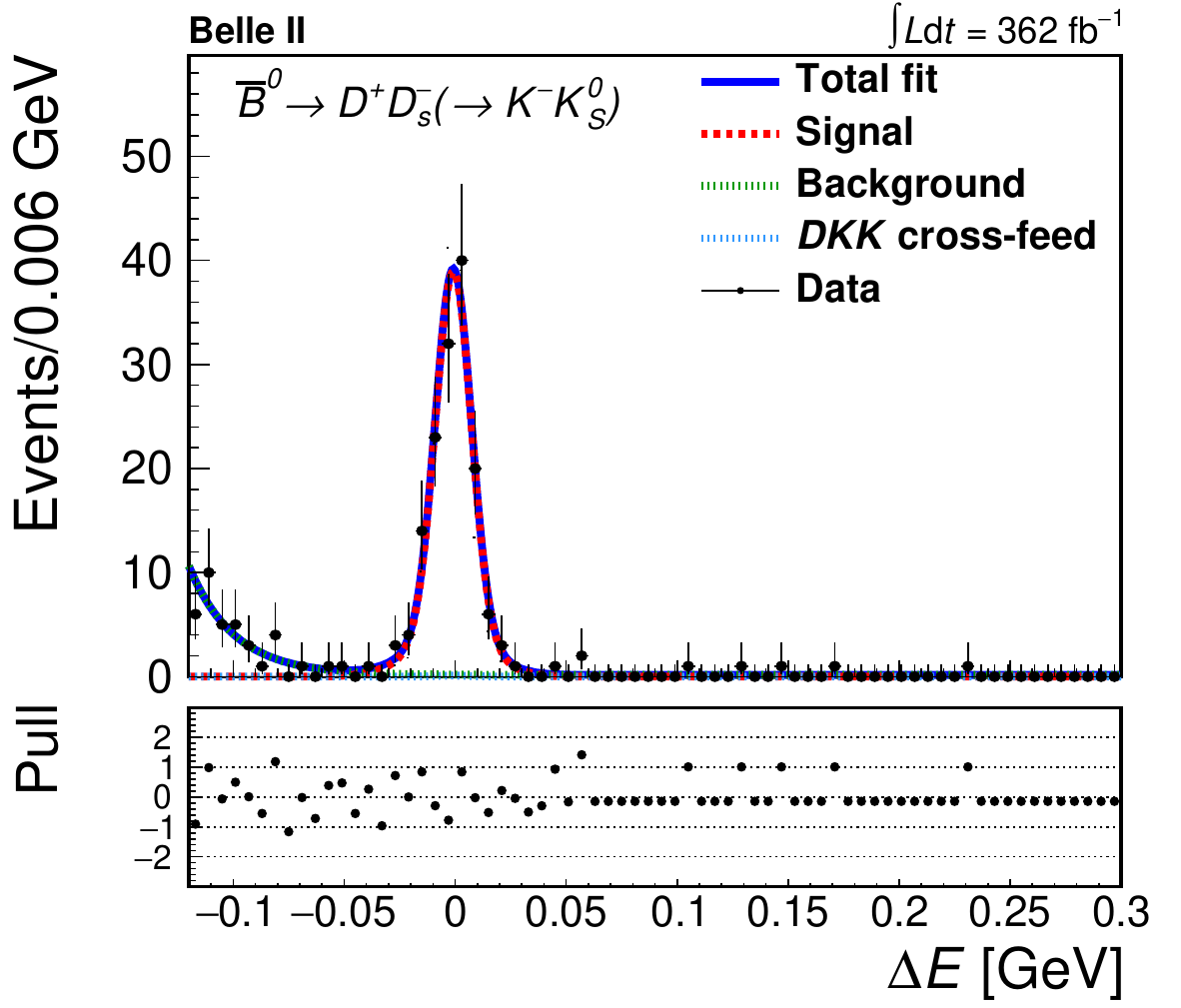}}
\subfigure{\includegraphics[width=0.45\columnwidth]{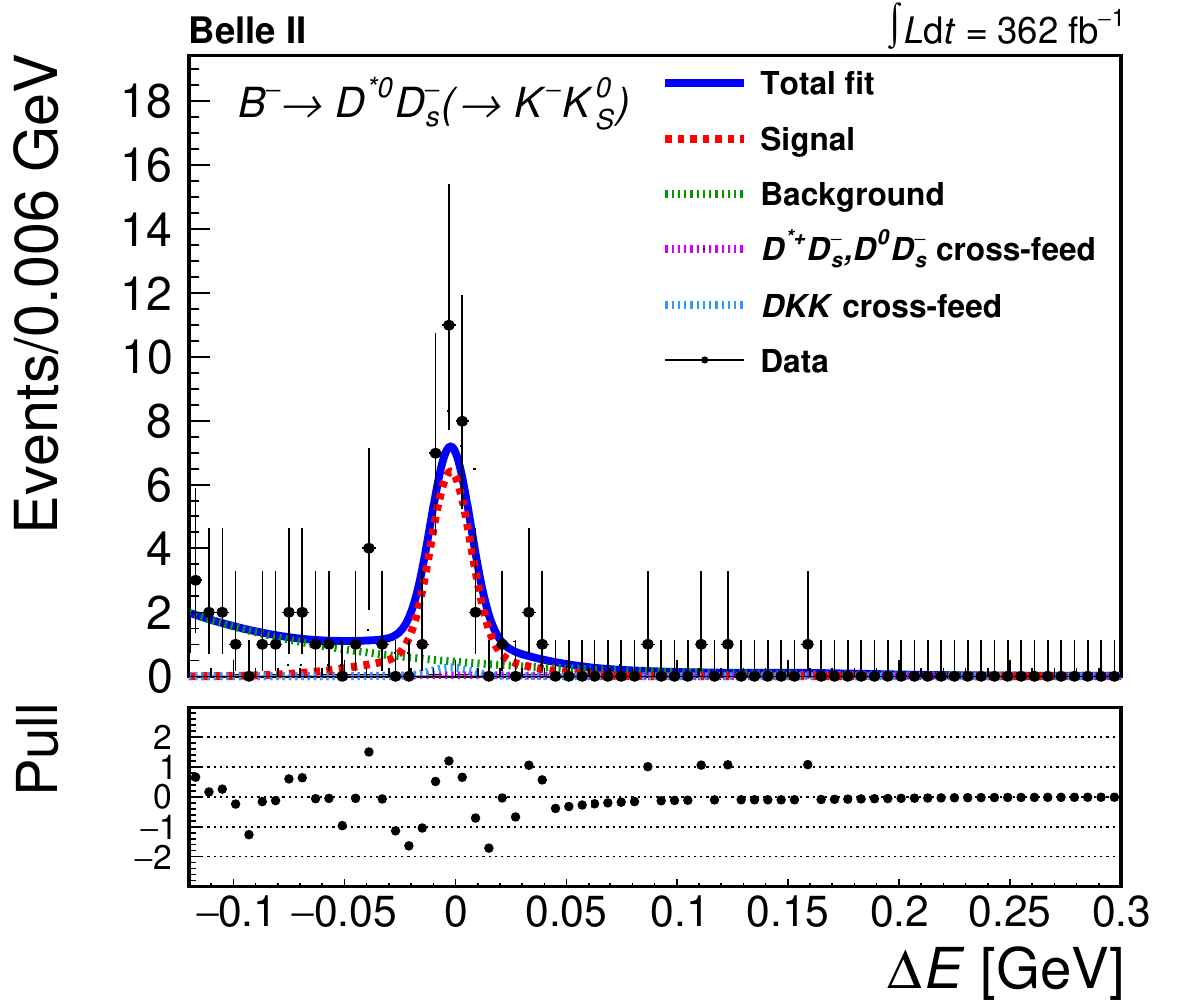}}
\subfigure{\includegraphics[width=0.45\columnwidth]{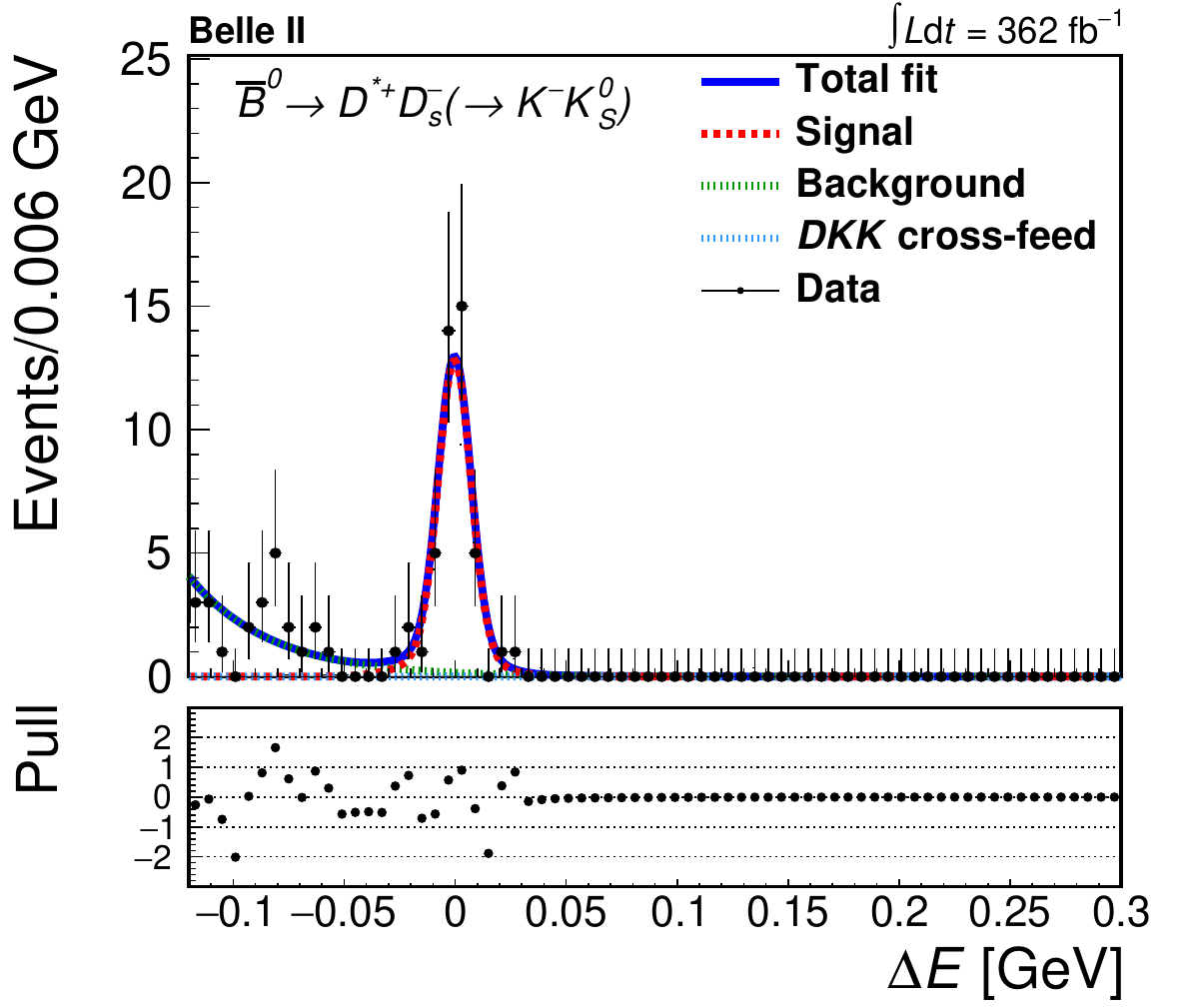}}
\caption{Distribution of $\Delta E$ for the $B^-\to D^0D_s^-(\to K^-K_S^0)$ (top left), $\overline B{}^0\to D^+D_s^-(\to K^-K_S^0)$ (top right), $B^-\to D^{*0}K^-K_S^0$ (bottom left),  and $\overline B{}^0\to D^{*+}D_s^-(\to K^-K_S^0)$ (bottom right) channels. The projections of the fits are overlaid, the fit components are highlighted, and the pulls between the fit and the data are shown in the bottom panel of each plot. }  \label{fig:deltaE_fit_DsD_KS0}
\end{figure}

\begin{figure}[!t]
\centering
\subfigure{\includegraphics[width=0.45\columnwidth]{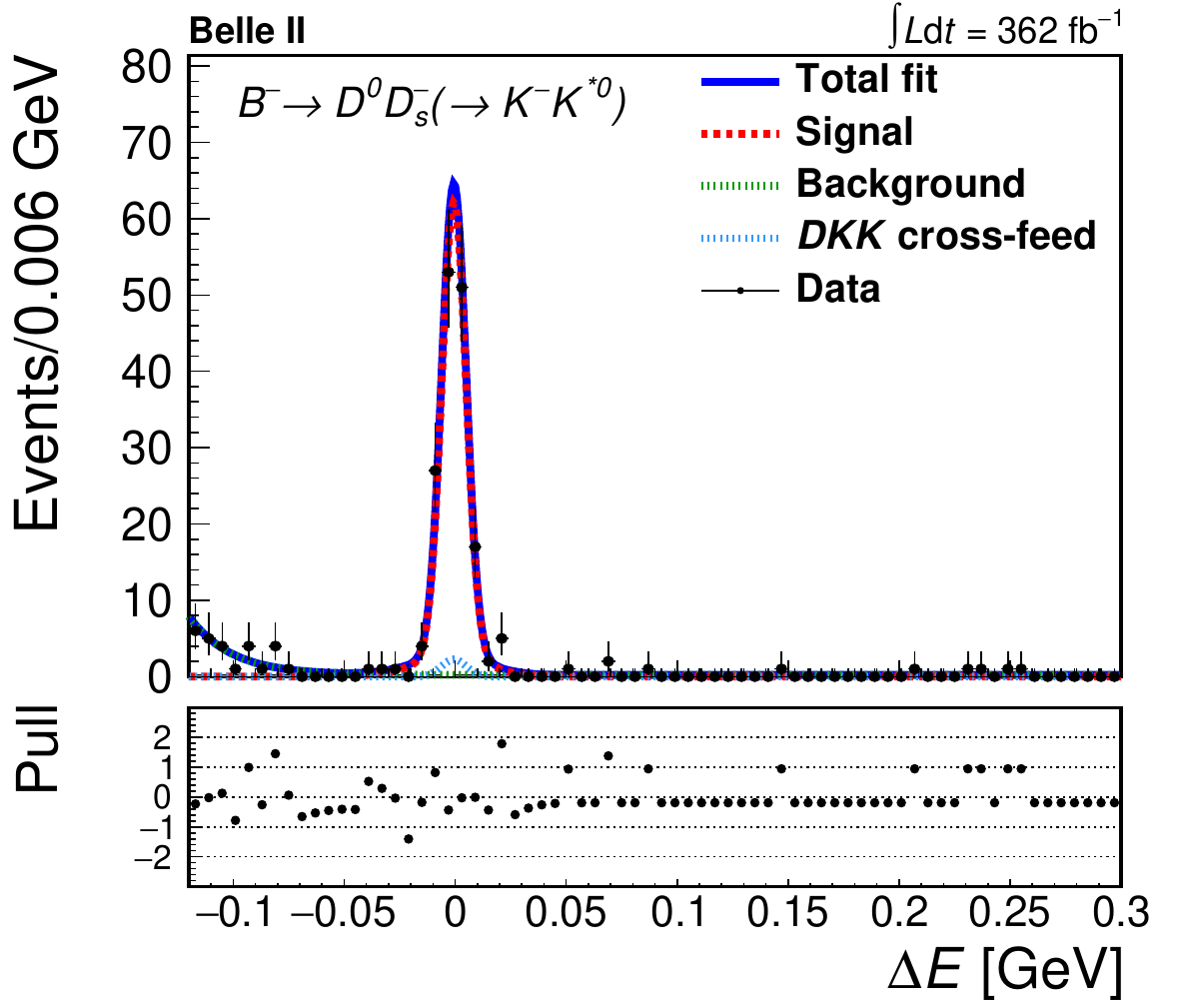}}
\subfigure{\includegraphics[width=0.45\columnwidth]{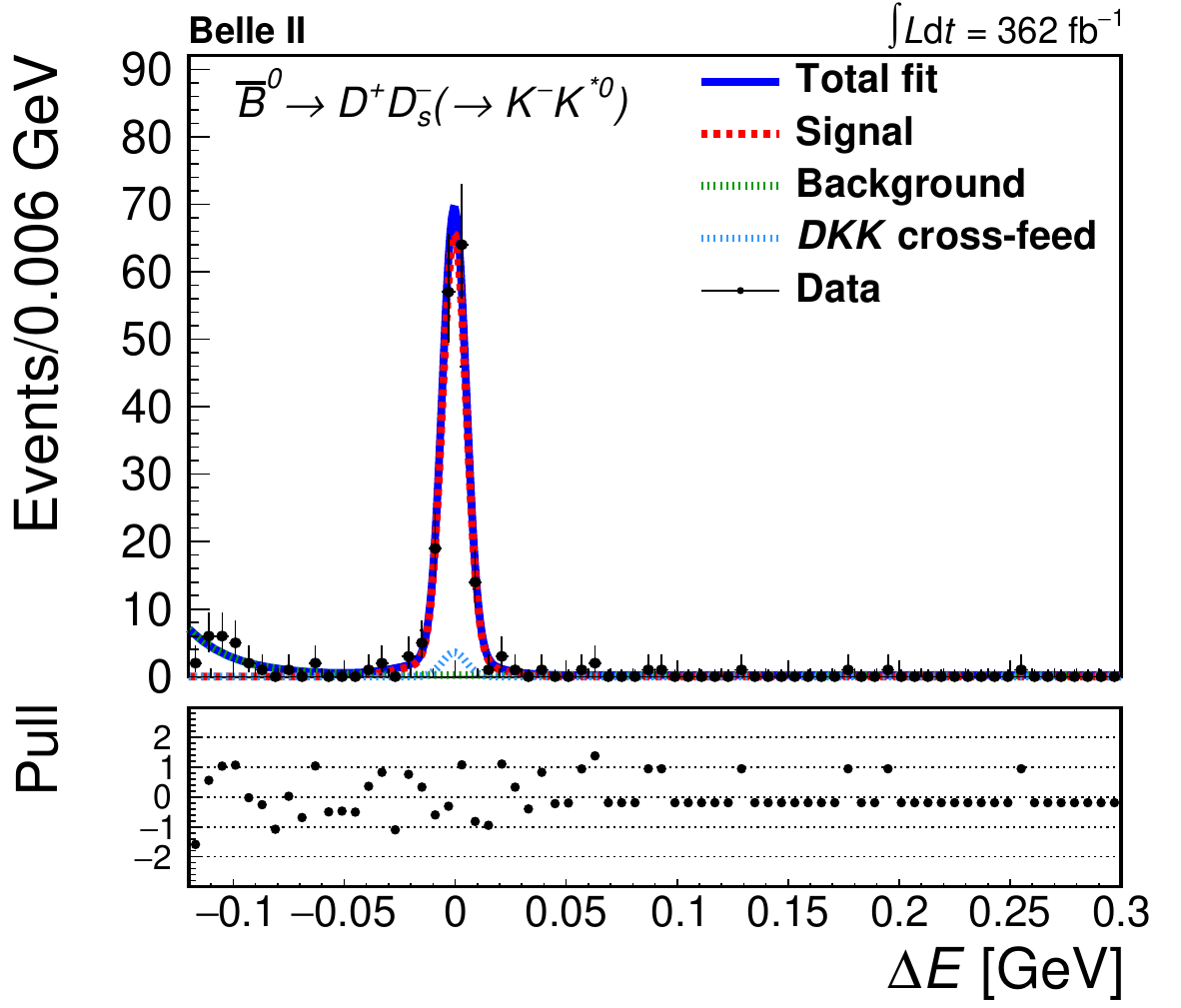}}
\subfigure{\includegraphics[width=0.45\columnwidth]{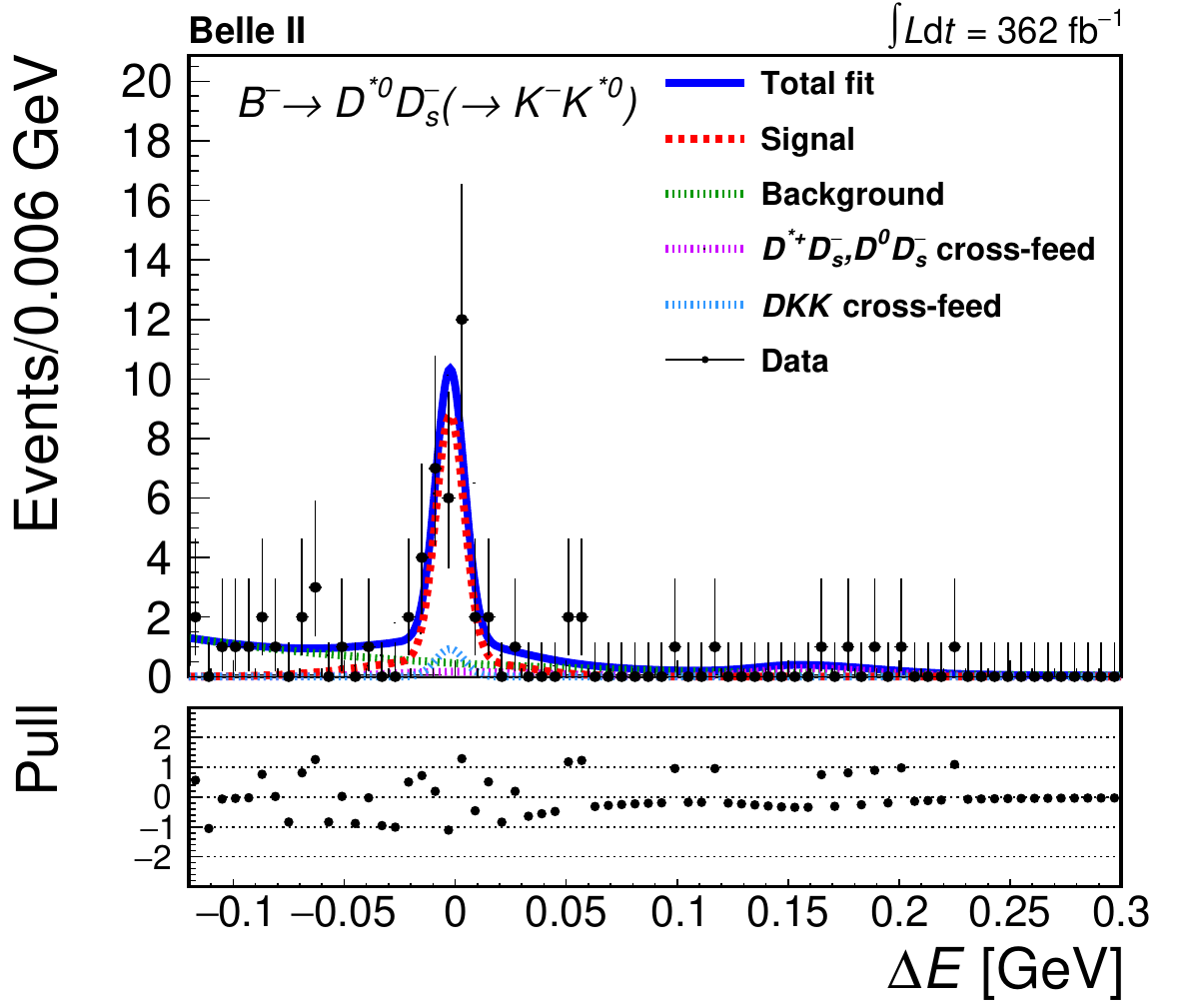}}
\subfigure{\includegraphics[width=0.45\columnwidth]{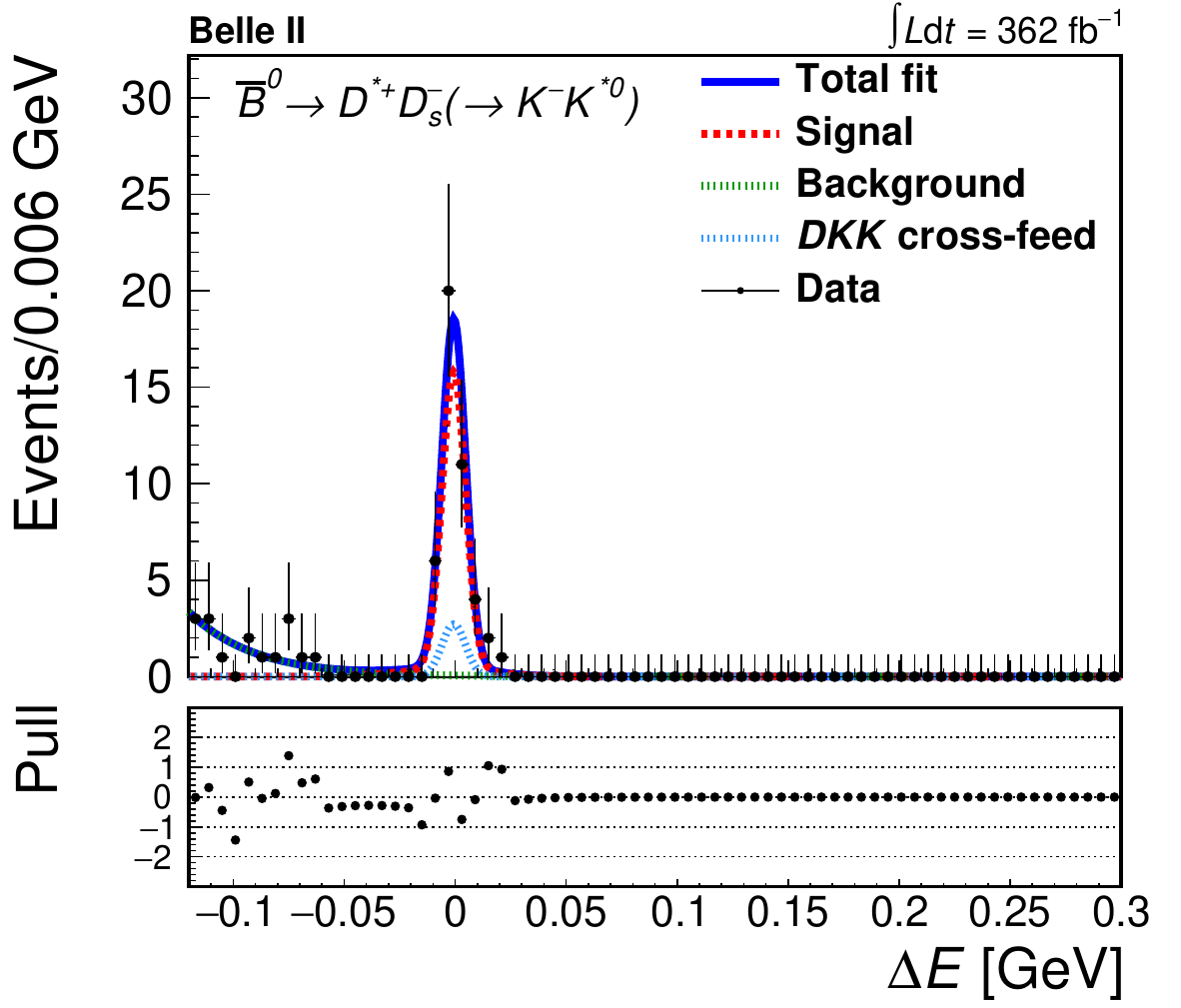}}
\caption{Distribution of $\Delta E$ for the $B^-\to D^0 D_s^-(\to K^-K^{*0})$ (top left), $\overline B{}^0\to D^+ D_s^-(\to K^-K^{*0})$, (top right), $B^-\to D^{*0}D_s^-(\to K^-K^{*0})$ (bottom left),  and $\overline B{}^0\to D^{*+} D_s^-(\to K^-K^{*0})$ (bottom right) channels. The projections of the fits are overlaid, the fit components are highlighted, and the pulls between the fit and the data are shown in the bottom panel of each plot. \vspace{-0.2cm}}  \label{fig:deltaE_fit_DsD_Kst0}
\vspace{-0.2cm}
\end{figure}
\vspace{-0.2cm}
\subsection{{$B\to DD_s^-$ channels}}
\vspace{-0.2cm}
The $B\to DD_s^-$ channels are fitted with the same strategy as the $B\to DK^-K$ channels. The fits are performed independently for {the} $D_s^-\to K^-K_S^0$ and $D_s^-\to K^-K^{*0}$ final states. The resolution scale factors are fixed in the $\overline B{}^0\to D^{*+}D_s^-$ and $ \overline B^-\to D^{*0}D_s^-$ fits  using the average of the resolution scale factors obtained from the $D^+$ and $D^0$ channels. The $D_s^-\to K^-K^+\pi^-$ yield is assumed to be zero in the  $D_s^-\to K^-K^{*0}$ channels, given the previous measurements of $D_s^-$ branching fractions~\cite{PDG}. The $B\to DK^-K$ background is described with an additional Gaussian component, with the same mean and width of the core Gaussian. The yield of the latter is fixed in the fit and is estimated from the $m(K^-K)$ distribution of the $B\to DK^-K$ channels measured in this work, assuming it uniform within the bin of interest.  Data with fit projections overlaid and the pulls between data and the fit are shown in Fig.~\ref{fig:deltaE_fit_DsD_KS0} and Fig.~\ref{fig:deltaE_fit_DsD_Kst0} for the four $K_S^0$ channels, and the four $K^{*0}$ channels, respectively. The eight channels are almost free from background, except for the cross-feed in the $D^{*0}$ channels and the $B\to DK^-K$ cross-feed, which are between {0.1\% and 13.0\%} of the signal. 

\vspace{0.4 cm}
The fits are validated by repeating the analysis on $10^3$ {Toy MC experiments}, i.e. {simplified} simulated pseudo-experiments {based on sampling the likelihood, {and show} no bias.}
The $B\to DK^-K$ fit is validated on data using the ${B\to DD_s^-(\to K^- K)}$ {decays} as control channels, which shares the same final state as the signal, and a very similar decay topology. Thus, the efficiency for the control channels is similar to that of $B\to DK^-K$ decays. In addition, the control {channels} are well measured~\cite{PDG}. 
The resulting control-channel branching fractions are in agreement with the world average{s}, validating the {signal extraction} on data. 

\section{Efficiency determination}\label{sec:efficiency}

The efficiency is obtained for each channel separately using signal simulation samples. An {admixture} of three-body $B\to DK^-K$  {decays} produced with a uniform distribution in phase-space, and $B\to DX^-$ events (where $X=a_1(1260)^-$ or $\rho(1450)^-$ for the $K^{*0}$ or $K_S^0$ channels, respectively) is used. 
We divide the $\bigl(m(K^-K), m(DK)\bigr)$ distribution into $20\times 20$ equally-spaced intervals (bins) with $0.9~\text{GeV}<m(K^-K)<3.5~\text{GeV}$ and $2~\text{GeV}<m(DK)<5~\text{GeV}$. 
The efficiency $\varepsilon\bigl(m(K^-K), m(DK)\bigr)$ is defined as the fraction of generated events that are reconstructed and selected in each bin of reconstructed $\bigl(m(K^-K), m(DK)\bigr)$. 
This allows the efficiency to be {much less dependent} {on} the $\bigl(m(K^-K), m(DK)\bigr)$ distribution of the data, which is a priori unknown and possibly different from the simulation. The reconstructed $\bigl(m(K^-K), m(DK)\bigr)$ spectrum is obtained by fitting the $\Delta E $ distribution of signal simulation, using the functional forms described in Sec.~\ref{sec:yield}. An $s$Plot is then used to obtain the $\bigl(m(K^-K), m(DK)\bigr)$ distribution. 

\begin{figure}[!b]
\vspace*{-0.3cm}
\centering
\subfigure{\includegraphics[width=0.45\columnwidth]{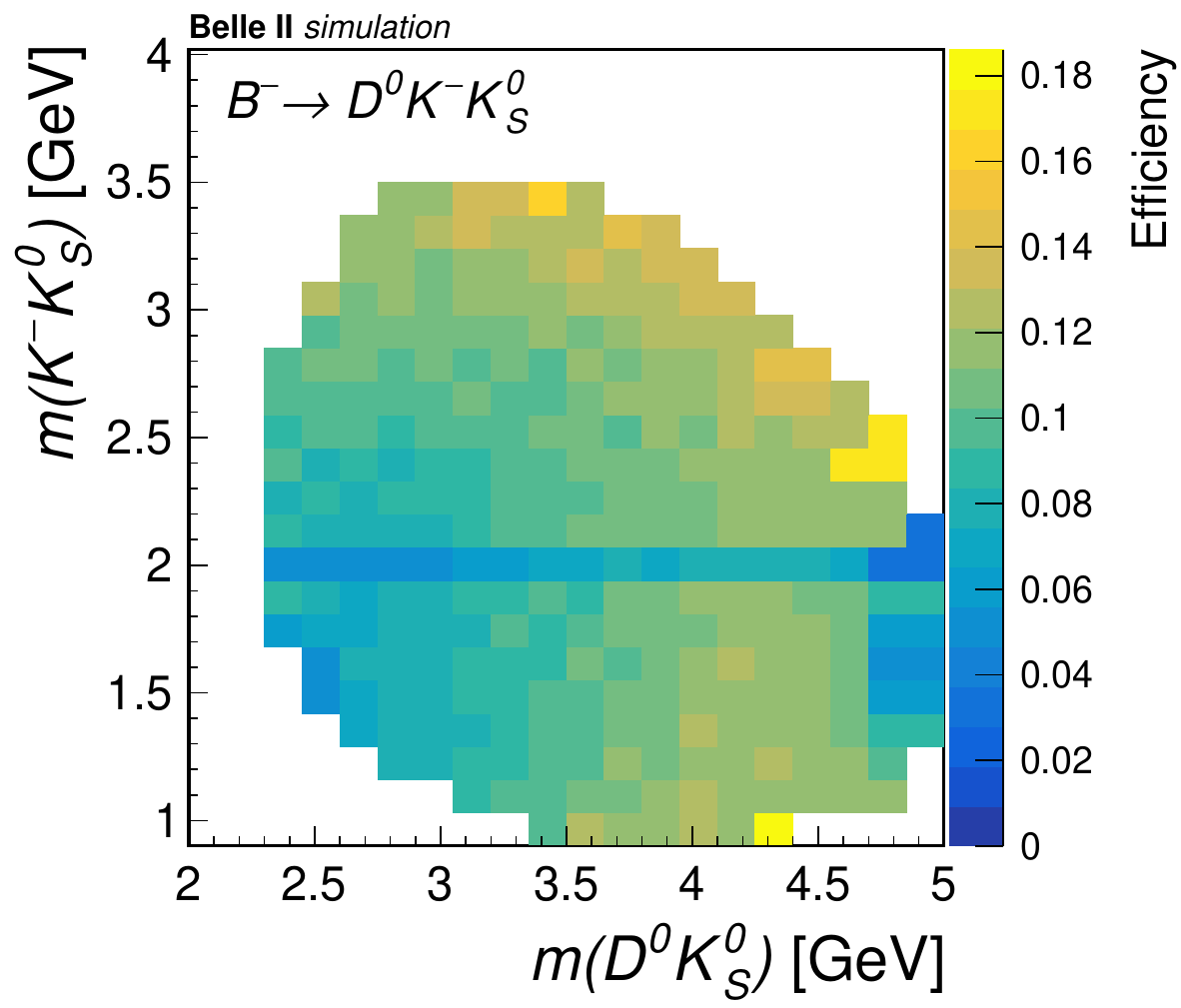}}
\subfigure{\includegraphics[width=0.45\columnwidth]{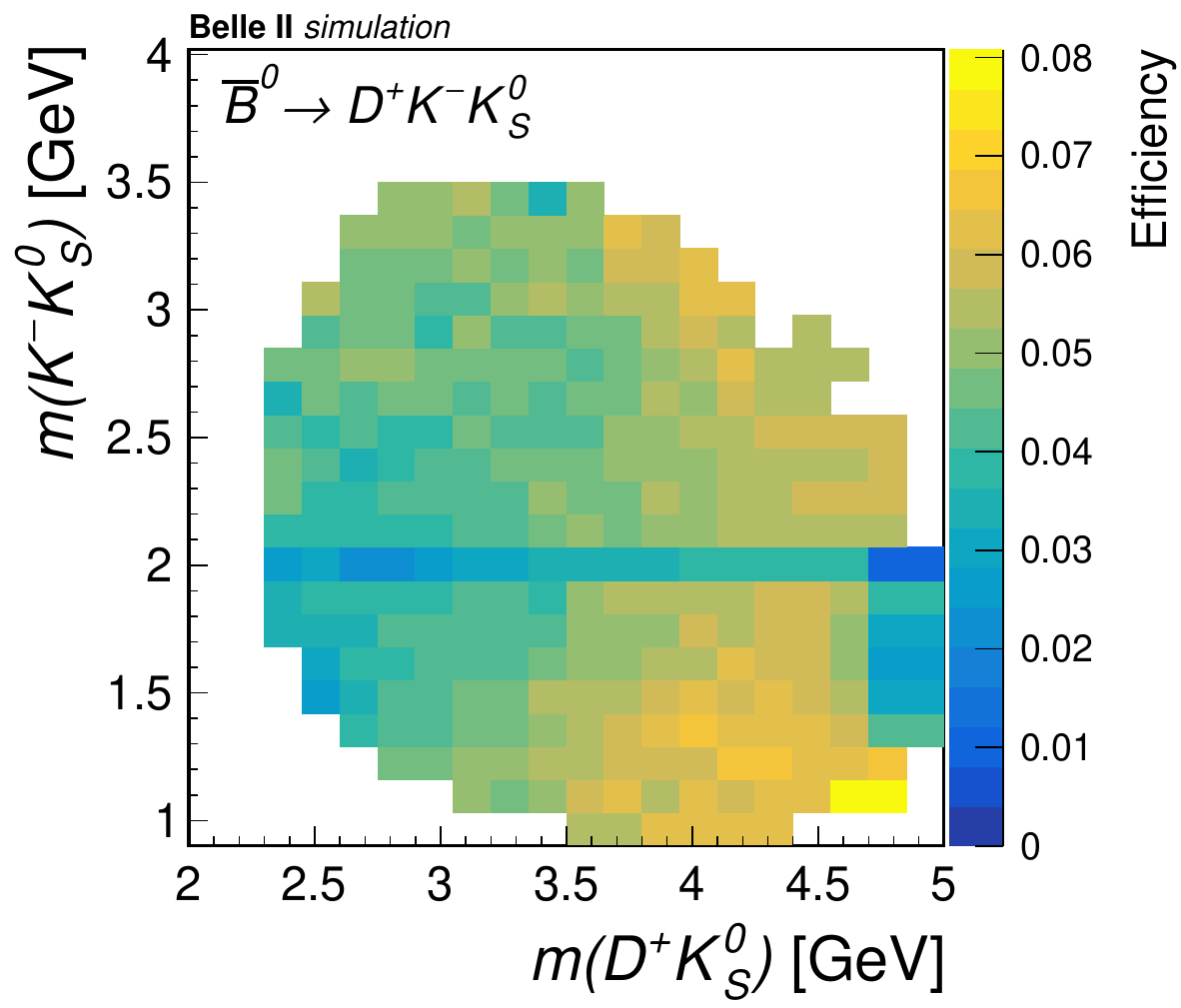}}
\subfigure{\includegraphics[width=0.45\columnwidth]{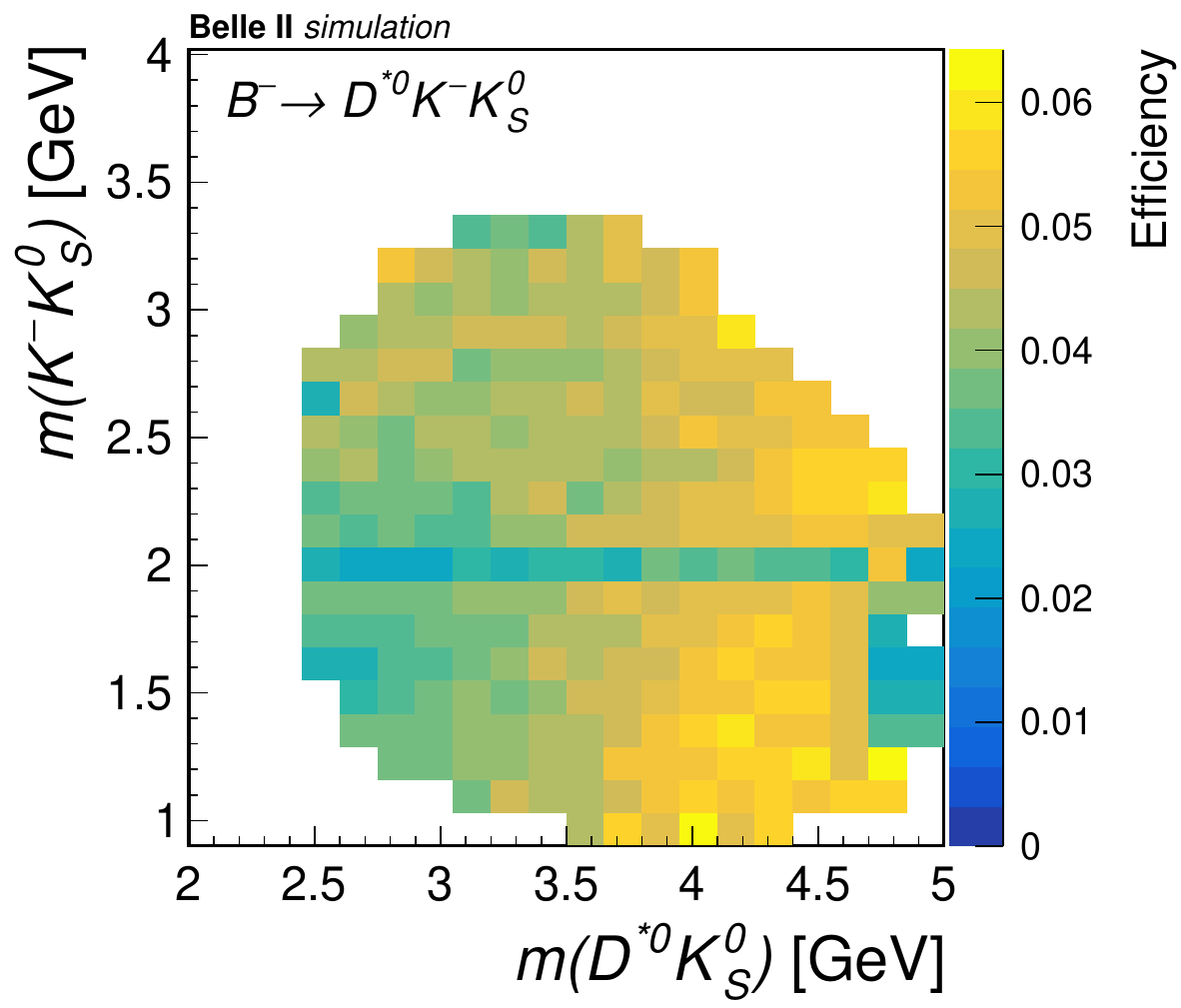}}
\subfigure{\includegraphics[width=0.45\columnwidth]{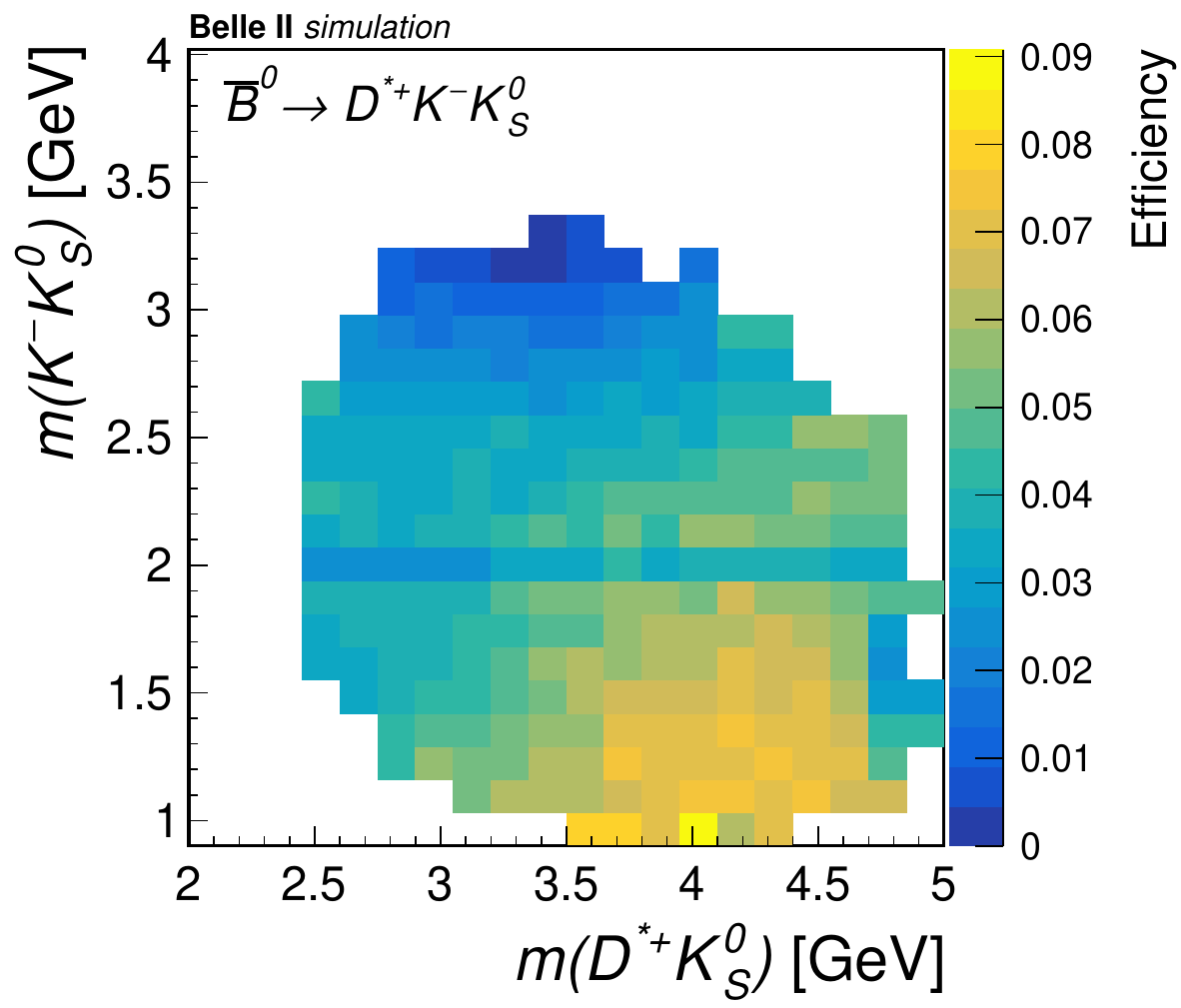}}
\caption{Efficiency as a function of $\bigl(m(K^-K),m(DK)\bigr)$ for the $B^-\to D^0K^-K_S^0$ (top left),  $\overline B{}^0\to D^+K^-K_S^0$ (top right), $B^-\to D^{*0}K^-K_S^0$ (bottom left), and $\overline B{}^0\to D^{*+}K^-K_S^0$ (bottom right) channels. }\label{fig:efficiency_differential_KS0}
\vspace*{-0.3cm}
\end{figure}

\begin{figure}[!t]
\vspace*{-0.3cm}
\centering
\subfigure{\includegraphics[width=0.45\columnwidth]{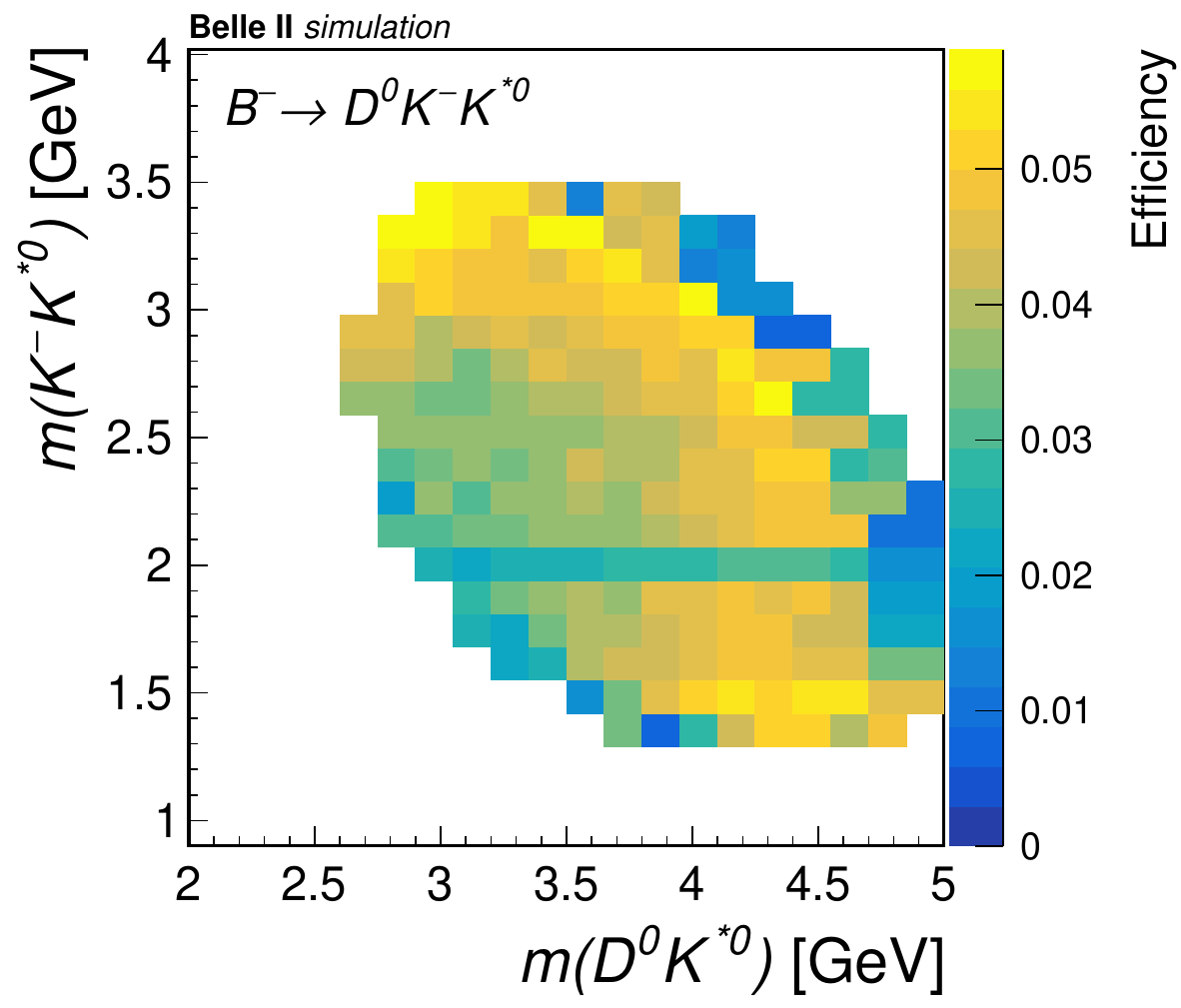}}
\subfigure{\includegraphics[width=0.45\columnwidth]{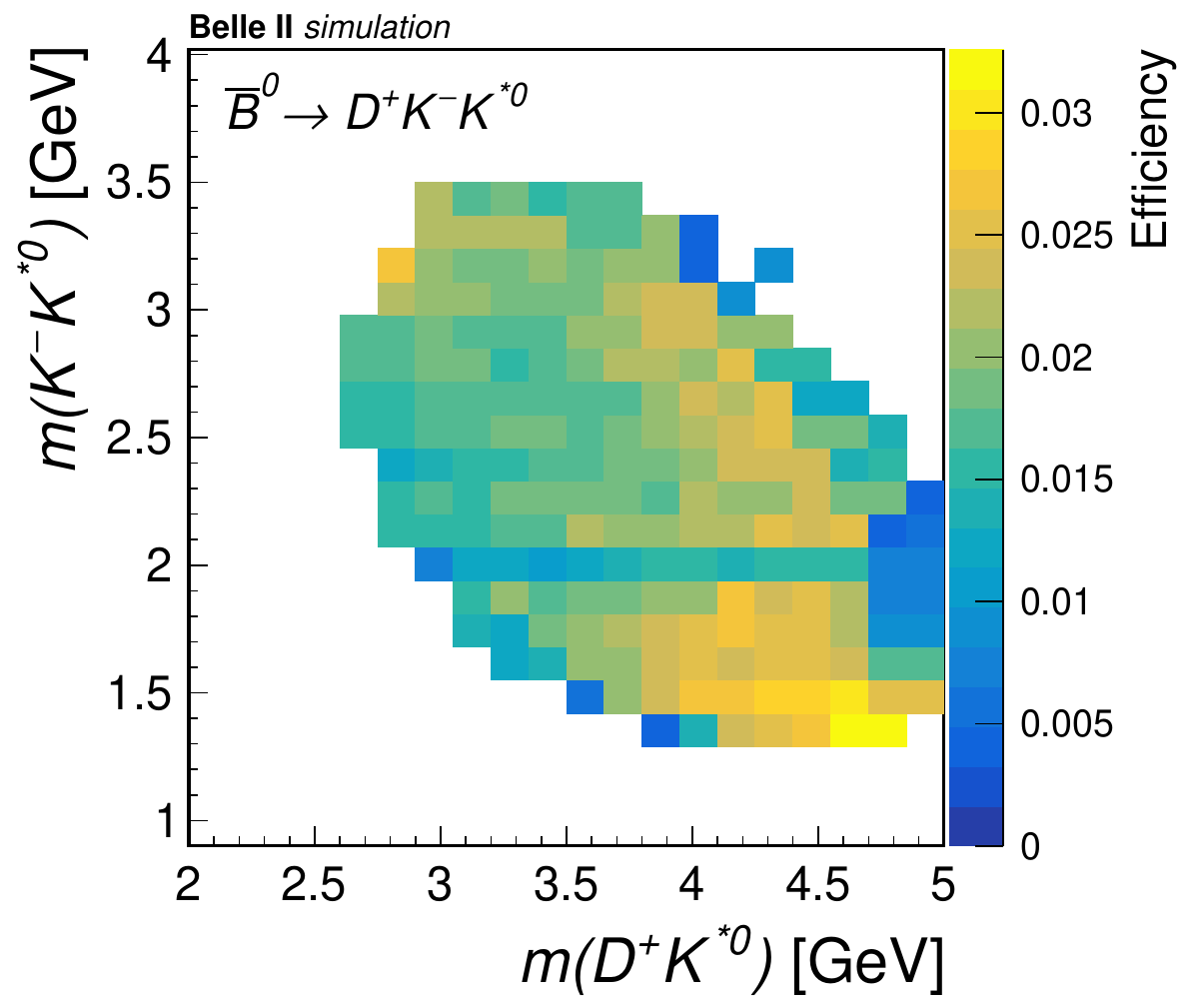}}
\subfigure{\includegraphics[width=0.45\columnwidth]{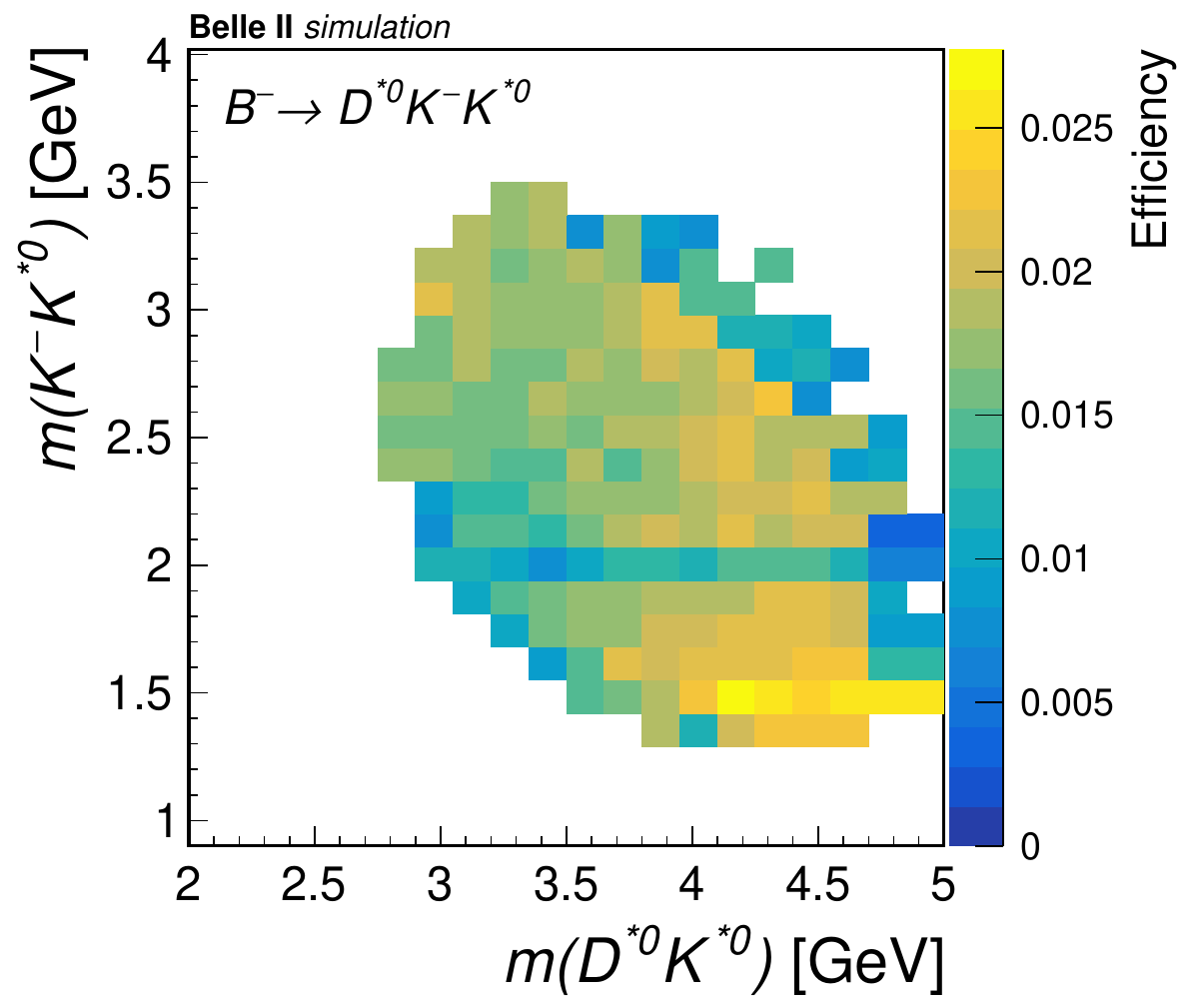}}
\subfigure{\includegraphics[width=0.45\columnwidth]{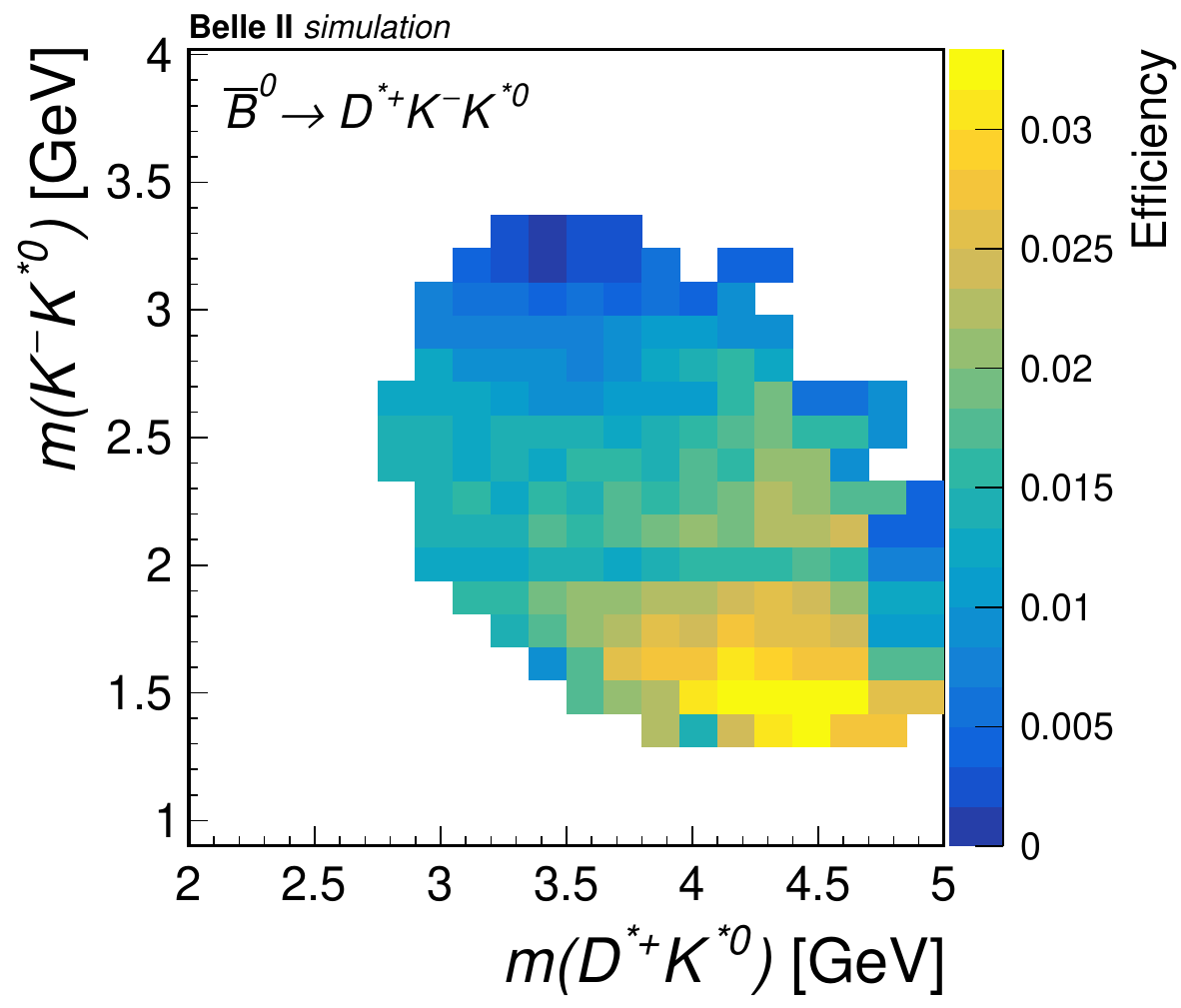}}
\caption{Efficiency as a function of $\bigl(m(K^-K),m(DK)\bigr)$ for the $B^-\to D^0K^-K^{*0}$ (top, left), $\overline B{}^0\to D^+K^-K^{*0}$ (top right), $B^-\to D^{*0}K^-K^{*0}$ (bottom left), and $\overline B{}^0\to D^{*+}K^-K^{*0}$ (bottom right) channels. }\label{fig:efficiency_differential_Kst0}
\vspace*{-0.3cm}
\end{figure}

The efficiency is corrected for known data-simulation mismodeling. In particular, the PID selection efficiency is calibrated with a scale factor as a function of momentum and polar angle for each track; the $K_S^0$ reconstruction and selection efficiency is calibrated with a scale factor as a function of the vertex separation from the IP, {and the}  momentum and polar angle of the $K_S^0$; the reconstruction and selection efficiencies of the low-momentum pions from the $D^{*+}$ and $D^{*0}$ decays are calibrated with two scale factors as function{s} of the pion momentum. The overall correction factor ranges between 1\% and 10\%.

The {resulting} efficiencies  $\varepsilon\bigl(m(K^-K), m(DK)\bigr)$ are shown in Fig.~\ref{fig:efficiency_differential_KS0} and Fig.~\ref{fig:efficiency_differential_Kst0} for the four $K_S^0$ channels and the four $K^{*0}$ channels, respectively. All channels show a drop in efficiency at {$m(K^-K)$ close to {the} $D_s^-$ mass} due to the ${B\to DD_s^-}$ veto. 
For the $D^{*+}$ channels, $\varepsilon\bigl(m(K^-K), m(DK)\bigr)$ decreases at high $m(K^-K)$ due to the {anticorrelation} between $m(K^-K^{0}_S)$ and the momentum of the pion from the $D^{*+}$ decay, for which the efficiency decreases at low momentum. These efficiencies have statistical uncertainties between 0.5\% and 5\%, {and increase up to 15\% at the edge of the phase-space}.  The allowed phase-space is properly populated using the admixture of simulated samples. The invariant mass resolutions are sufficiently small compared to the efficiency bin width to allow neighbour-bin migration effects to be neglected.

The {efficiencies} for the $B\to DD_s$ channels are defined as the fraction of the generated events that are reconstructed and selected from the $B\to DD_s(\to K^-K)$ signal simulation. The efficiencies, listed in Table~\ref{tab:BR_data}, are estimated separately for the $D_s^-\to K^-K_S^0$ and $D_s^-\to K^-K^{*0}$ final states. 

\section{Branching fraction extraction}\label{sec:BR}

For the $B\to DK^-K$ channels, the branching fraction can be expressed as 
\begin{equation}\label{eq:BR}
    \mathcal{B}=\frac{N_\text{reco}^{\varepsilon \text{ corr}}}{2 f_{+-,00} N_{B\overline B}  \mathcal{B}_{D^{(*)}}\mathcal{B}_{K_{(S)}^{(*)0}}}
\end{equation}
where $N_\text{reco}^{\varepsilon \text{ corr}}$ is the background-subtracted and efficiency-corrected signal yield, $\mathcal{B}_{D^{(*)}}\mathcal{B}_{K_{(S)}^{(*)0}}$ is the product of the  branching fractions of the {relevant intermediate} $D^{(*)}$ and $K_{(S)}^{(*)0}$ decays in the reconstructed decay chain, $N_{B\overline B}$ is the total number of $B\overline B$ pairs, and $f_{+-,00}$ is the fraction of charged or neutral $B\overline B$ pairs.  

The signal yield as a function of $\bigl(m(K^-K), m(DK)\bigr)$ is extracted using an $s$Plot technique. The $\Delta E$ distribution, fitted as described in Sec.~\ref{sec:yield}, is used as a discriminating variable to obtain the $s$Weights. The $s$Weights for the signal are then used to obtain the  $\bigl(m(K^-K), m(DK)\bigr)$ distribution for pure signal. The prerequisites for $s$Plot validity are satisfied as the $\Delta E$ distribution is independent of $m(K^-K)$ and $m(DK)$.  The efficiency correction is obtained by applying a $1/\varepsilon\bigl(m(K^-K),m(DK)\bigr)$ weight event-by-event, according to the efficiency map described in Sec.~\ref{sec:efficiency}. 

The $f_{+-,00}$ values used in Eq.~\eqref{eq:BR} are evaluated from the ratio ${f_{+-}/f_{00}}$ from Ref.~\cite{Belle:fpm00},
while ${N_{B\overline B}}$ is reported in Sec.~\ref{sec:Belle2}. 
We obtain $2f_{\pm} N_{B\overline B}=(399\pm 11)\times 10^6$ and $2f_{0} N_{B\overline B}=(375\pm 10)\times 10^6$. The branching fractions of the intermediate $D^{(*)}$ and $K_S^0$ decays are taken from the world-average values in Ref.~\cite{PDG}. 

The $B\to DD_s^-$ branching fractions are first {determined} independently for {the} $D_s^-\to K^-K_S^0$ and $D_s^-\to K^-K^{*0}$ sub-channels using Eq.~\eqref{eq:BR}, where {the product of the branching fractions is replaced by} $\mathcal{B}_{D^{(*)}}\mathcal{B}_{D_s^-}\mathcal{B}_{K_{(S)}^{(*)0}}$ and $N_\text{reco}^{\varepsilon \text{ corr}}= N_\text{reco}^K/\varepsilon_K$, where $N_\text{reco}^K$ and $\varepsilon_K$ are the reconstructed yield and the efficiency in the specific sub-channel. 
For each of the four channels, the two $D_s^-$ sub-channels give compatible results. The $B\to DD_s^-$ branching fractions are then obtained from the weighted average of {the results from} the two $D_s^-$ sub-channels, {including the statistical uncertainty only in the weights}.  

The branching fractions are given in Table~\ref{tab:BR_data} for each decay channel.

\begin{table}[!htb]
\small
\caption{ \small{Observed yields for the eight $B\to DK^-K$ channels and the four $B\to DD_s^-$ channels and their statistical uncertainty; average efficiency with the correction described in Sec.~\ref{sec:efficiency}; measured branching fractions with statistical (first)  and systematic (second) uncertainties; statistical significances. The first/second value for yield, efficiency, and significance of $B\to DD_s^-$ channels refer to $K_S^0/K^{*0}$ sub-channels}}\label{tab:BR_data} 
\begin{tabularx}{1.0\linewidth}{l c c c c}
\toprule
 Channel                               & Yield         & Average $\varepsilon$  & $\mathcal{B}$ [$10^{-4}$]  & {Stat. significance} [$\sigma$]      \\ 
\midrule
\footnotesize{$B^- \to D^0 K^- K^0_S$}                         & $209 \pm 17$  & $0.098 $                 & $1.82 \pm 0.16 \pm 0.08$    & {$>10$}\\ 
\footnotesize{$\overline B{}^0 \to D^{+} K^- K^{0}_S$}           & $105 \pm 14$ & $0.048 $                & $0.82 \pm 0.12 \pm 0.05$    &  10             \\
\footnotesize{$B^- \to D^{*0} K^- K^0_S$}                      & $51 \pm 9$  & $0.044 $                & $1.47 \pm 0.27 \pm 0.10$       &  8         \\
\footnotesize{$\overline B{}^0 \to D^{*+} K^- K^0_S$}            & $36 \pm 7$  & $0.046 $                & $0.91 \pm 0.19 \pm 0.05$     &  9               \\ 
\footnotesize{$ B^{-}\rightarrow D^{0}K^{-}K^{*0}$}            & $325\pm19$  & $0.043$         & $7.19\pm0.45\pm0.33$   &  {$>10$}\\ 
\footnotesize{$ \overline B{}^{0}\rightarrow D^{+}K^{-}K^{*0}$}       & $385\pm 22$  & $0.021$        & $7.56\pm0.45\pm0.38$  & {$>10$}\\ 
\footnotesize{$ B^{-}\rightarrow D^{*0}K^{-}K^{*0}$}           & $160\pm 15$  & $0.019$        & $11.93\pm1.14\pm0.93$  & {$>10$}\\ 
\footnotesize{$ \overline B{}^{0}\rightarrow D^{*+}K^{-}K^{*0}$}      & $193\pm 14$ & $0.020$         & $13.12\pm1.21\pm0.71$  & {$>10$}\\ 
\midrule
\footnotesize{$B^- \to D^0 D_s^-$}                       &   $ 144 \pm 12  \,\,/\,\,153 \pm 13 $     & $0.09\,\,/\,\,0.04  $    & $95 \pm 6 \pm 5$ & {$>10\,\,/\,\,>10$} \\ 
\footnotesize{$\overline B{}^0 \to D^{+} D_s^-$ }          &   $  145 \pm 12 \,\,/\,\, 159 \pm 13 $    & $0.05\,\,/\,\,0.02  $    & $89 \pm 5 \pm 5$  & {$>10\,\,/\,\,>10$} \\ 
\footnotesize{$B^- \to D^{*0} D_s^-$}                    &   $ 30\pm 6  \,\,/\,\,29 \pm 7  $   & $0.04\,\,/\,\,0.02 $    & $65 \pm 10 \pm 6$   &  $7\,\,/\,\,8 $                                \\
\footnotesize{$\overline B{}^0 \to D^{*+}D_s^-$}           &   $  43 \pm 7 \,\,/\,\,37 \pm 7$      & $0.04 \,\,/\,\,0.02  $    & $83 \pm 10 \pm 6$  &   {$>10\,\,/\,\,>10$} \\ 
\bottomrule
\end{tabularx}
\end{table}

\section{Systematic uncertainties}\label{sec:syst}

The  contributions to the systematic uncertainties for each $B\to DK^-K$ channel are summarized in Table~\ref{tab:systematicsRel}, expressed as relative uncertainties {to the corresponding branching fractions.}

The first group of systematic uncertainties affects the efficiency estimation. They are related to the difference{s} between efficiencies in data and in simulation, and are derived from the uncertainties on the corresponding corrections. 
An uncertainty associated with the limited size of the MC simulation sample is included (``Eff - MC sample size'' in Table~\ref{tab:systematicsRel}), which gives the statistical uncertainty in the efficiency determination. 
The uncertainty related to the tracking efficiency (``Eff - tracking'') is estimated using a $e^+e^-\to \tau^+\tau^-$ data control sample.  No correction is applied, but a per-track uncertainty of 0.24\% is included assuming full correlation between the tracks. 
For the uncertainty associated with the  ${K_S^0}$ efficiency (``Eff - $K_S^0$''), the nominal scale factor is varied by its uncertainty. The uncertainty is assumed to be fully correlated in the vertex separation from the IP, momentum and polar angle space. The scale factors and the uncertainties are determined in $D^{*+}\to D{}^0(\to K_S^0\pi^+\pi^-) \pi^+$ decays. 
{A similar approach is adopted} for the uncertainty associated with the PID-efficiency correction (``Eff - PID''). The uncertainty is assumed to be fully correlated between tracks and as a function of momentum, and polar angle of the tracks. 
The scale factors and the associated uncertainty are evaluated using $D^{*+}\to \overline D{}^0(\to K^-\pi^+) \pi^+$ and $K_S^0\to \pi^+\pi^-$ data control samples. 
For the uncertainty associated with the {efficiency for reconstructing the low-momentum charged pion}  (``Eff - $\pi^+$ from $D^{*+}$''), the scale factors are evaluated with statistical (partially correlated in momentum) and systematic (correlated) uncertainties estimated on a $B{}^0\to D^{*-}\pi^+$ data control sample. A fully correlated 2.7\% uncertainty is used for the scale factors, equal to the sum in quadrature of both uncertainties. This uncertainty affects only the $D^{*+}$ channels. 
For the uncertainty associated with the ${\pi^0}$-efficiency correction (``Eff - $\pi^0$'') {in the $D^{*0}$ channels}, the scale factor is varied according to the uncertainties evaluated on a $B^+\to D^{*0}\pi^+$ data control sample.

A systematic uncertainty is assigned to the efficiency (``Eff - modeling'') due to the possible mismodeling of the efficiency distribution in the $\bigl(m(K^-K),m(DK)\bigr)$ plane and the integration over additional degrees of freedom beyond the two adopted dimensions, due to polarization of vector particles ($D^*$, $K^*$, and possible vector resonances). 
The systematic uncertainty is determined by validating the efficiency map correcting distributions generated with different resonant and polarization schemes ($B\to D\rho^\prime, B\to Da_1^-,\dots)$.

The signal model used in {the} $\Delta E$ fit is modified to check its robustness.  Alternative branching fractions are evaluated, and a systematic uncertainty is quoted as the difference between the nominal and the alternative branching fraction (``Signal model''). 
A variation is considered for the channels in which the resolution scale factors are fixed using auxiliary channels. The corresponding systematic uncertainty is evaluated by shifting the resolution scale factor{s within their uncertainties} ($\delta r = 0.1$ for $D^{(*)+}$ and $\delta r = 0.03$ for $D^{*0}$) and repeating the fit. This variation is applied to the $D^{+}$, $D^{*0}$, and $D^{*+}$ channels. 
A second variation is obtained from a fit to the $B\to DD_s^-$ data sample allowing the widths of the tail Gaussian  to float in the fit. The difference between the fixed values (i.e., those fixed from signal simulation samples) and the values obtained from the fit are used as variations of the fixed values of the widths in the $B\to DK^-K_S^0$ fit, to obtain an alternative value of the branching fraction. 
Given their small signal yields, the latter approach is not feasible for the $\overline B{}^0\to D^+D_s^-$ and $B\to D^{*}D_s^-$ {channels}. Therefore, the variation observed in the $B^-\to D^0D_s^-$  mode is included for the $D^+$ and $D^{*+}$ {channels}, {but not} for the $D^{*0}$ {channels}, {for which a different approach is used as described below}. 

The $B^-\to D^{*0}K^-K$ signal includes a 29\% self-cross-feed component i.e.,\ misreconstructed signal events, mostly due to incorrect $\pi^0$ associations. Since the self-cross-feed component is {treated as} signal, it does not artificially increase the branching ratio of the signal channels. However, it does degrade the resolution in $\Delta E$. Given the possible disagreement in the description of the $\pi^0$ misreconstruction between data and simulation, a dedicated systematic uncertainty is assigned (included in ``Signal model''). 
A fit to the $B^-\to D^{*0}\pi^-$ data control channel is performed allowing the widths of the tail Gaussian $\sigma_L$ and $\sigma_R$ to vary in the fit. {A second fit is performed} on the simulation fixing the values of the widths. The differences in widths from the two fits is used to obtain an alternative value of the branching fractions. 
The systematic uncertainty is the difference between the nominal and the {alternative value}. The self-cross-feed is small in the other six channels and the fit-model systematic uncertainty covers possible mismodelings.
    
A systematic uncertainty related to the specific choice of background model is assigned (``Bkg model''). In the nominal fit, the background for all channels is described by the sum of an exponential function and a constant. Two alternative fits are performed by adding a linear term or removing the constant term. The differences between the nominal branching fractions and the results obtained with the alternative background models are quoted as systematic uncertainties.

A systematic uncertainty related to {the possible mismodeling of the} non-$K^{*0}$-resonant fraction is assigned to the four $K^{*0}$ {channels} (``$DKK\pi$ bkg'').  Four alternative non-$K^{*0}$-resonant background fractions compared to the nominal one are {assumed}, and the branching fraction is estimated again with the varied fractions. First, two alternative non-$K^{*0}$-resonant fractions are obtained by varying the fractions according to the statistical uncertainties of the fractions propagated from the fit components. Second, an alternative non-$K^{*0}$-resonant fraction is determined by replacing the signal template with a relativistic Breit-Wigner distribution corrected for the two-body phase-space factor. Third, the parameters of the Chebyshev polynomial are fixed from a simulation produced with \texttt{PYTHIA} instead of \texttt{EvtGen}. The total systematic uncertainty is the sum in quadrature of the four variations. 
    
A systematic uncertainty is assigned  (``$D^{*0}$ peaking bkg'') to account for the uncertainties in Eq.~\eqref{eq:DstpBkg_yield}, used to assess the yield of the cross-feed from ${\overline B{}^0\to D^{*+}K^-K}$ channels in the $D^{*0}$ channels. Equation~\eqref{eq:Dst0_hypothesis} is verified with signal simulation, and holds with a maximum deviation of 10\%. The yield of the peaking background {$N_{D^{*+}}^\text{bkg}$} is scaled to 110\% and 90\% and two resulting signal yields are extracted. 
In addition, the branching fractions of $D^{*+}$ and $D^{0}$ channels are scaled by their uncertainties, to take into account the correlation between the branching fractions and the $N_{D^{*+}}$ yield, producing additional variations of $\mathcal B(\overline B{}^{0}\to D^{*0}K^-K$). This systematic uncertainty also {addresses} the correlation between the $D^{*0}$ channel and the $D^{*+}$ and $D^{0}$ channels. The systematic uncertainty is the sum in quadrature of the differences between the {resulting} alternative branching fractions and the nominal one. This systematic uncertainty affects only the $D^{*0}$ channels. 

A systematic uncertainty related to the knowledge of the total number of $B\overline B$ pairs $\delta N_{B\overline B}=5.6\times 10^6$, which enters in Eq.~\eqref{eq:BR},  is included (``$N_{B\overline B}$''). Similarly, a the systematic uncertainty related to $f_{+-,00}$ from Ref.~\cite{Belle:fpm00} (``$f_{+-,00}$'') is included.  

A systematic uncertainty related to the intermediate branching fraction used in Eq.~\eqref{eq:BR} is also included (``Intermediate $\mathcal B$s'') by propagating the uncertainties of the known intermediate branching fractions to the {final results}.

\begin{table}[!b]
\small
\caption{Relative statistical uncertainties and breakdown of the relative systematic uncertainties {to the corresponding branching fractions} for the eight channels. All values are in percent. A dash is {shown when} the uncertainty is not applicable.  }\label{tab:systematicsRel}
\hspace{-1cm}
\begin{tabularx}{1.1\linewidth}{l c c c c c c c c}
\toprule
Source  & \tiny{$D^0 K^- K^0_S$}  & \tiny{$D^{+} K^- K^{0}_S$}& \tiny{$D^{*0} K^- K^0_S$}  & \tiny{$D^{*+} K^- K^0_S$}  & \tiny{$D^0 K^- K^{*0}$} & \tiny{$D^{+} K^- K^{*0}$} & \tiny{$D^{*0} K^- K^{*0}$}  & \tiny{$D^{*+} K^- K^{*0}$}  \\
\midrule
Eff. - MC sample size  & 0.5 & 0.8 & 1.1 & 0.9 & 0.5 & 0.7 & 0.9 & 1.2 \\

Eff. -  tracking & 0.7 & 1.0 & 0.7 & 1.0 & 1.0 & 1.2 & 1.0 & 1.2  \\

Eff. - $\pi^+$ from $D^{*+}$  & - & - & - & 2.7 & - & - & - & 2.7 \\ 

Eff. - $K_S^0$  & 2.4 & 2.7 & 2.3 & 2.3 & - & - & - & - \\

Eff. -  PID  & 1.3 & 1.7 & 0.5 & 0.6 & 2.5 & 2.6 & 1.6 & 1.7   \\

Eff. -  $\pi^0$ & - & - &  5.1 & - & - & - & 5.1 & - \\ 

Eff. - modeling & 0.2 & 0.3 & 0.6 & 0.7 & 1.3 & 2.0 & 3.1 & 2.4   \\

Signal model & 1.5 & 3.6 & 2.3 & 2.7 & 0.8 & 1.0 & 2.5 & 0.6 \\

Bkg model & 0.8 & 1.1 & 0.8 & 0.8 & 1.1 & 0.4 & 0.2 & 0.1    \\

 $DKK\pi$ bkg & - & - & - & - & 1.4 & 0.7 & 0.7 & 0.8 \\

$D^{*0}$ peaking bkg & - & - & $<0.1$ & - & - & - & 2.0 & -\\
 
$N_{B\overline B}$  & 1.4 & 1.4 & 1.4 & 1.4 & 1.4 & 1.4 & 1.4 & 1.4   \\

$ f_{+-,00}$  & 2.4 & 2.5 & 2.4 & 2.5 & 2.4 & 2.5 & 2.4 & 2.5   \\
 
Intermediate $\mathcal{B}$s  & 0.8 & 1.7 & 1.6 & 1.1 & 0.8 & 1.7 & 0.6 & 1.1 \\
\midrule
Total systematic  & 4.4 & 6.1 & 7.1 & 5.7 & 4.6 & 5.1 & 7.8 & 5.4  \\
 \midrule
 Statistical        & 8.8 & 14.4 & 18.1 & 20.5 & 6.2  & 6.0 & 9.6 & 9.2  \\
\bottomrule
\end{tabularx}
\end{table}

The  contributions to the systematic uncertainties for each $B\to DD_s^-$ channel are summarized in Table~\ref{tab:systematicsRel_DsD}, expressed as relative uncertainties {to the corresponding branching fractions}. When feasible, the same procedure used for the $B\to DK^-K$ channels is applied. An additional systematic uncertainty is assigned for the $B\to DK^-K$ background estimation (``$DKK$ bkg''). This uncertainty is obtained by varying the background yield by the {statistical} uncertainty of the {relevant bin in the} measured $m(K^-K)$ distribution. The systematic uncertainties are evaluated separately for the $D_s^-\to K^-K_S^0$ and $D_s^-\to K^-K_S^{*0}$ sub-channels and then a weighted average is provided. 

The ``Eff-MC sample size'' systematic uncertainty is not correlated between the channels. The ``Intermediate $\mathcal B$s'' are partially correlated between the channels, according to the shared branching fractions. The uncertainty in the factor $f_{+-,00}$ is fully correlated between $B^0$ channels, and fully anti-correlated between $B^-$ and $B^0$ channels. All other systematic uncertainties are fully correlated between all the relevant channels.

\begin{table}[!hb]
\small
\caption{Relative statistical uncertainties and breakdown of the relative systematic uncertainties {to the corresponding branching fractions} for the four $B\to DD_s^-$ channels. All values are in percent. A dash is {shown when} the uncertainty is not applicable.}\label{tab:systematicsRel_DsD}
\begin{tabularx}{1\linewidth}{X>{\hsize=1.2\hsize} c c c c }
\toprule
Source  & $B^- \to D^0 D_s^-$ & $\overline B{}^0 \to D^{+}D_s^-$& $B^- \to D^{*0}D_s^-$  & $\overline B{}^0 \to D^{*+} D_s^-$  \\
\midrule
Eff. - MC sample size  & $<0.1$ & $<0.1$ & $<0.1$ & $<0.1$ \\

Eff. -  tracking & 0.8 & 1.0 & 0.8 & 1.0 \\

Eff. - $\pi^+$ from $D^{*+}$  & - & - & - & 2.7 \\ 

Eff. - $K_S^0$  & 1.2 & 1.2 & 1.2 & 1.2\\

Eff. -  PID  & 1.9 & 2.1 & 1.1 & 1.3 \\

Eff. -  $\pi^0$ & - & - &  5.1 & - \\ 

Signal model & $<0.1$ & $<0.1$ & 1.1 & 0.3\\

 Bkg model & 0.7 & 0.7 & 1.6 & 0.1 \\

$DKK$ bkg & 1.7 & 2.1 & 6.1 & 4.5 \\
 
 $D^{*0}$ peaking bkg & - & - & 0.6 & - \\
 
$N_{B\overline B}$  & 1.4 & 1.4 & 1.4 & 1.4    \\

$ f_{+-,00}$  & 2.4 & 2.5 & 2.4 & 2.5    \\ 
Intermediate $\mathcal{B}$s  & 2.5 & 2.9 & 2.8 & 2.9\\
\midrule
Total systematic  & 5.0 & 5.6 & 9.5 & 7.0\\
 \midrule
 Statistical        & 6.0 & 5.9 & 15.2 & 11.9 \\
\bottomrule
\end{tabularx}
\end{table}

\section{Invariant mass and helicity angles analysis for $B\to DK^-K$ channels}\label{sec:mKK}

The signal yield as a function of $m(K^-K)$ is extracted using the signal $s$Weights  as described in Sec.~\ref{sec:BR} and applying the event-by-event efficiency correction using the map described in Sec.~\ref{sec:efficiency}.

The distribution{s} of the $\theta_{KK}$ and $\theta_{K^{*}}$ helicity angles are extracted using the same approach. 
The {former} is defined as the angle between the $K^{*0}$ momentum and the direction opposite to the $B$ momentum, both calculated in the $K^-K^{*0}$ system rest frame. 
The {latter} is defined as the angle between the {momentum of the} $K^+$ from $K^{*0}$ decay and the direction opposite to the $K^-K^{*0}$ system momentum, both in the $K^{*0}$ rest frame. 
The angles $\theta_{KK}$ and $\theta_{K_S}$ are defined {analogously} for the $K_{S}^0$ {channels}. There is no significant correlation between these angles and $\Delta E$, as required for the $s$Plot technique. 

The expected angular distributions $dN/d\theta$ are shown in Table~\ref{tab:angles_expected} for different hypotheses for the spin-parity $J^P$ of the  $K^-K^{(*)0}_{(S)}$ system, assuming a single $J^P$ for the transition. Only the $J^P$ states allowed by the factorization hypothesis {and assuming exact isospin symmetry} are shown. In the case of $J^{P}=1^-$, and scalar to vector-vector decay ($D^*$ {channels}), the distribution depends on the fraction of the longitudinal polarization in the $B$ decay, so it is a mixture of different contributions with unknown strengths. Hence, the $dN/d\theta$ distribution is not known a priori for this hypothesis. In the case of $J^{P}=1^+$, the D-wave contribution is assumed to be negligible compared to the S-wave~\cite{PDG_a1}, otherwise the angular distribution will be a mixture of different contributions {as well}. The $\theta_{K_S}$ distribution is not sensitive to {the spin-parity of the} $K^-K^{0}_{S}$ system and is expected to be uniform; thus, it is used to verify the absence of bias. 

\begin{figure}[!hp]
\centering
\subfigure{\includegraphics[width=0.36\columnwidth]{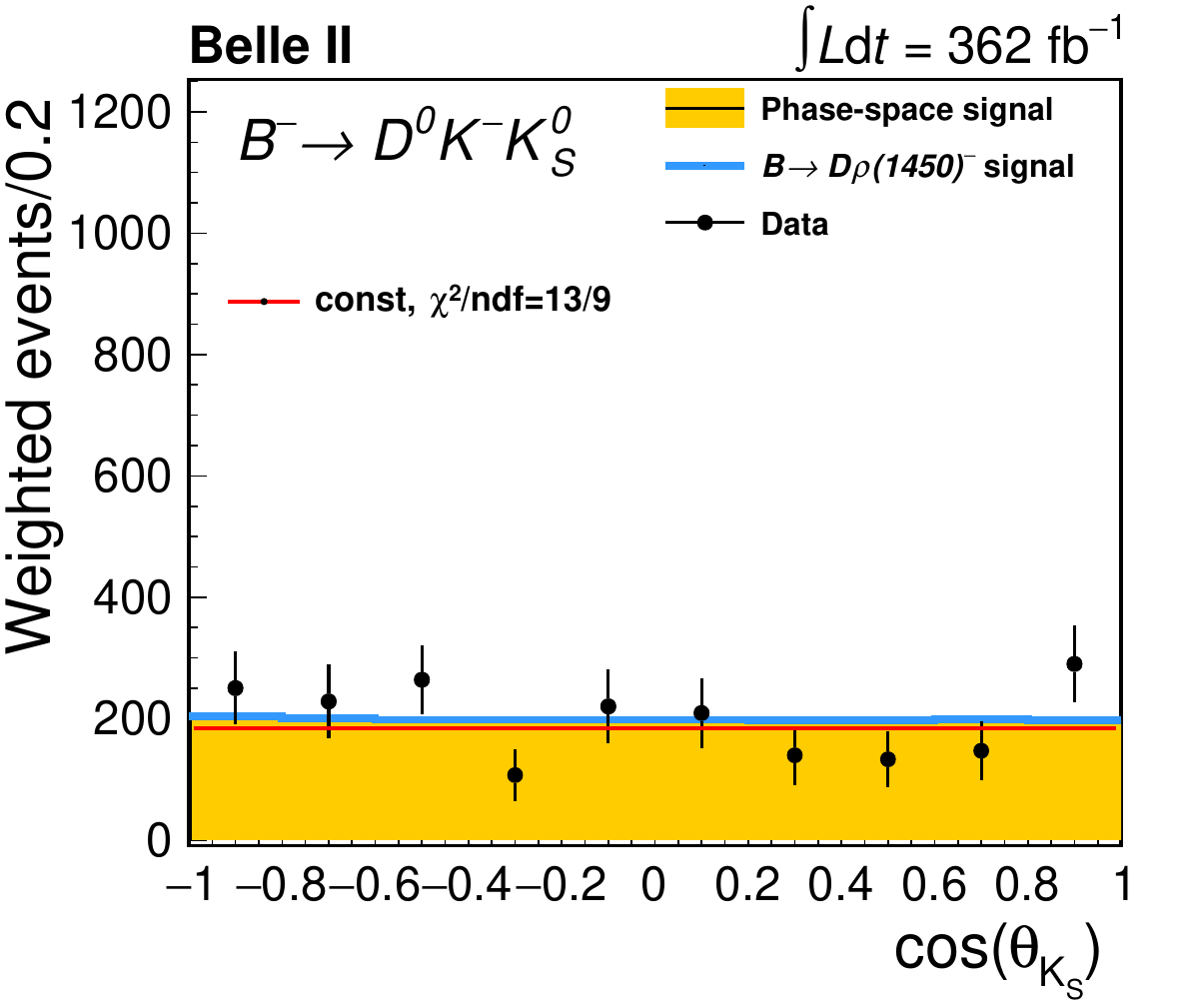}}
\subfigure{\includegraphics[width=0.36\columnwidth]{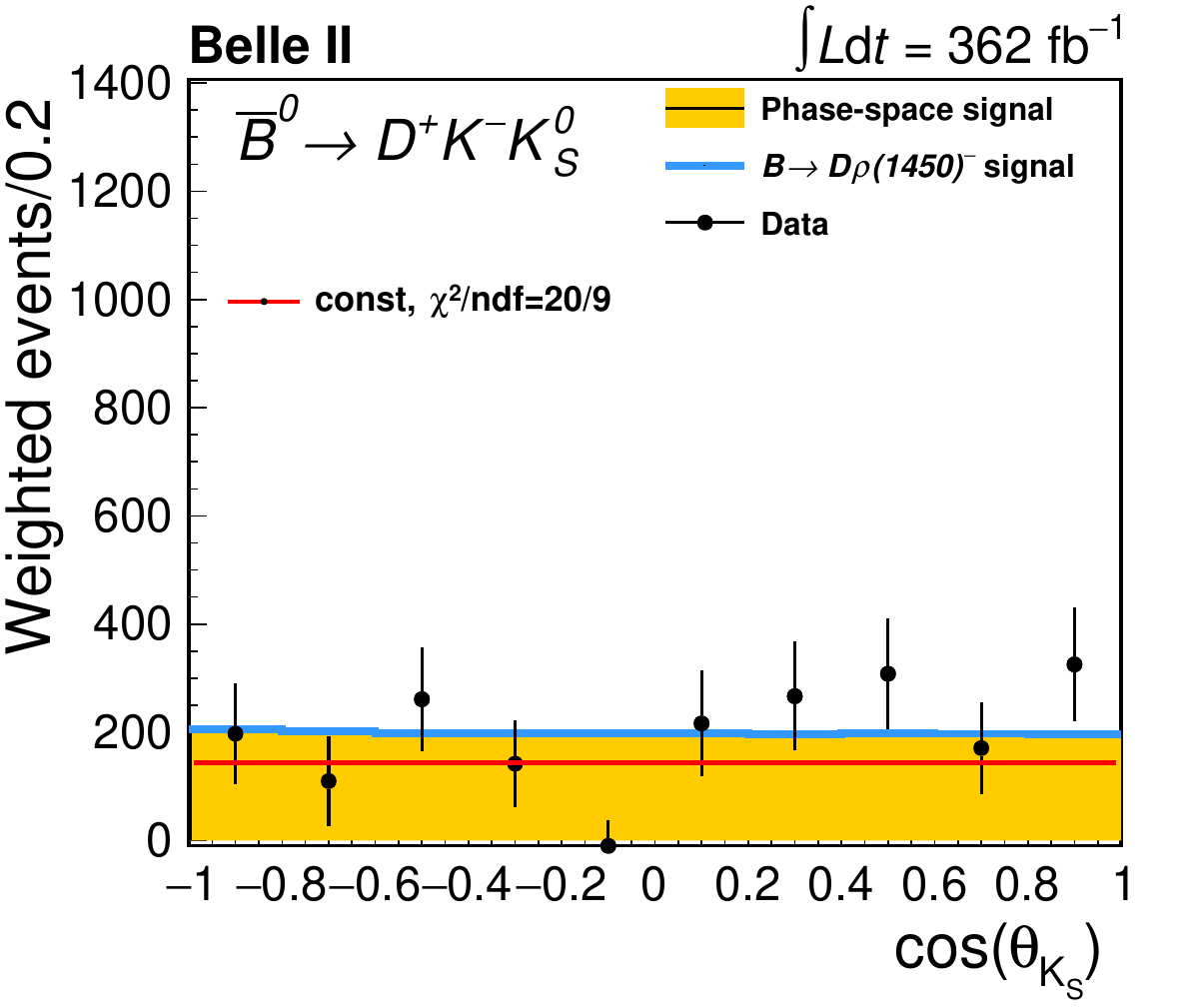}}
\subfigure{\includegraphics[width=0.36\columnwidth]{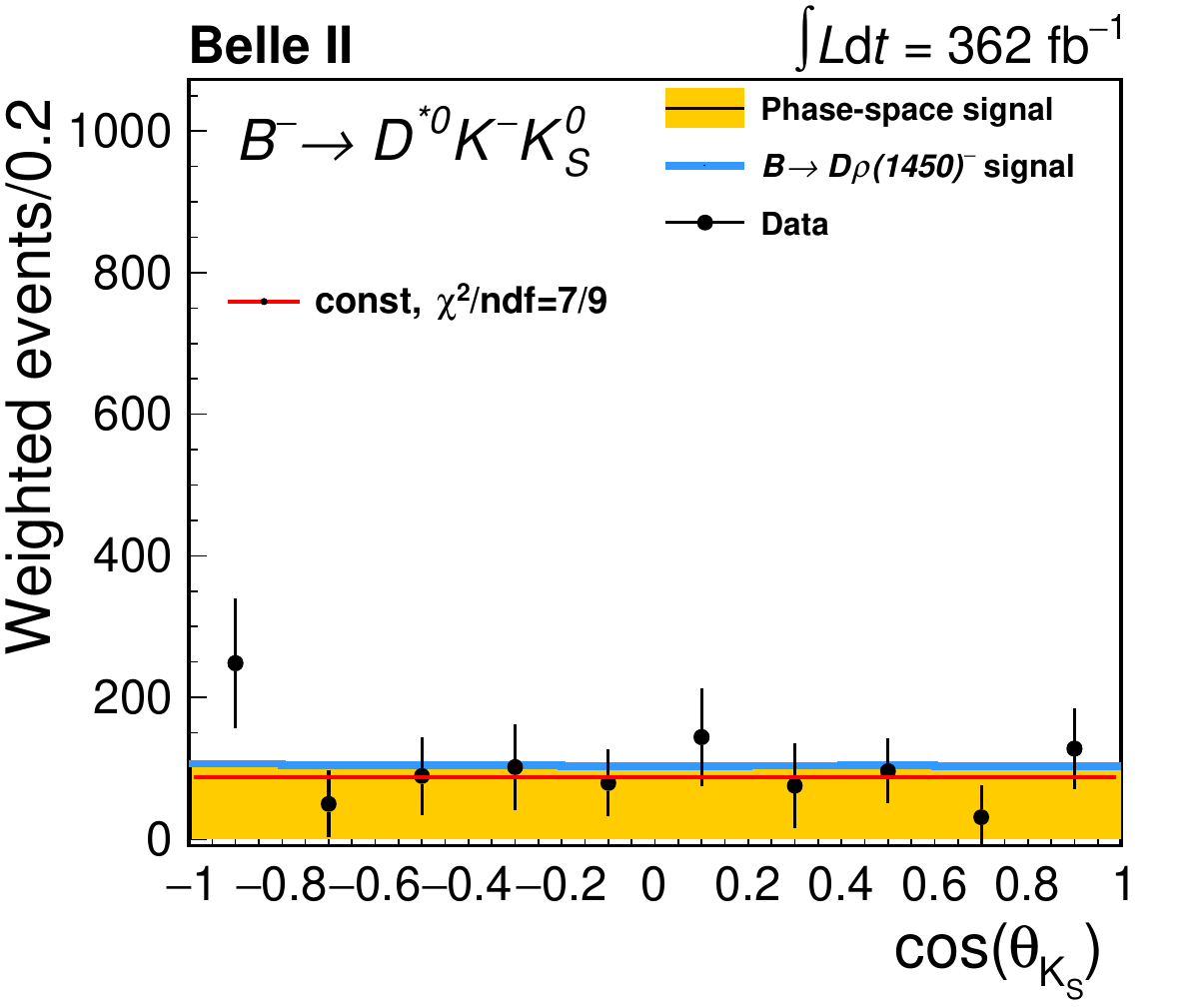}}
\subfigure{\includegraphics[width=0.36\columnwidth]{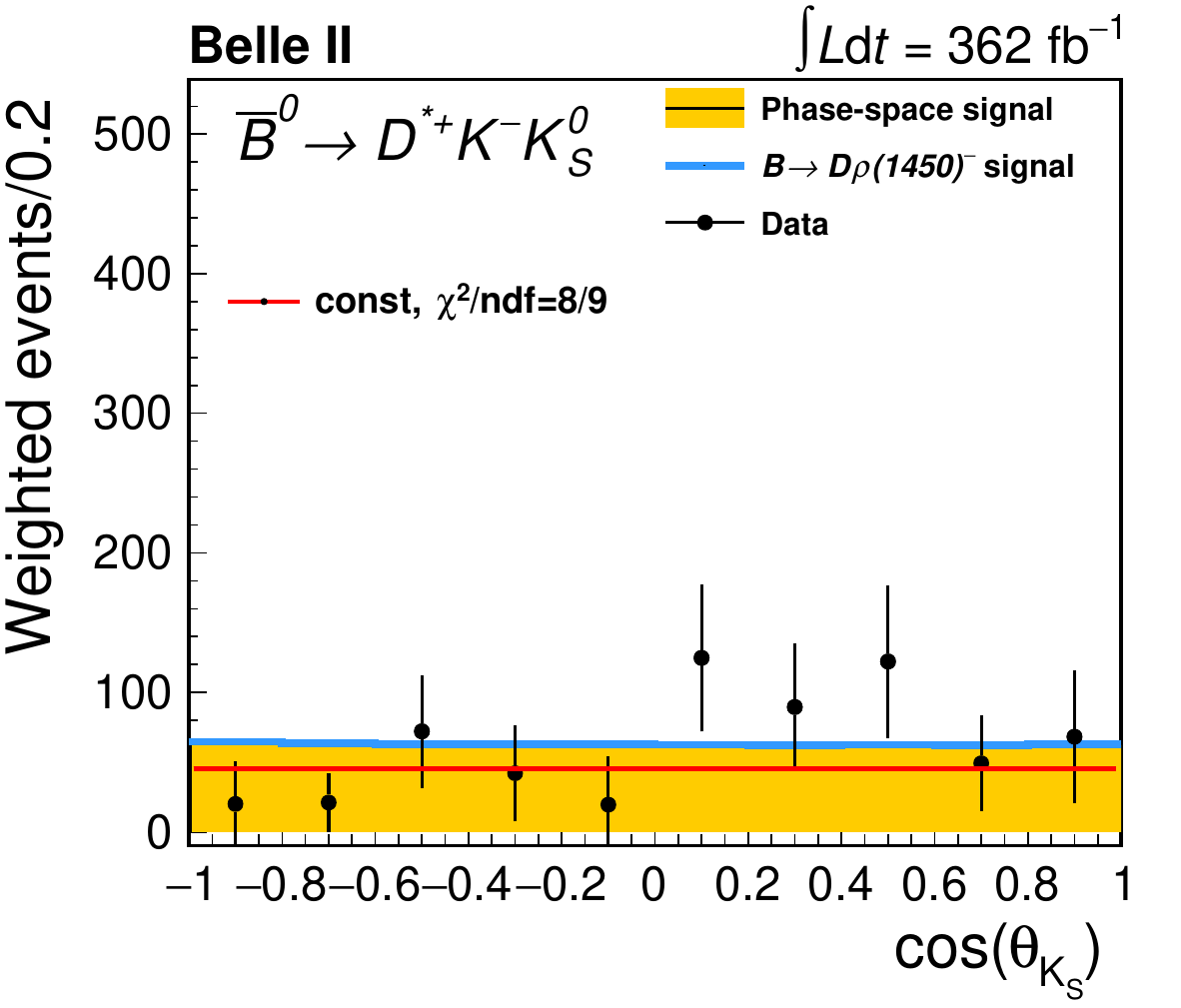}}
\subfigure{\includegraphics[width=0.36\columnwidth]{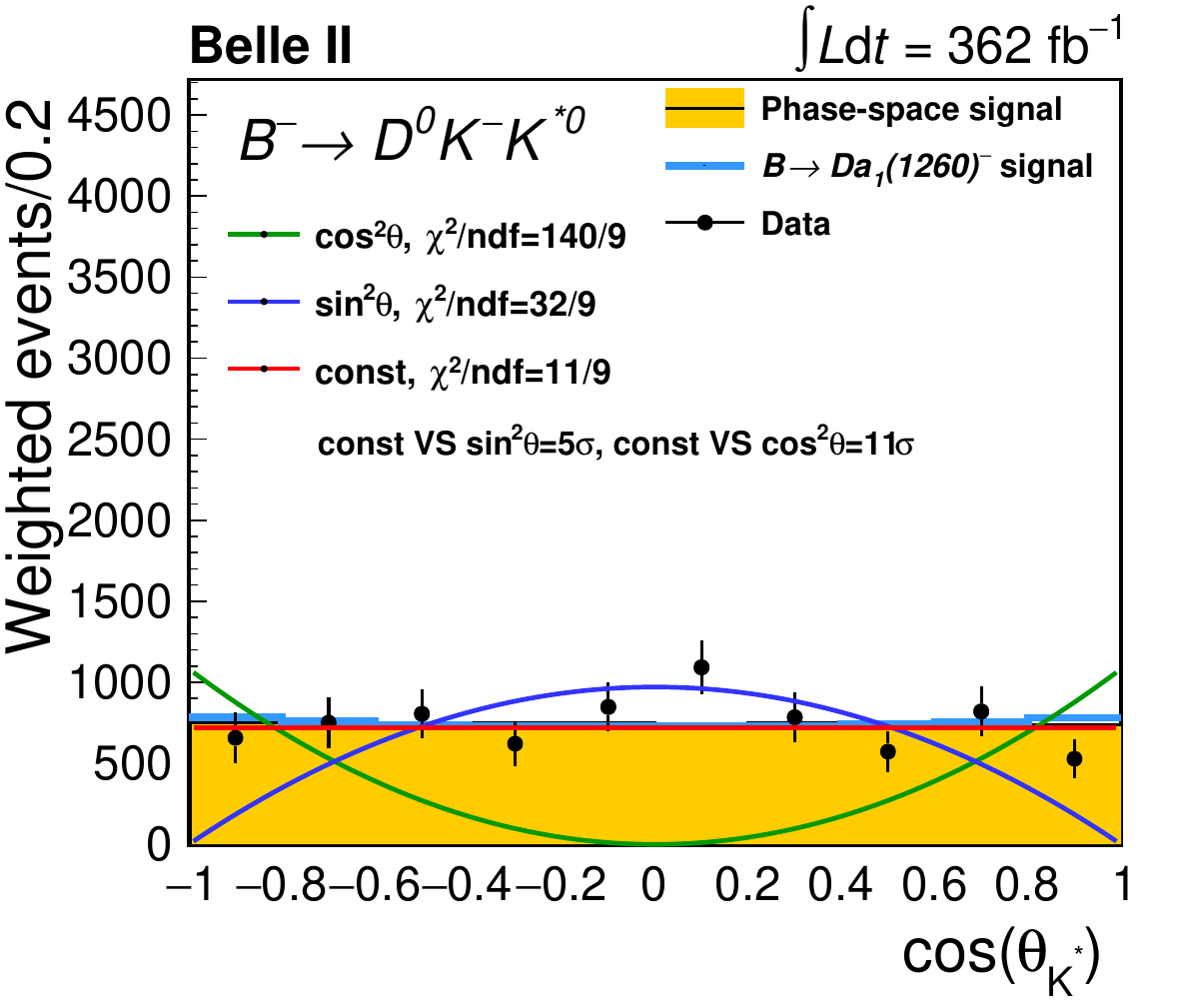}}
\subfigure{\includegraphics[width=0.36\columnwidth]{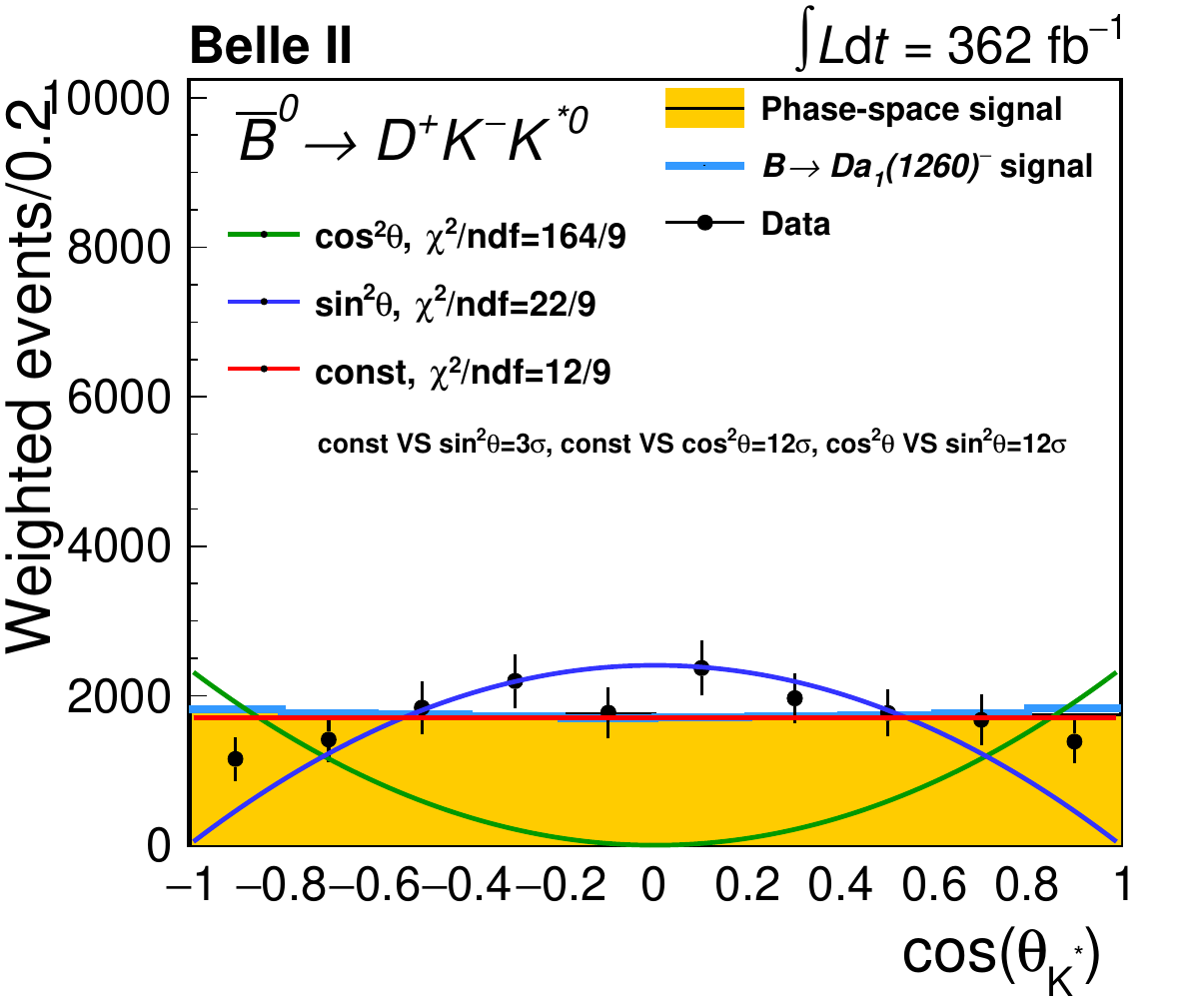}}
\subfigure{\includegraphics[width=0.36\columnwidth]{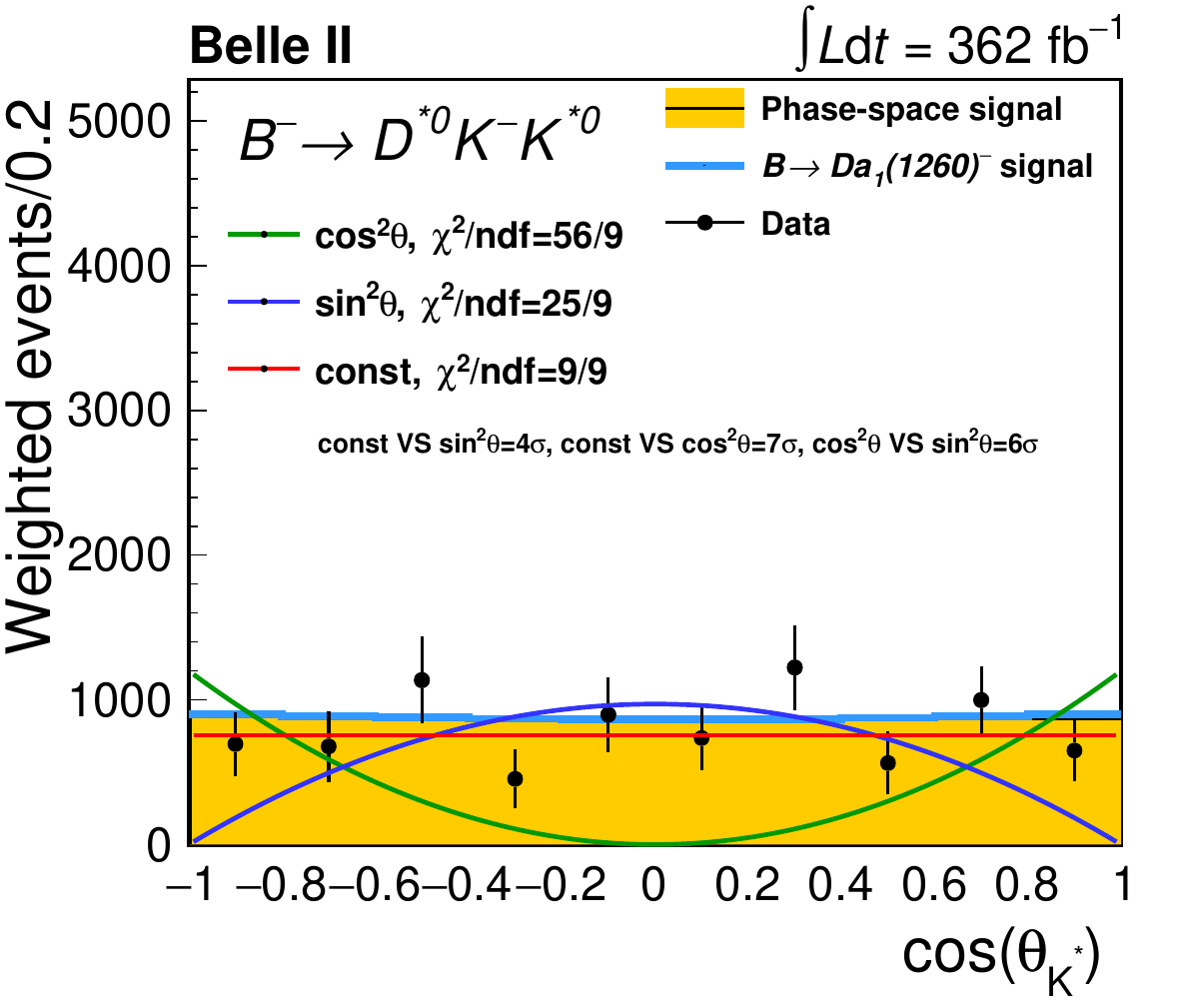}}
\subfigure{\includegraphics[width=0.36\columnwidth]{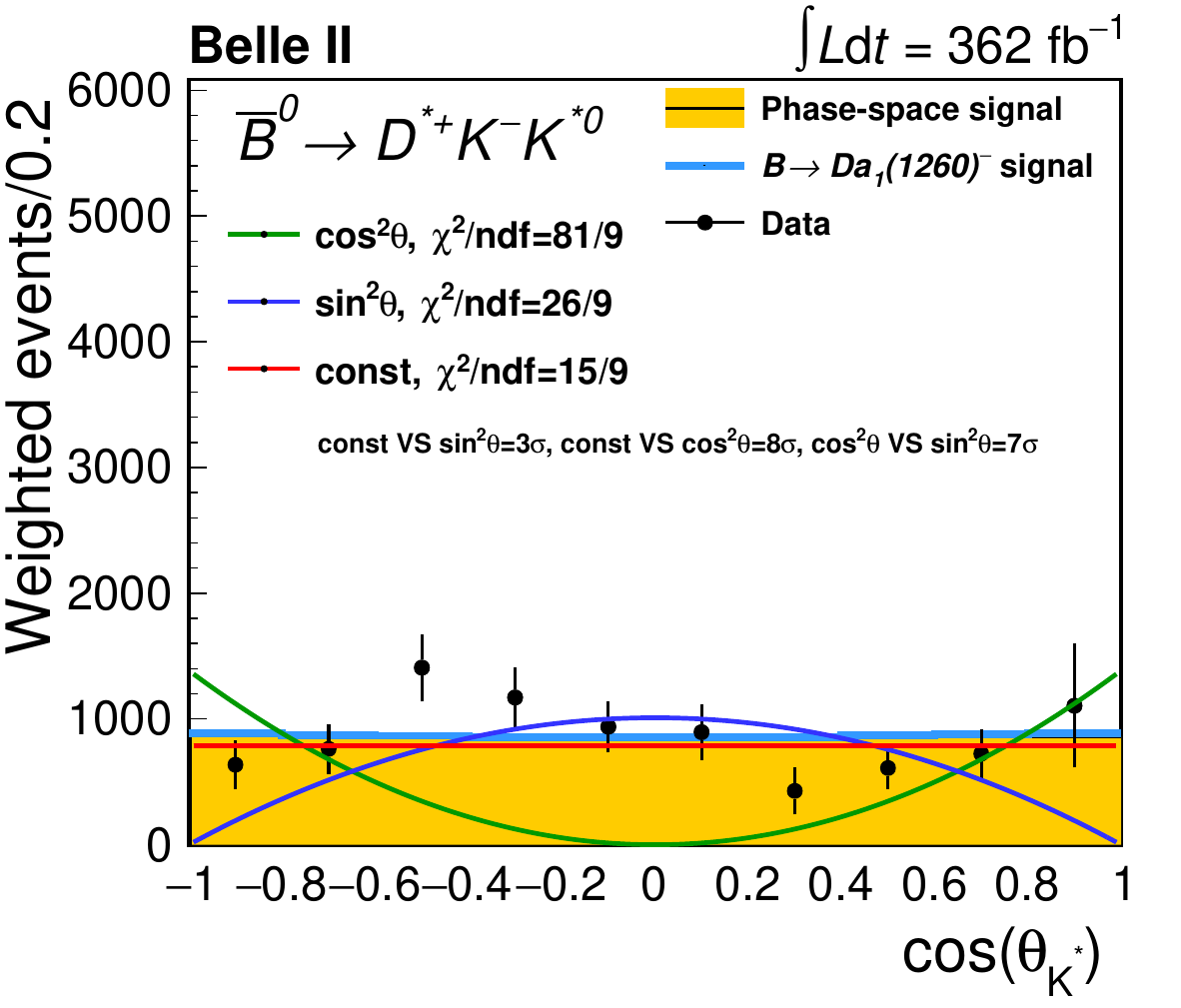}}
\caption{Background-subtracted and efficiency-corrected distribution of $dN/d\cos\theta_{K^{(*)}_{(S)}}$ for the $B^-\to D^0K^-K_S^0$ (first line, left),  $\overline B{}^0\to D^+K^-K_S^0$ (first line, right), $B^-\to D^{*0}K^-K_S^0$ (second line, left), $\overline B{}^0\to D^{*+}K^-K_S^0$ (second line, right), $B^-\to D^0K^-K^{*0}$ (third line, left), $\overline B{}^0\to D^+K^-K^{*0}$ (third line, right), $B^-\to D^{*0}K^-K^{*0}$ (fourth line, left), and $\overline B{}^0\to D^{*+}K^-K^{*0}$ (fourth line, right) channels. The error bars represent the statistical uncertainty. A phase-space MC simulation and a resonant MC simulation at generator level, rescaled to the integral of the data distribution, are also shown for comparison. The three fit-hypotheses and the respective $\chi^2$ are overlaid, together with the significance of the difference between each fit and the other two.} \label{fig:angles_fit_K}
\end{figure}

\begin{figure}[!hp]
\centering
\subfigure{\includegraphics[width=0.36\columnwidth]{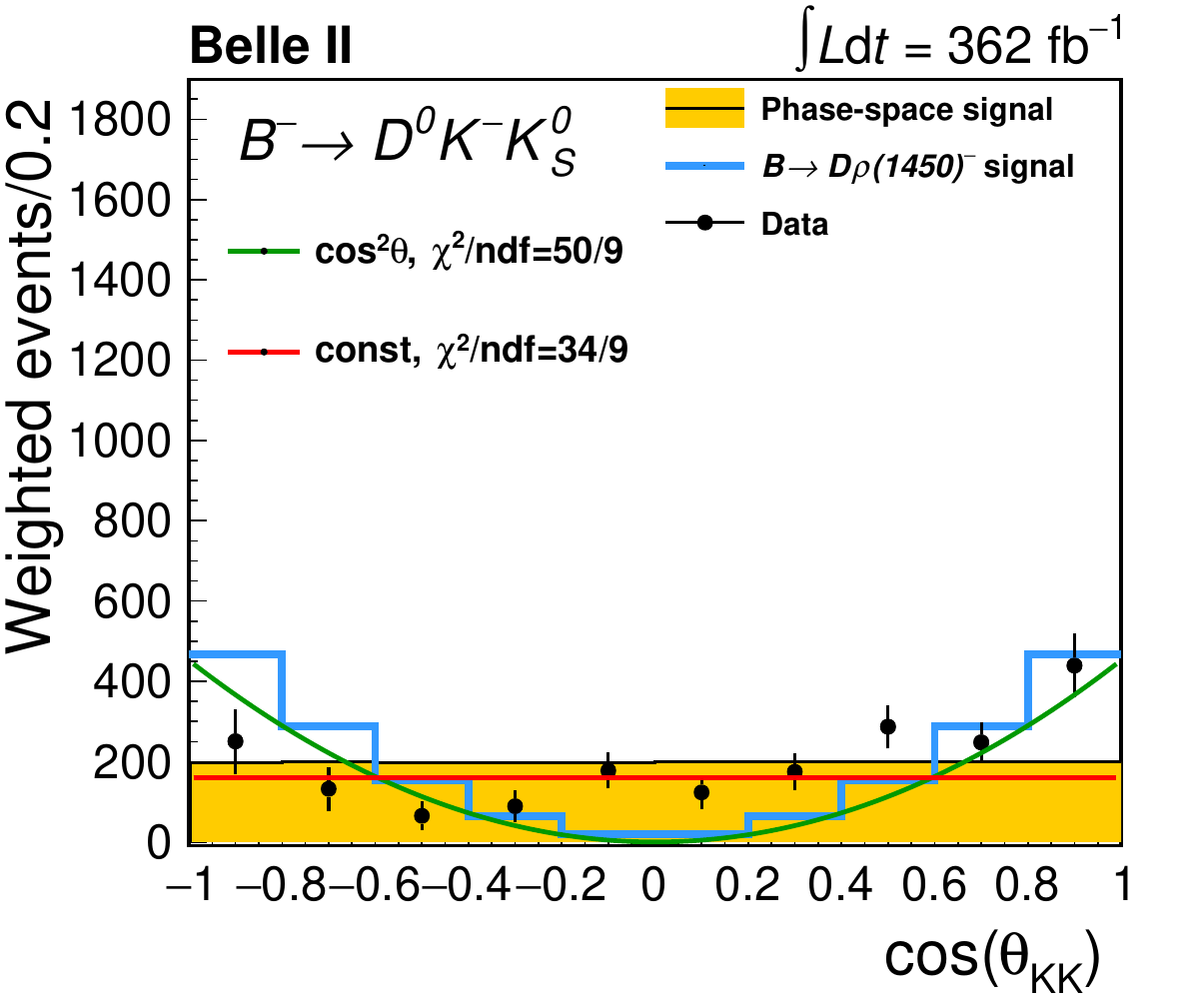}}
\subfigure{\includegraphics[width=0.36\columnwidth]{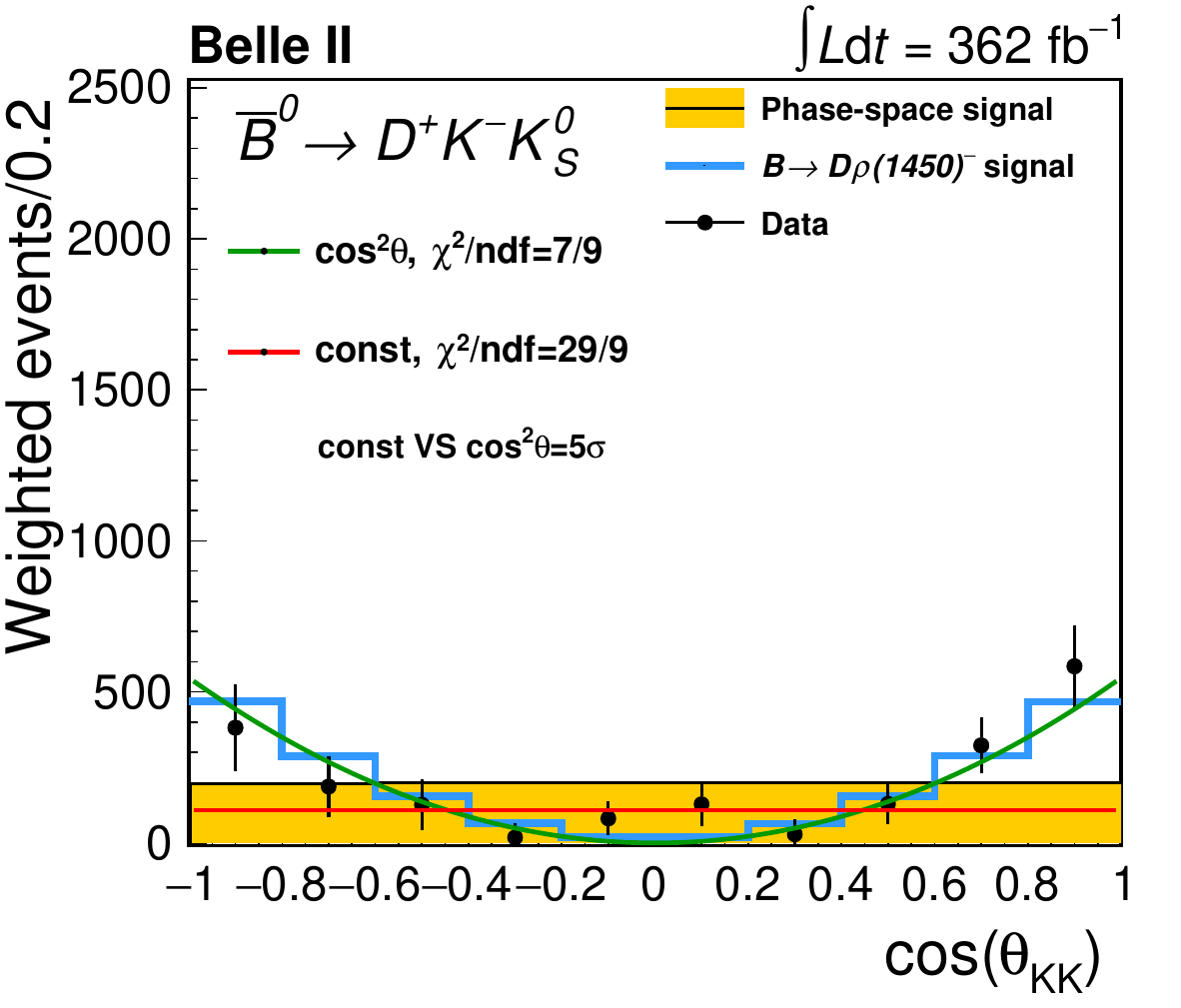}}
\subfigure{\includegraphics[width=0.36\columnwidth]{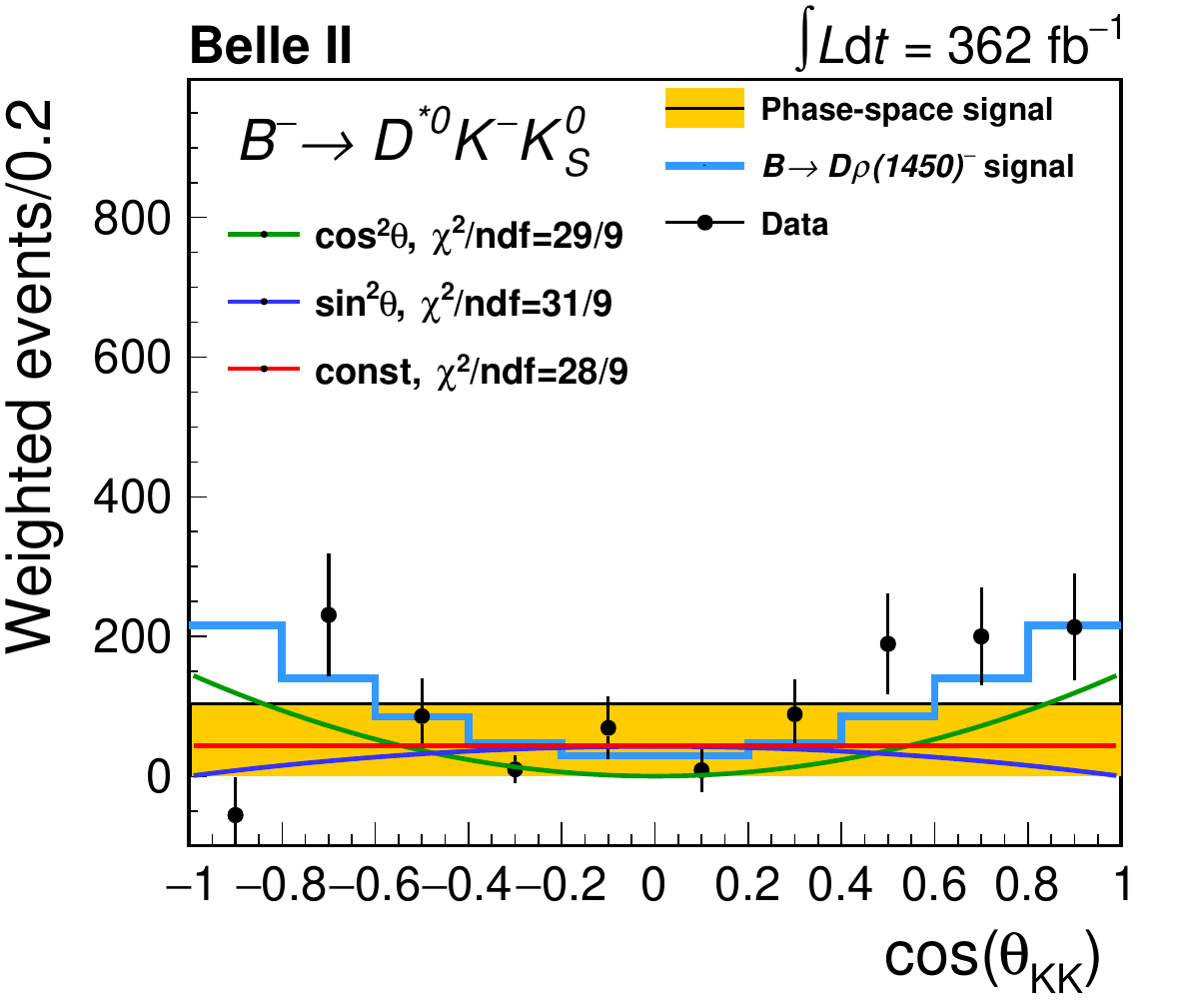}}
\subfigure{\includegraphics[width=0.36\columnwidth]{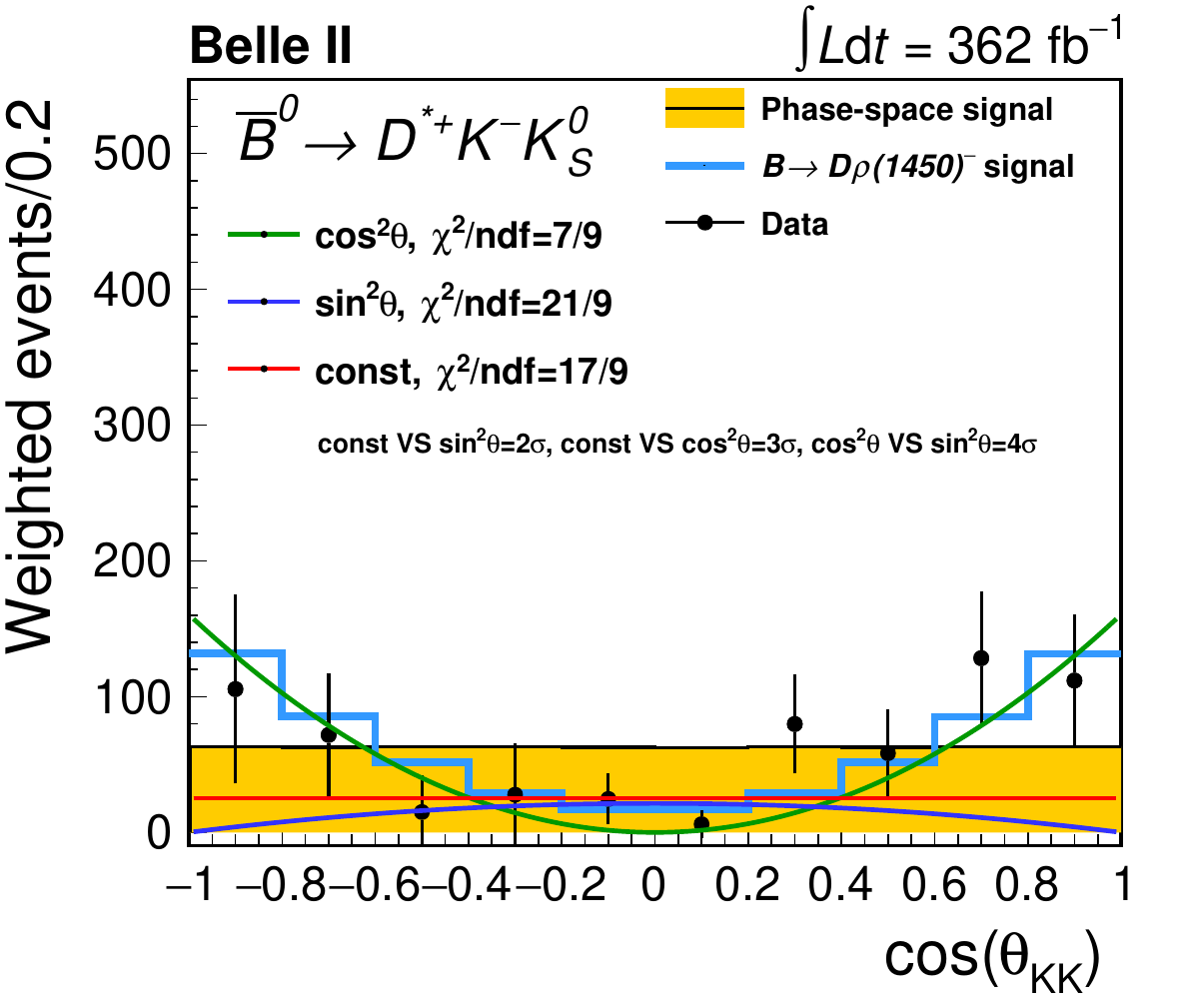}}
\subfigure{\includegraphics[width=0.36\columnwidth]{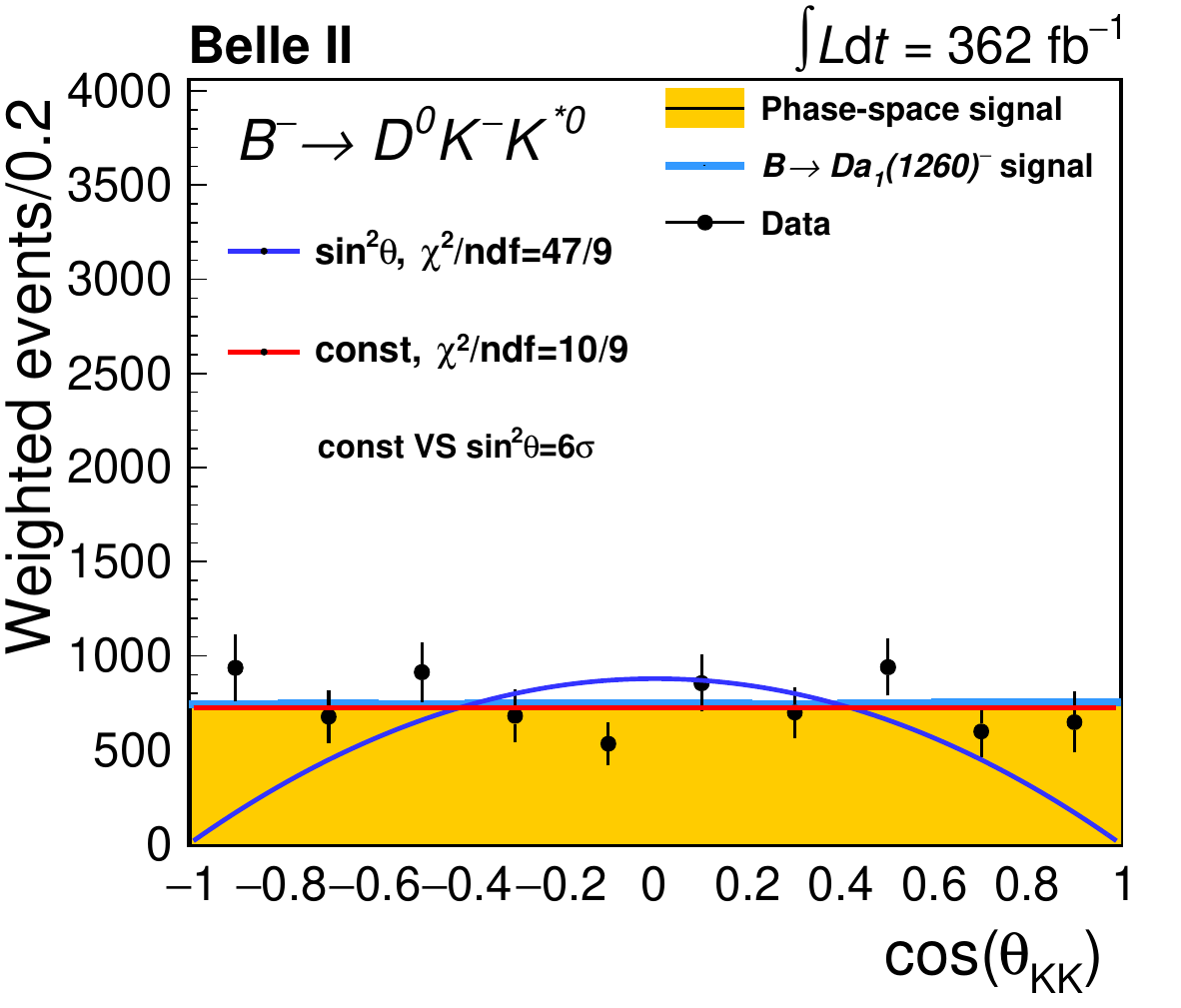}}
\subfigure{\includegraphics[width=0.36\columnwidth]{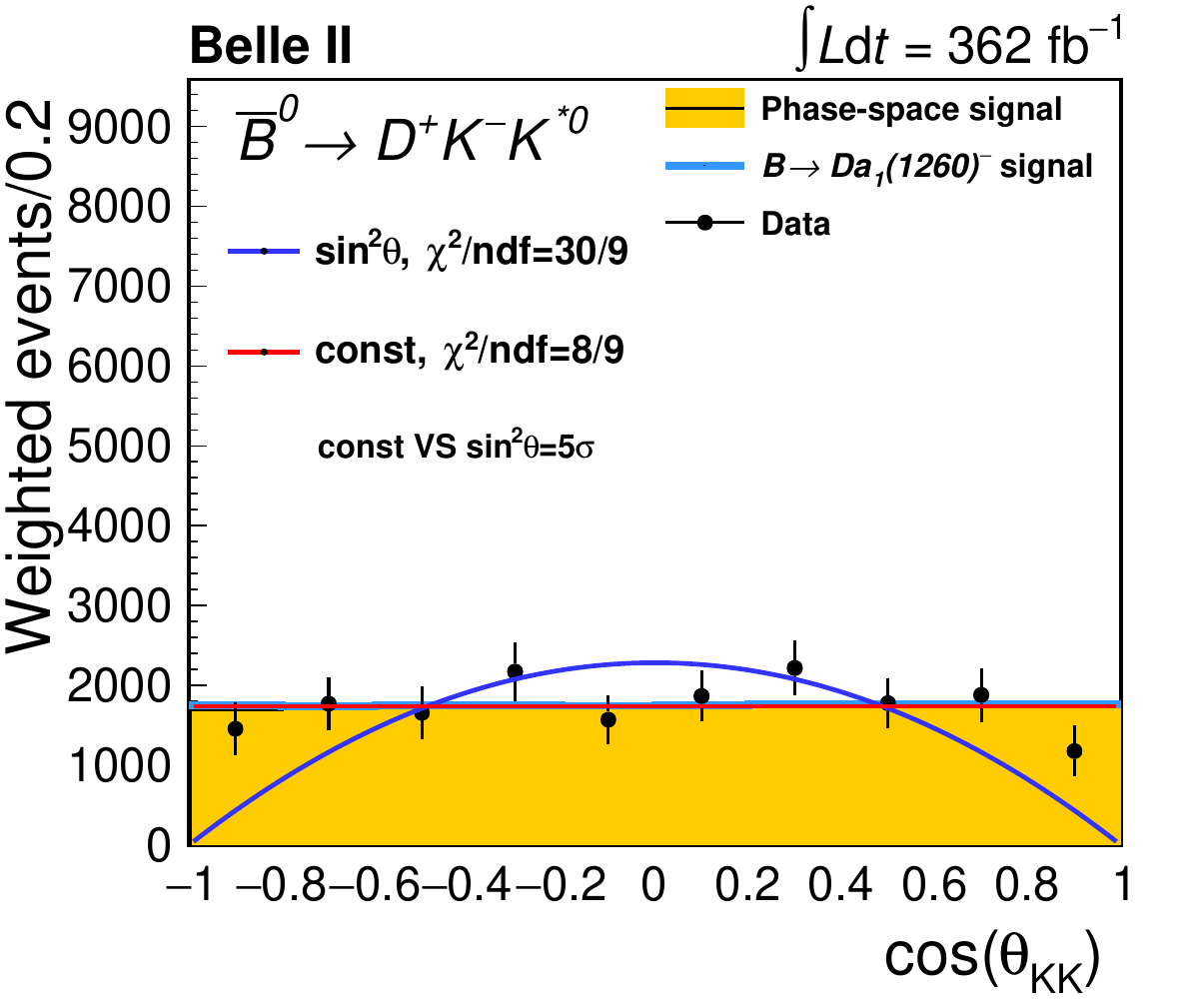}}
\subfigure{\includegraphics[width=0.36\columnwidth]{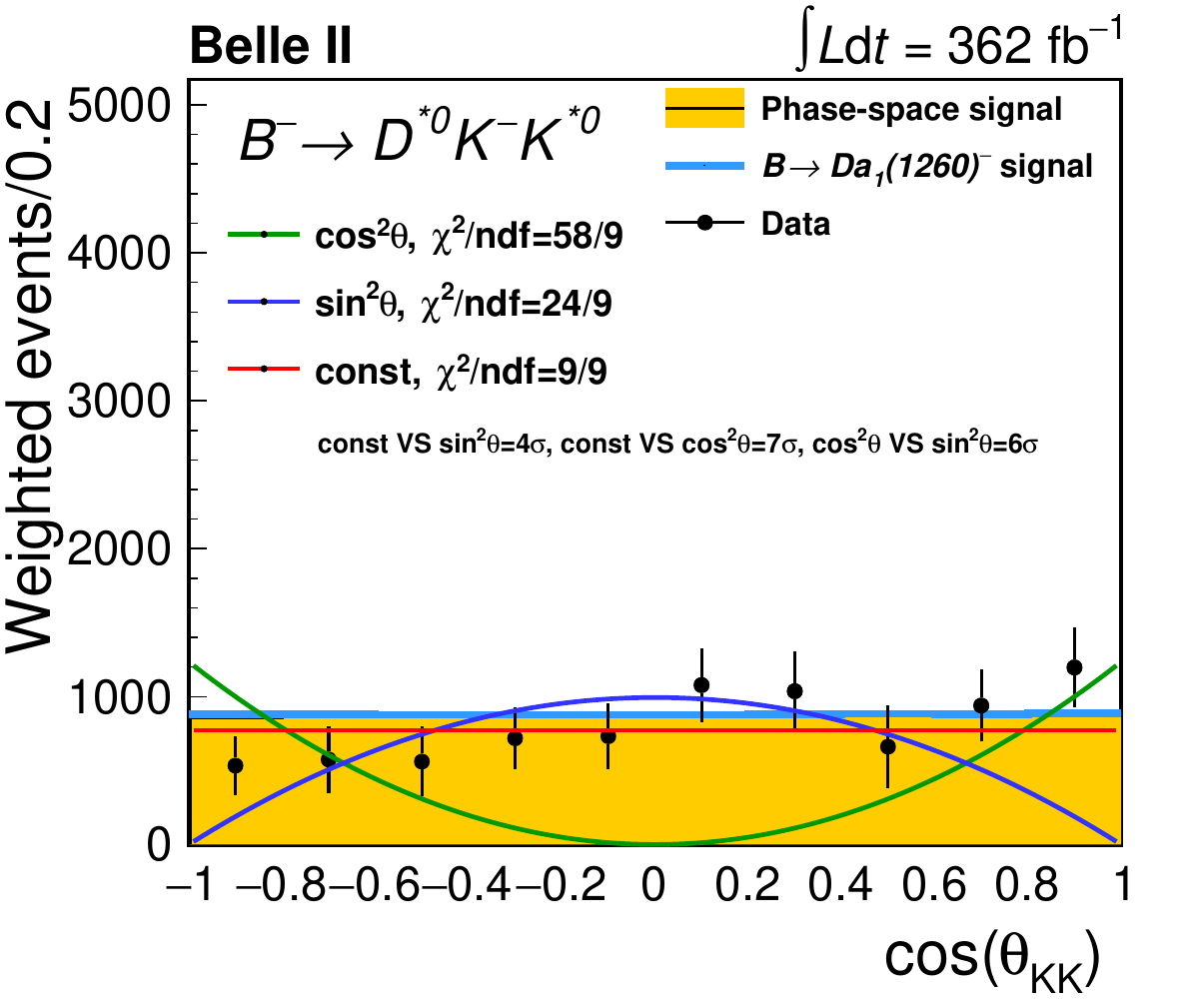}}
\subfigure{\includegraphics[width=0.36\columnwidth]{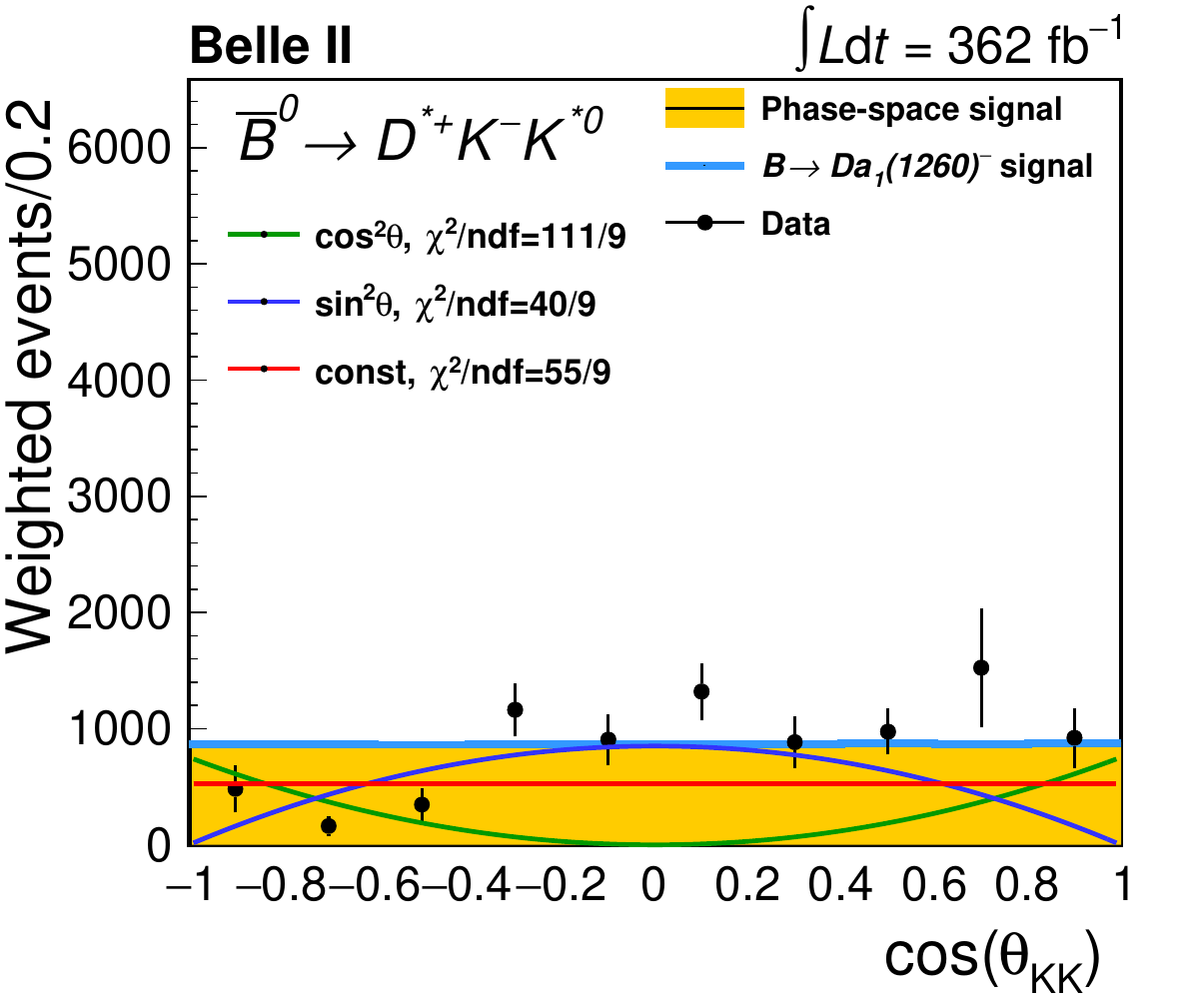}}
\caption{Background-subtracted and efficiency-corrected distribution of $dN/d\cos\theta_{KK}$ for the $B^-\to D^0K^-K_S^0$ (first line, left),  $\overline B{}^0\to D^+K^-K_S^0$ (first line, right), $B^-\to D^{*0}K^-K_S^0$ (second line, left), $\overline B{}^0\to D^{*+}K^-K_S^0$ (second line, right), $B^-\to D^0K^-K^{*0}$ (third line, left), $\overline B{}^0\to D^+K^-K^{*0}$ (third line, right), $B^-\to D^{*0}K^-K^{*0}$ (fourth line, left), and $\overline B{}^0\to D^{*+}K^-K^{*0}$ (fourth line, right) channel. The error bars represent the statistical uncertainty. A phase-space MC simulation and a resonant MC simulation at generator level, rescaled to the integral of the data distribution, are also shown for comparison. The three fit-hypotheses and the respective $\chi^2$ are overlaid, together with the significance of the difference between each fit and the other two.} \label{fig:angles_fit_KK}
\end{figure}

\vspace{-0.4cm}
\begin{table}[!hb]
\small
\caption{Possible angular distributions given a specific spin-parity state of the $K^-K^{(*)0}_{(S)}$ system, subdividing between pseudoscalar {channels} ($D^0, D^+$) and vector {channels} ($D^{*0}, D^{*+}$).  The hyphen (-) stands for a forbidden spin-parity {assuming factorization and exact isospin symmetry}; mix stands for a polarization dependent distribution; const stands for a uniform distribution; the ${}^\dagger$ symbol indicates that the uniform distribution requires S-wave dominance.  }\label{tab:angles_expected}
\hspace{-1.0 cm}
\begin{tabular}{c cccc cccc}
\toprule
      &                        \multicolumn{4}{c}{$K^-K^{*0}$ {channels}} &                                                                 \multicolumn{4}{c}{$K^-K_S^0$ {channels}} \\
       &                \multicolumn{2}{c}{$D^0, D^+$ {channels}} &         \multicolumn{2}{c}{$D^{*0}, D^{*+}$ {channels}}     &          \multicolumn{2}{c}{$D^0, D^+$ {channels}} &  \multicolumn{2}{c}{$D^{*0}, D^{*+}$ {channels}} \\
$J^{P}$ & $dN/d\theta_{KK}$ &      $dN/d\theta_{K^*}$ &   $dN/d\theta_{KK}$ &      $dN/d\theta_{K^*}$     & $dN/d\theta_{KK}$ &     $dN/d\theta_{K_S}$ & $dN/d\theta_{KK}$ &     $dN/d\theta_{K_S}$  \\
\midrule
{Three-body}           & const                    &const                       & const                       &const           & const                       &const                  & const                       &const             \\
$0^-$            & const                    & $\cos^2 \theta$                  & const                      & $\cos^2 \theta$      &   - & - & - & -                      \\
$1^-$            & $\sin^2 \theta$               & $\sin^2\theta$                                     & mix   & $\sin^2\theta$             &   $\cos^2\theta$  &const                 &  mix     & const                         \\
$1^+$            & const$^\dagger$                         & const$^\dagger$                    & const$^\dagger$                         & const$^\dagger$        &    - & - & - & -  \\
\bottomrule
\end{tabular}
\label{tab:my-table}
\end{table}
\vspace{-0.2cm}
The {various} spin-parity hypotheses are tested by fitting the $dN/\cos\theta_{KK}$ and $dN/\cos\theta_{K^{(*)}_{(S)}}$  distributions, for each channel, with the {functional forms corresponding to} individual $J^P$ hypotheses and comparing the $\chi^2$ of the fits. The significance of the difference between the fit hypotheses is also estimated using the Akaike information criterion to compare non-nested models~\cite{AIC}. For the $K_S^0$ channels $dN/\theta_{KK}\propto \cos^2\theta$ and $dN/\theta_{K_S}\propto \text{const}$ are preferred, suggesting $J^P=1^-$. For the $K^{*0}$ channels $dN/\theta_{KK}\propto \text{const}$ and $dN/\theta_{K^*}\propto \text{const}$ are preferred, suggesting $J^P=1^+$ or a non-resonant decay. The distributions of $dN/\cos\theta_{K^{(*)}_{(S)}}$ and $dN/\cos\theta_{KK}$ for the eight channels are shown in {Fig.~\ref{fig:angles_fit_K} and Fig.~\ref{fig:angles_fit_KK}; the fits projections are overlaid. Additional details about the fits are provided in the appendix~\ref{sec:App}.}

\begin{figure}[!t]
\centering
\subfigure{\includegraphics[width=0.45\columnwidth]{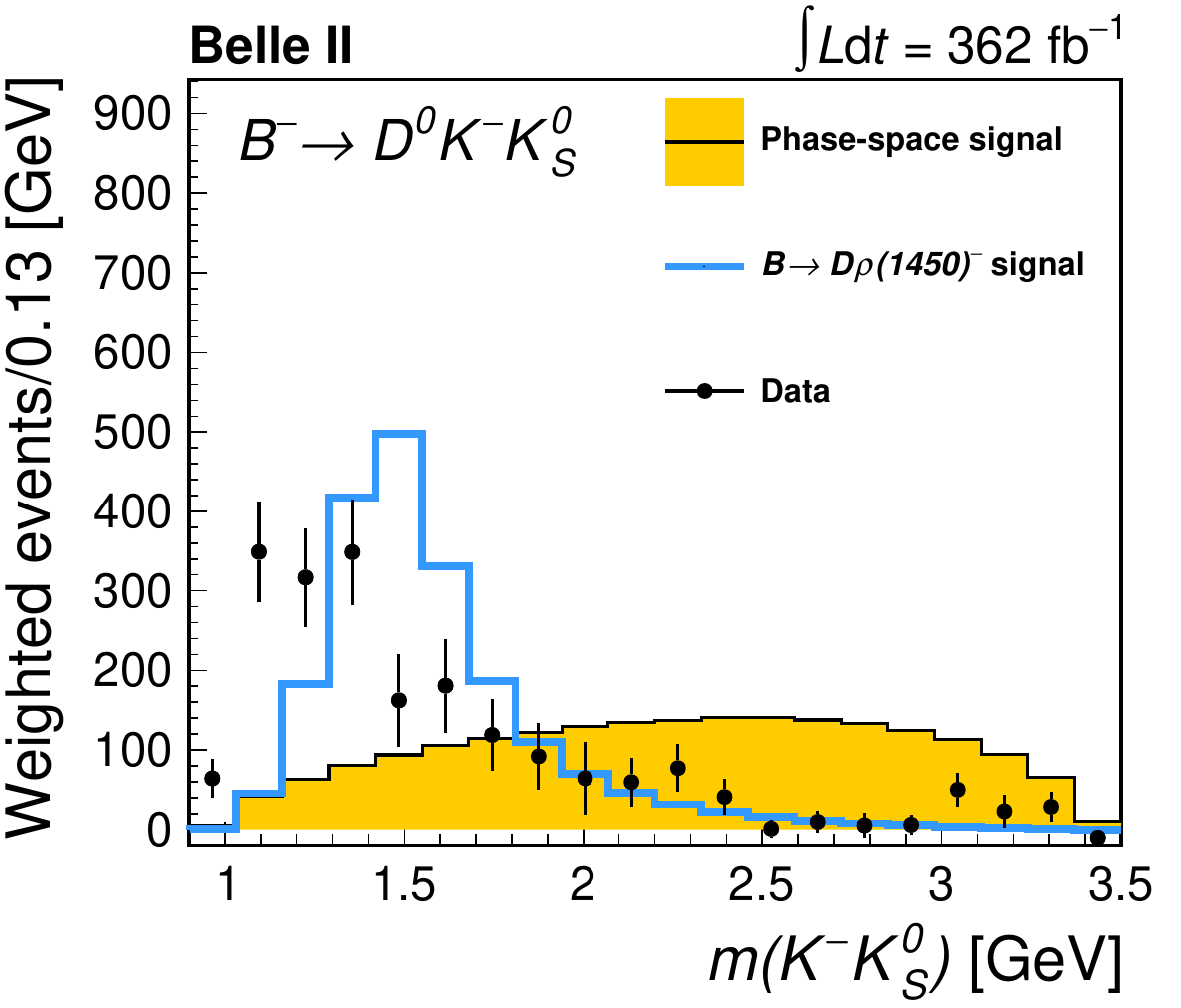}}
\subfigure{\includegraphics[width=0.45\columnwidth]{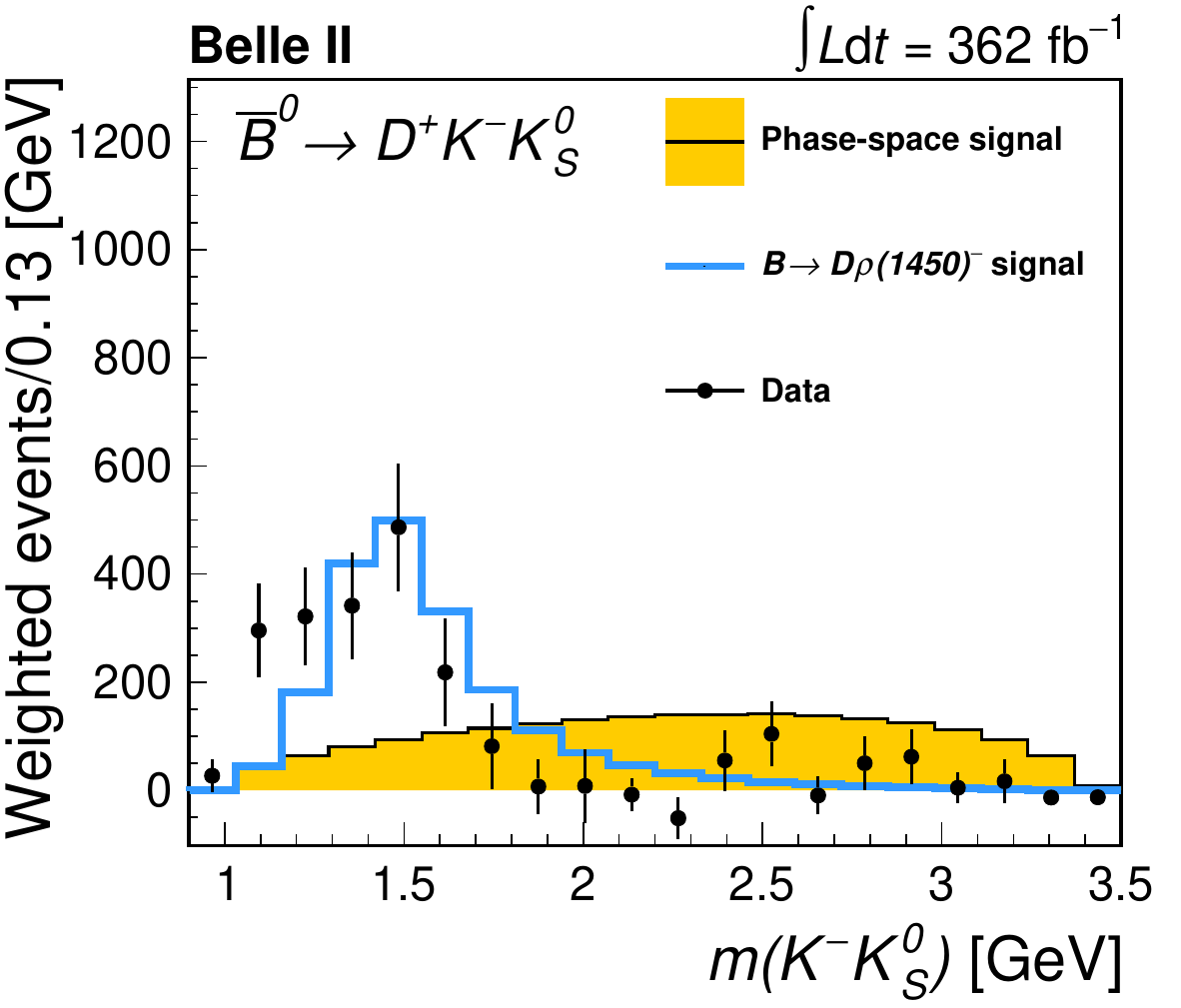}}
\subfigure{\includegraphics[width=0.45\columnwidth]{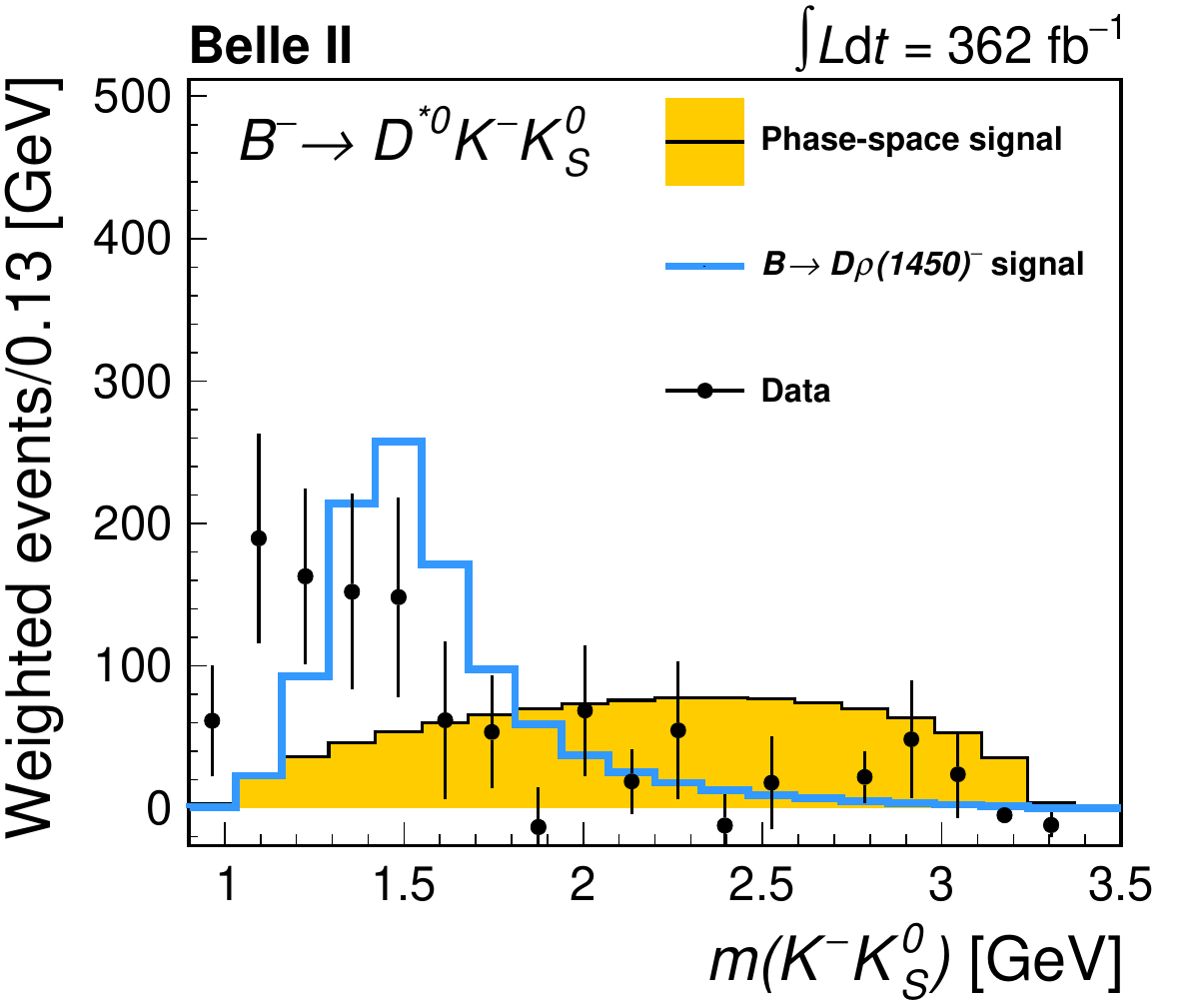}}
\subfigure{\includegraphics[width=0.45\columnwidth]{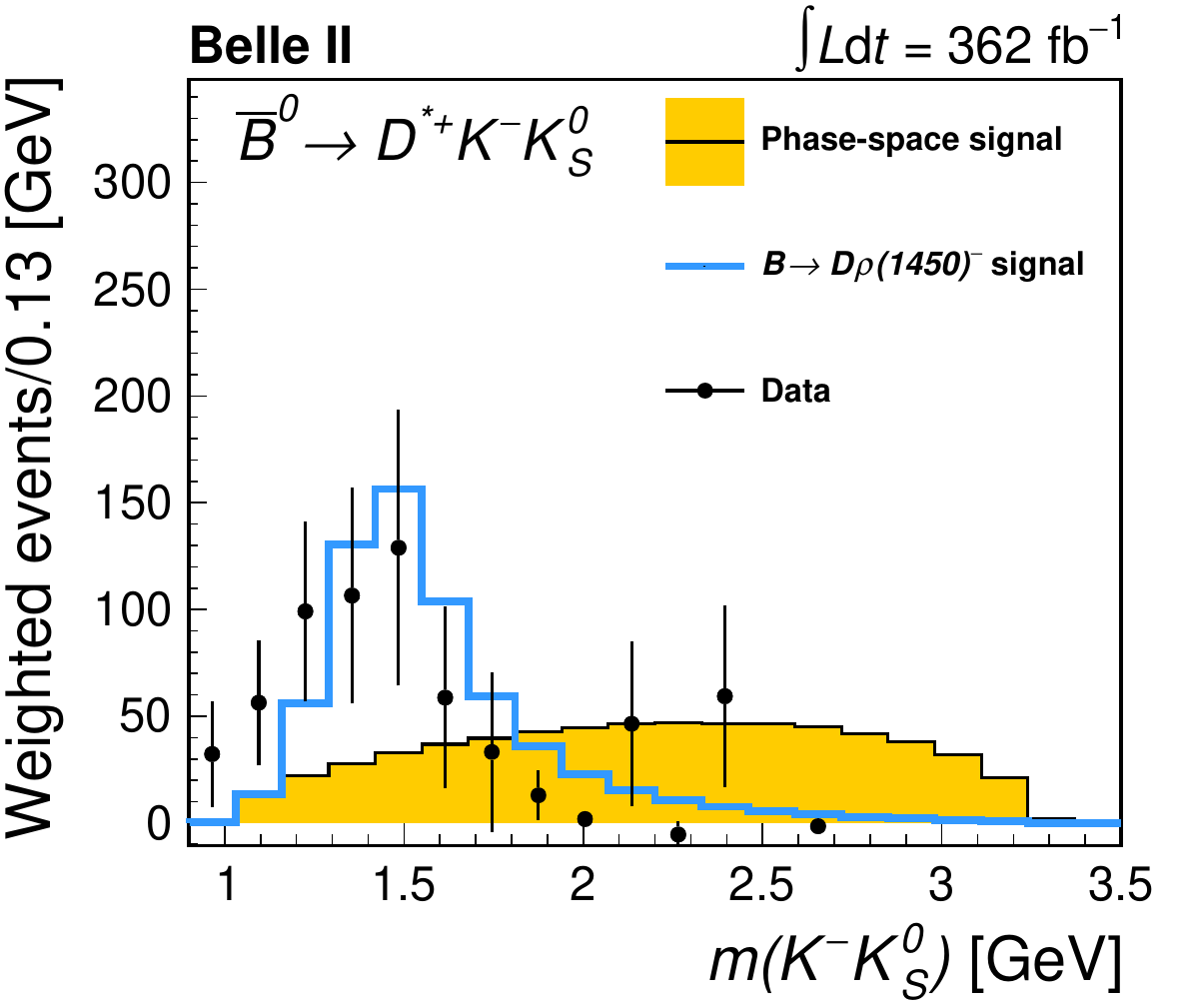}}
\caption{Background-subtracted and efficiency-corrected distribution of $m(K^-K_S^0)$ for the $B^-\to D^0K^-K_S^0$ (top left),  $\overline B{}^0\to D^+K^-K_S^0$ (top right), $B^-\to D^{*0}K^-K_S^0$ (bottom left), and $\overline B{}^0\to D^{*+}K^-K_S^0$ (bottom right)  channels.  The error bars represent the statistical uncertainties. A phase-space MC simulation and a resonant MC simulation at generator level, rescaled to the integral of the data distribution, are also shown for comparison. }\label{fig:mKK_effCorr_KS0}
\end{figure}

\begin{figure}[!h]
\centering
\subfigure{\includegraphics[width=0.45\columnwidth]{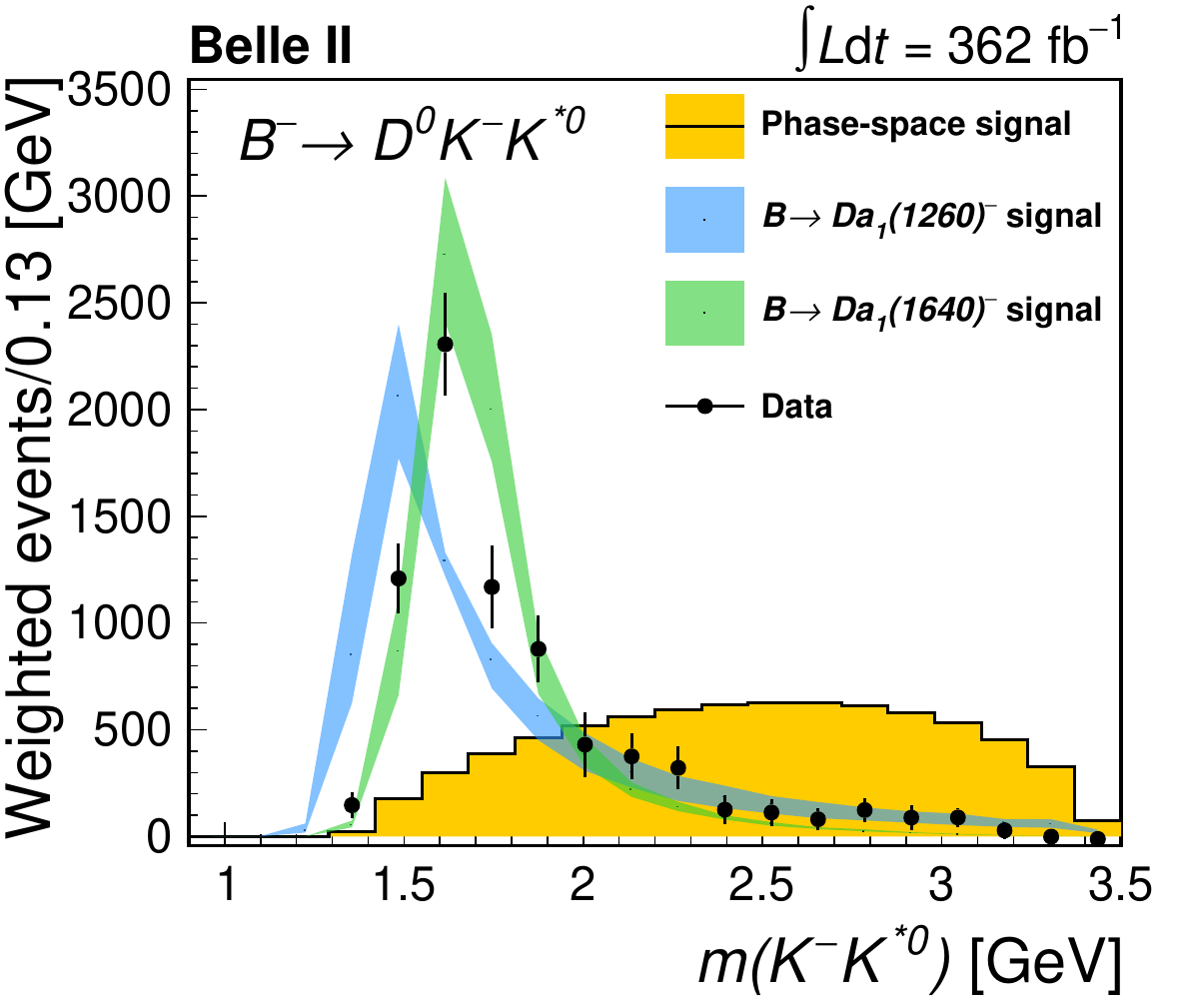}}
\subfigure{\includegraphics[width=0.45\columnwidth]{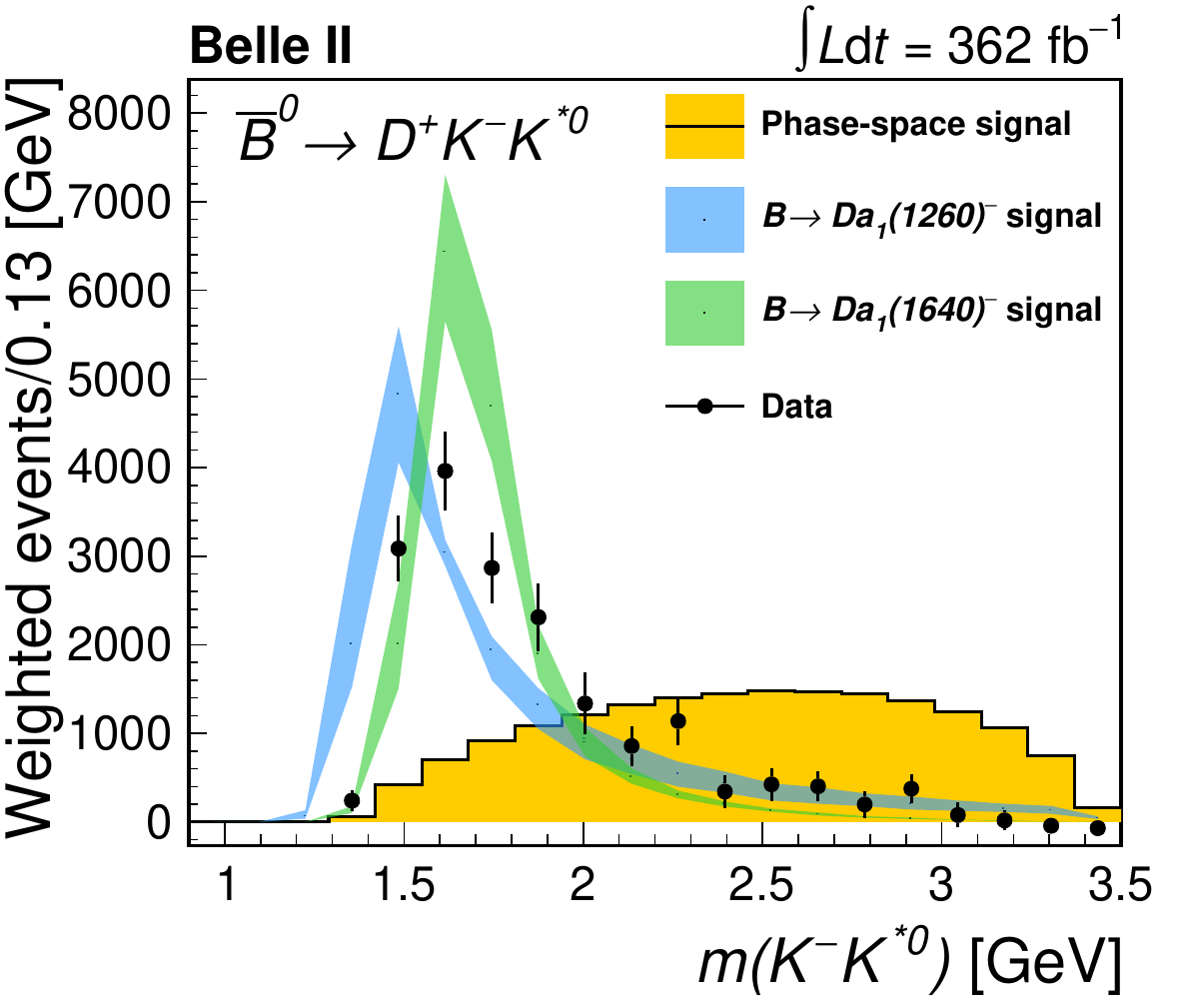}}
\subfigure{\includegraphics[width=0.45\columnwidth]{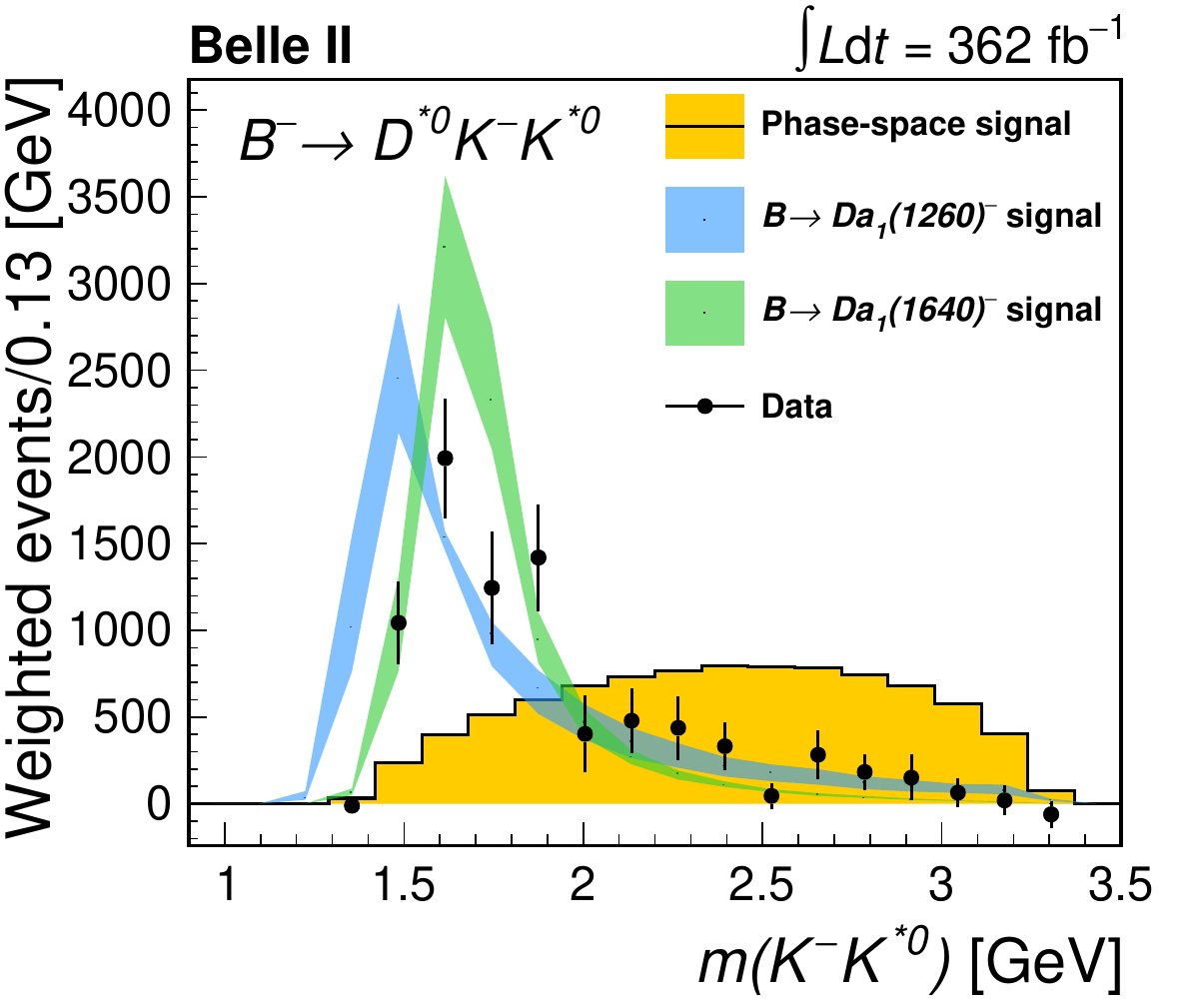}}
\subfigure{\includegraphics[width=0.45\columnwidth]{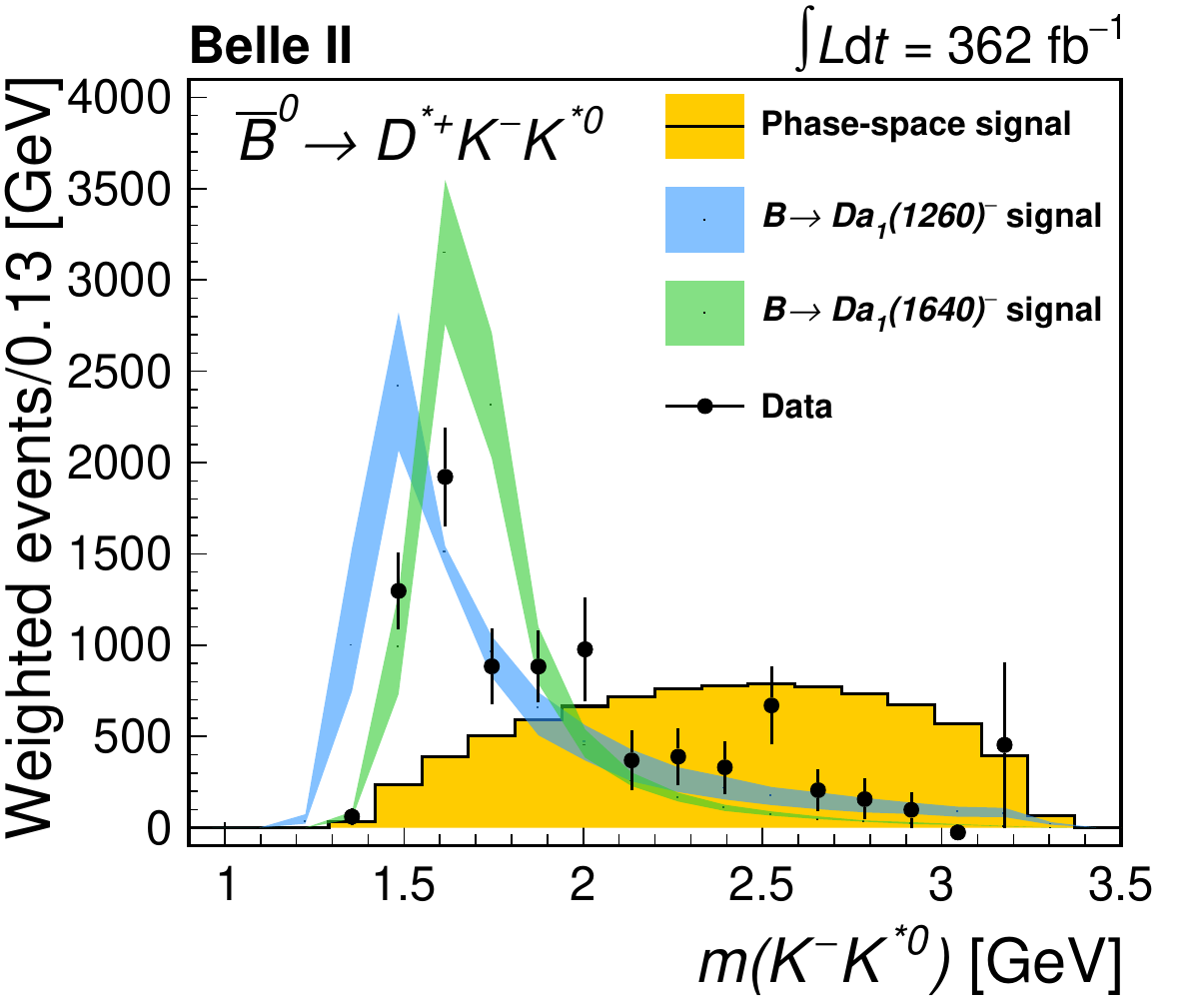}}
\caption{Background-subtracted and efficiency-corrected distribution of $m(K^-K^{*0})$ for the $B^-\to D^0K^-K^{*0}$ (top, left), $\overline B{}^0\to D^+K^-K^{*0}$ (top right), $B^-\to D^{*0}K^-K^{*0}$ (bottom left), and $\overline B{}^0\to D^{*+}K^-K^{*0}$ (bottom right) channels. The error bars represent the statistical uncertainties. A phase-space MC simulation and two resonant MC simulation at generator level, rescaled to the integral of the data distribution, are also shown for comparison. The uncertainties on the resonance parameters simulations are shown as shaded areas.} \label{fig:mKK_effCorr_Kst0}
\end{figure}

The $m(K^-K)$ distributions for data are shown in Fig.~\ref{fig:mKK_effCorr_KS0} and Fig.~\ref{fig:mKK_effCorr_Kst0}. The $m(K^-K)$ distribution from the signal simulation is also overlaid to show a comparison with phase-space and specific resonant distributions. In all channels, the bulk of the observed  $m(K^- K_S^0)$ distributions is located at low $m(K^- K_S^0)$ {values}, {and show} structures different from phase-space distributions. The distributions vanish above 2.0--2.5~GeV, disfavoring the presence of a significant phase-space component. The overlay with the $\rho(1450)^-$ meson lineshape shows partial agreement of the peak position.  Therefore, combining the observed $m(K^-K)$ distribution and the helicity angle constraint, we conclude that the four $K_S^0$ {channels} proceed {predominantly} via $B\to D\rho^{\prime-}(\to K^-K_S^0)$, where $\rho^\prime$ indicates one or more $\rho-$like resonance. This interpretation is supported by Ref.~\cite{Theory:DKKS0_new}, where the $m(K^- K_S^0)$ distribution is {predominantly} described as the combination of the   $\rho(1450)^-$ and $\rho(770)^-$ contributions. 
Moreover, we cannot exclude the interference of even-spin states, which are allowed if factorization is not assumed. {In this case, the transition can proceed via {a} color-suppressed diagram with internal $W$ boson emission, and {could} result in a $J^P=0^+$ state. However, from Ref.~\cite{Theory:DKKS0_new} this contribution is expected to be small or negligible.} For the $K^{*0}$ channels, the observed $m(K^-K^{*0})$ distribution also differs from phase-space, with the peak position and the high mass tail in good agreement with the simulated $a_1(1640)^-$ meson lineshape, thus strongly {disfavoring} a non-resonant transition. 
Given the imperfect agreement with a pure $a_1(1640)^-$ lineshape, we cannot exclude the possibility that the observed lineshape is due to the superposition of multiple $a_1^-$ resonances. However, the $a_1(1260)^-$ is unlikely to be the dominant contribution given the observed peak position, despite the large uncertainties related to the knowledge of the $a_1(1260)^-$ width. Combining the observed  $m(K^-K^{*0})$ distribution and the information from the helicity angles, we conclude that the four $K^{*0}$ {channels} proceed via  $B\to D a_1^{\prime-}(\to K^-K^{*0})$, where $a_1^{\prime}$ stands for one or multiple $a_1$-like resonances. 

\clearpage
\section{Conclusions}

Using an electron-positron data sample collected by Belle~II on the $\Upsilon(4S)$ resonance with an integrated luminosity of $362~\text{fb}^{-1}$, we report the measurement of the eight branching fractions
\begin{align*}
    \mathcal{B}(B^-\to D^0K^-K_S^0)=                &(1.82\pm 0.16\pm0.08)\times 10^{-4},\\
    \mathcal{B}(\overline B{}^0\to D^+K^-K_S^0)=    &(0.82\pm 0.12\pm0.05)\times 10^{-4},\\
    \mathcal{B}(B^-\to D^{*0}K^-K_S^0)=             &(1.47\pm 0.27\pm0.10)\times 10^{-4},\\
    \mathcal{B}(\overline B{}^0\to D^{*+}K^-K_S^0)= &(0.91\pm 0.19\pm0.05)\times 10^{-4}\\
    \mathcal{B}(B^-\to D^0K^-K^{*0})=               &(7.19\pm 0.45\pm0.33)\times 10^{-4},\\
    \mathcal{B}(\overline B{}^0\to D^+K^-K^{*0})=   &(7.56\pm 0.45\pm0.38)\times 10^{-4},\\
    \mathcal{B}(B^-\to D^{*0}K^-K^{*0})=            &(11.93\pm 1.14\pm0.93)\times 10^{-4},\\
    \mathcal{B}(\overline B{}^0\to D^{*+}K^-K^{*0})=&(13.12\pm 1.21\pm0.71)\times 10^{-4}
\end{align*} 
where the first uncertainty is statistical and the second {uncertainty is} systematic. 
The decays $\overline B{}^0\to D^+K^-K_S^0$,  $B^-\to D^{*0}K^-K_S^0$, and $\overline B{}^0\to D^{*+}K^-K_S^0$ are observed for the first time. 
The precision of the eight measurements is between 5\% and 20\% and is limited by statistical uncertainties. 
The precision of $\mathcal{B}(B^-\to D^0K^-K_S^0)$ and the four {$\mathcal B(\overline B\to D K^-K^{*0})$ are} improved by more than a factor of three compared to previous measurements~\cite{Belle:DKK}.  

The observed $m(K^-K^{(*)0}_{(S)})$ invariant mass and helicity angles distributions suggest the presence of a dominant resonant component in the $K^-K^{(*)0}_{(S)}$ system as has been previously reported by Belle~\cite{Belle:DKK}. In $B\to DK^-K_S^0$ decays, a $B\to D\rho^{\prime-}(\to K^-K_S^0)$ transition is {favored}, where $\rho^\prime$ stands for a $J^P=1^-$ state. The transitions are likely to proceed via multiple resonant states, and we cannot exclude the presence of an even-spin intermediate state. In $B\to DK^-K^{*0}$ decays a $B\to Da_1^{\prime-}(\to K^-K^{*0})$ {transition} is {favored}, where $a_1^{\prime}$ stands for a  $J^P=1^+$ state. The observed $m(KK^{*0})$ distributions favor the contribution of not only the $a_1(1260)^-$, but also excited $a_1^-$ states such as the $a_1(1640)^-$, in contrast to the earlier Belle measurement~\cite{Belle:DKK}.

These eight {channels} can be exploited in the FEI $B$-tagging algorithm, given their high purity. They can provide a few percent improvement in efficiency for hadronic FEI. 

We also provide measurements of the branching fractions of the four channels {$B\to DD_s^- $}, obtained from the same data sample. They are found to be
\begin{align*}
    \mathcal{B}(B^-\to D^0D_s^-)=                     &(95\pm6\pm5)\times 10^{-4},\\
    \mathcal{B}(\overline B{}^0\to D^+D_s^-)=         &(89\pm5\pm5)\times 10^{-4},\\
    \mathcal{B}(B^-\to D^{*0}D_s^-)=                  &(65\pm10\pm6)\times 10^{-4},\\
    \mathcal{B}(\overline B{}^0\to D^{*+}D_s^-)=      &(83\pm10\pm6)\times 10^{-4}
\end{align*} 
where the first uncertainty is statistical and the second one systematic. These measurements have a precision similar to or competitive with the current world averages~\cite{PDG}.

\appendix
\section{Additional material}\label{sec:App}

\begin{figure}[!hbt]
\centering
\subfigure{\includegraphics[width=0.45\columnwidth]{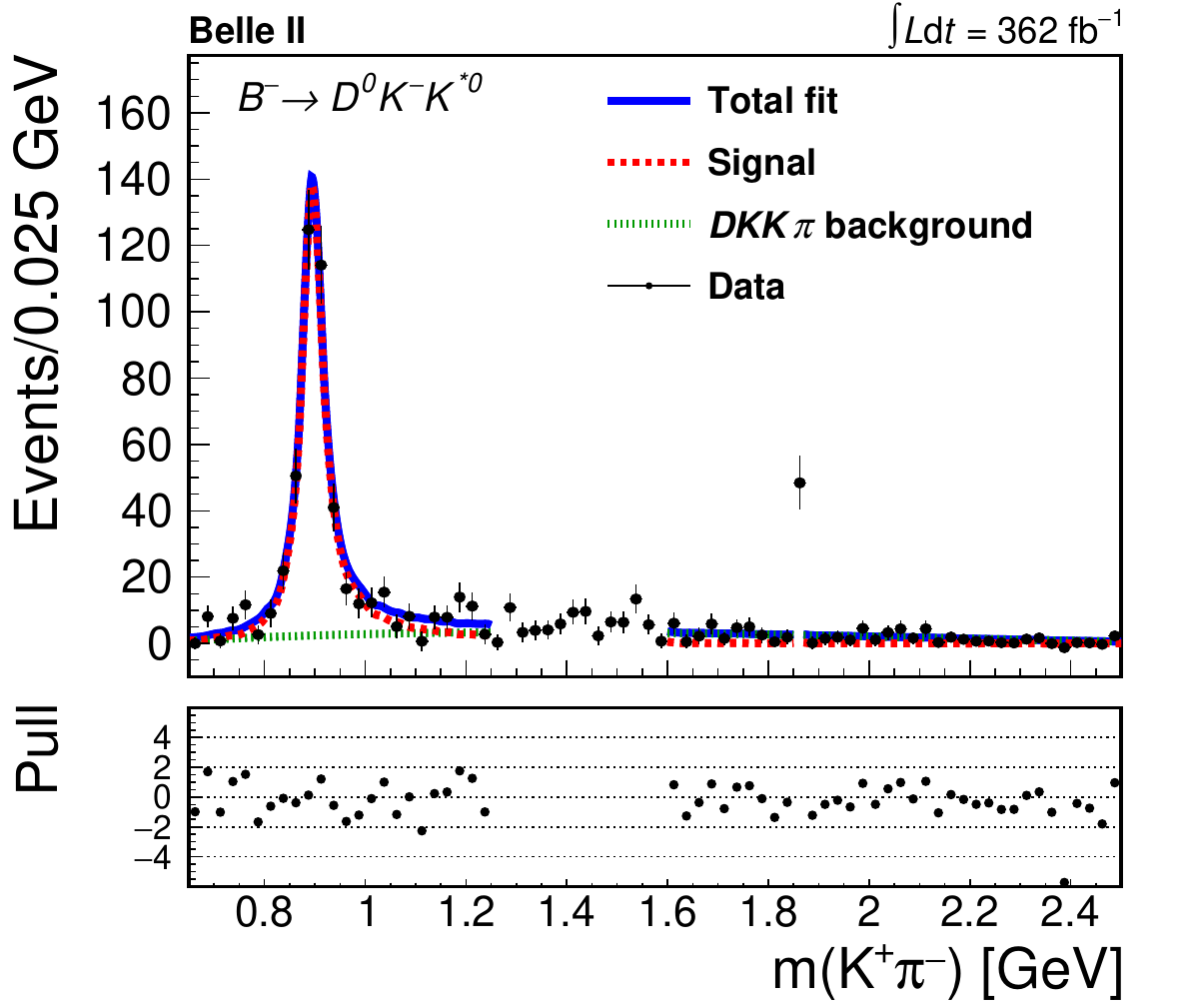}}
\subfigure{\includegraphics[width=0.45\columnwidth]{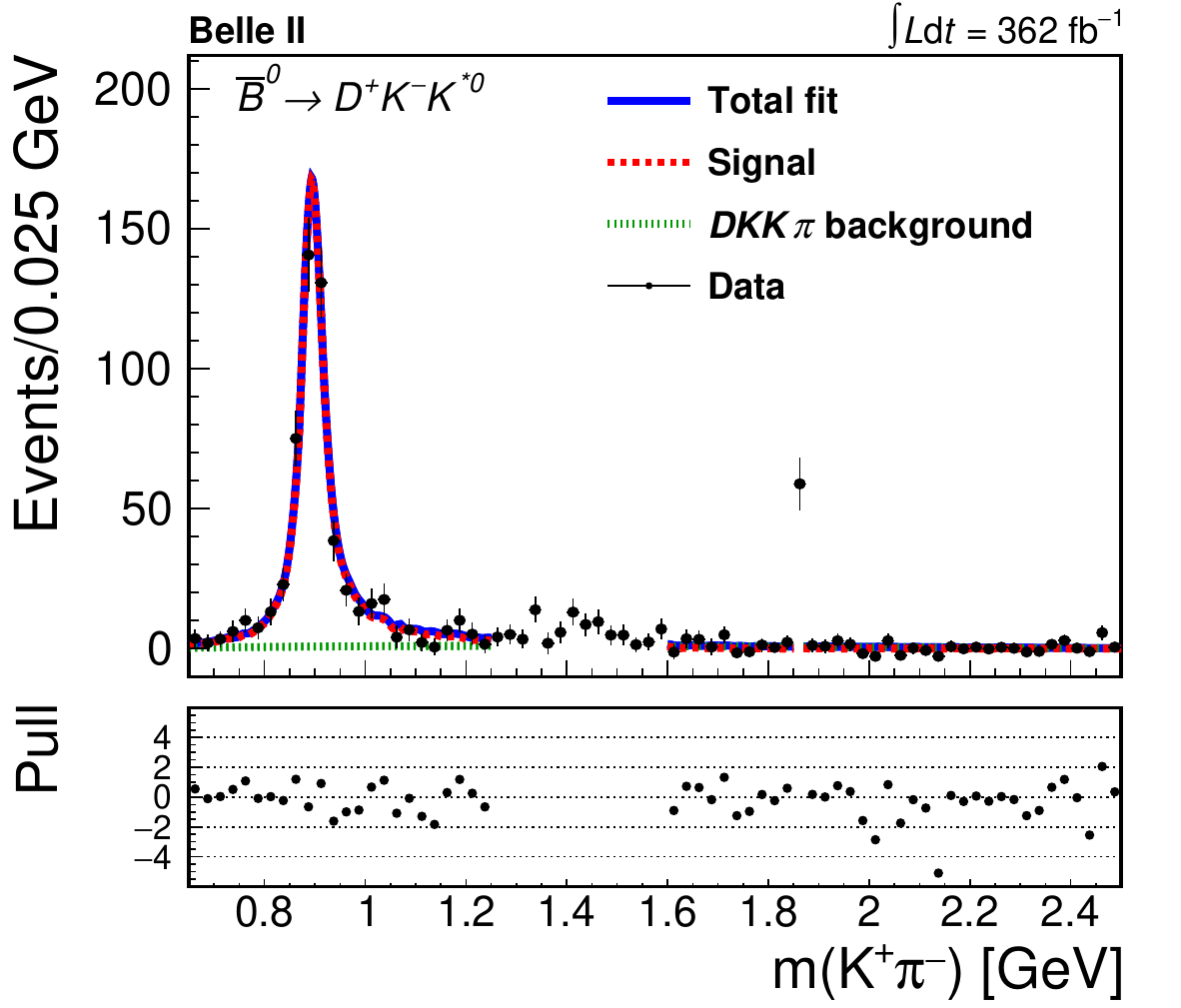}}
\subfigure{\includegraphics[width=0.45\columnwidth]{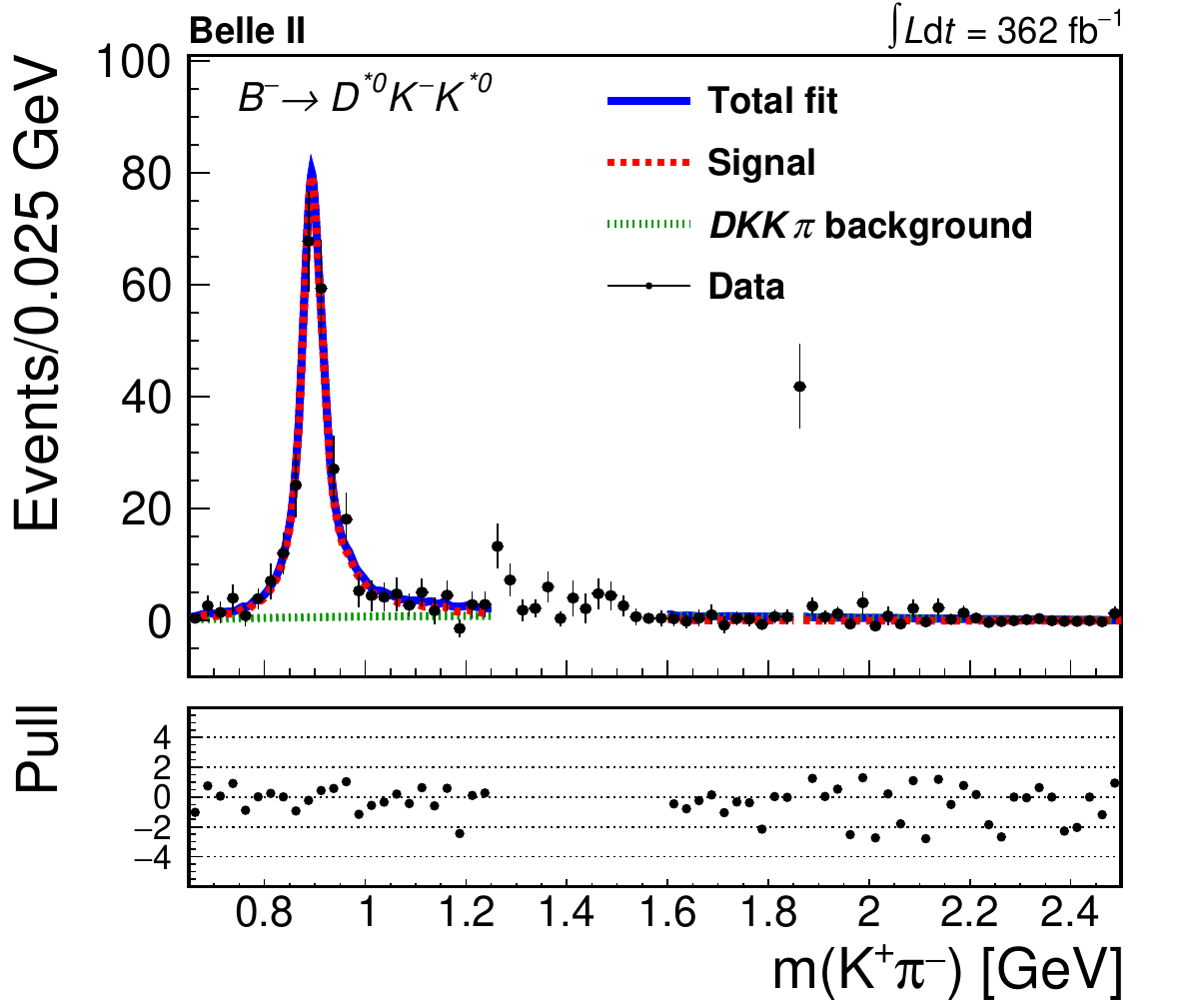}}
\subfigure{\includegraphics[width=0.45\columnwidth]{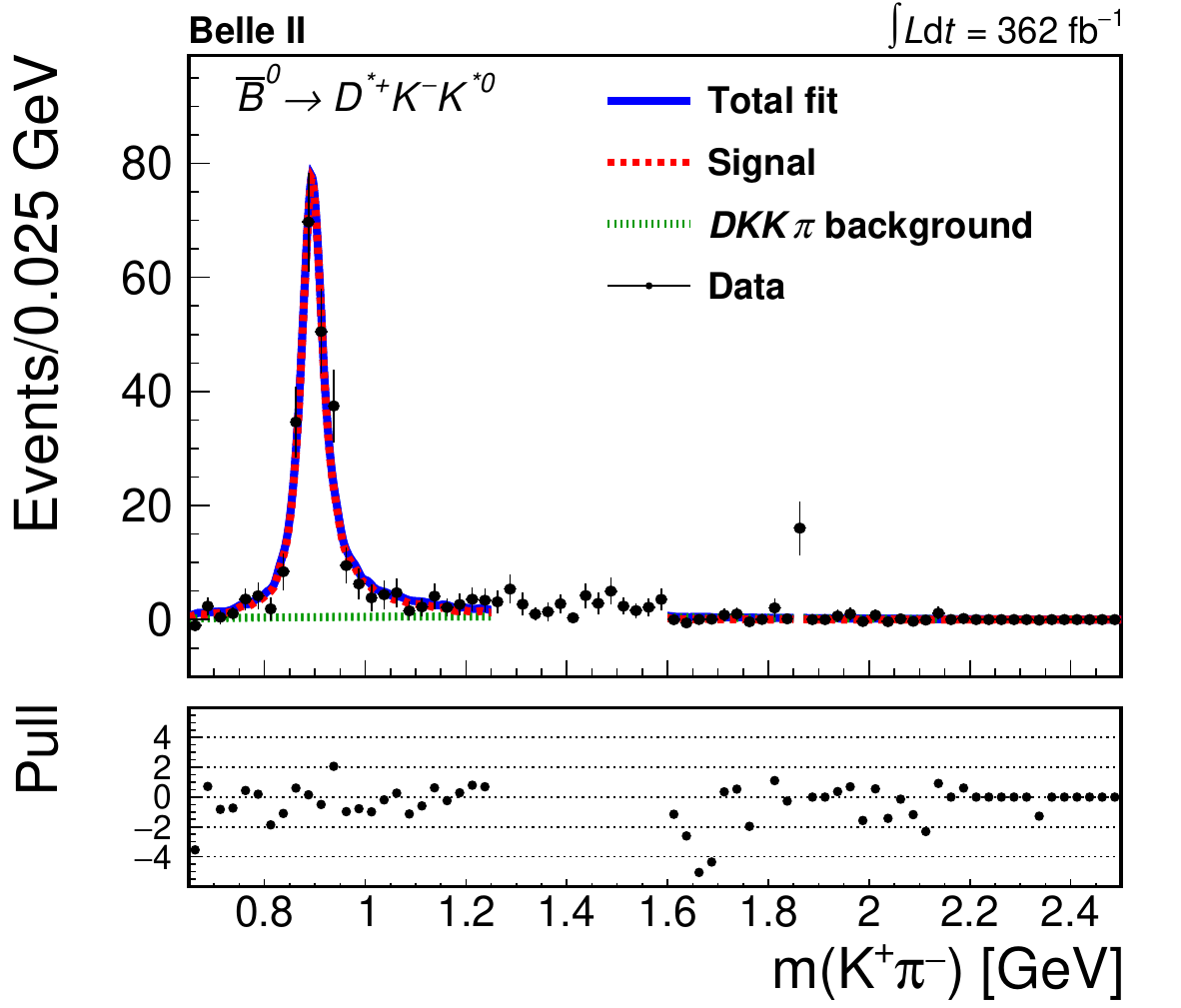}}
..\caption{Distribution of $m(K^+\pi^-)$  for the $B^-\to D^0K^-K^{*0}$ (top left), $\overline B{}^0\to D^+K^-K^{*0}$ (top right), $B^-\to D^{*0}K^-K^{*0}$ (bottom left),  and $\overline B{}^0\to D^{*+}K^-K^{*0}$ (bottom right) channels. The projections of the fits are overlaid, the fit components are highlighted, and the pulls between the fit and the data are shown in the bottom panel of each plot. The background $B\overline B$ and continuum background are subtracted by applying the signal $s$Weights.} \label{fig:mKpi}
\end{figure}

In Fig.~\ref{fig:mKpi} the distribution of $m(K^+\pi^-)$ invariant mass for the four $K^{*0}$ channels is shown. The projection of the fit, as described in Sec.~\ref{sec:yield}, is overlaid. The non-$K^{*0}$-resonant fraction is listed in Table~\ref{tab:NRfraction} for the four $K^{*0}$ channels. The systematic uncertainties are estimated as described in Sec.~\ref{sec:syst}.

In Fig.~\ref{fig:angles_fit_K} and Fig.~\ref{fig:angles_fit_KK} the distributions of $dN/d\cos\theta_{K^{(*)}_{(S)}}$ and $dN/d\cos\theta_{KK}$ respectively, are shown for the eight $B\to DK^-K$ channels. The tested fit hypotheses, according to Table~\ref{tab:angles_expected}, are also shown. These fits models have only the total yield as a free parameter and they consider statistical uncertainty only. A phase-space MC simulation and a specific resonant MC are overlaid as a reference.  For the $K_S^0$ channels, the four $\cos(\theta_{K_S})$ fit show good agreement with the uniform distribution, as expected. The $\cos(\theta_{KK})$ distributions for the $\overline B{}^0\to D^{+}K^-K_S^0$ and $\overline B{}^0\to D^{*+}K^-K_S^0$ channels follow a $\cos^2\theta$ distribution, in agreement with $J^P=1^-$. On the other hand, the $\cos(\theta_{KK})$ distributions for the $ B^-\to D^{0}K^-K_S^0$ and $ B^-\to D^{*0}K^-K_S^0$ channels disagree with all the tested hypotheses. 
The unknown polarization of the $D^{*0}$ justifies the observed distribution for the $B^-\to D^{*0}K^-K_S^0$ channel. The observed asymmetric $\cos(\theta_{KK})$ distribution for the $B^-\to D^{0}K^-K_S^0$ channel supports the hypothesis of interference between the $J^P=1^-$ state and spin-even states. 
For the $K^{*0}$ channels, both the $\cos(\theta_{K^*})$and $\cos(\theta_{KK})$ distributions are in good agreement with the uniform distribution, as expected for $J^P=1^+$, but for the $\cos(\theta_{KK})$ of the $B{}^0\to D^{*+}K^-K^{*0}$ channel.

\begin{figure}[!t]
\centering
\subfigure{\includegraphics[width=0.37\columnwidth]{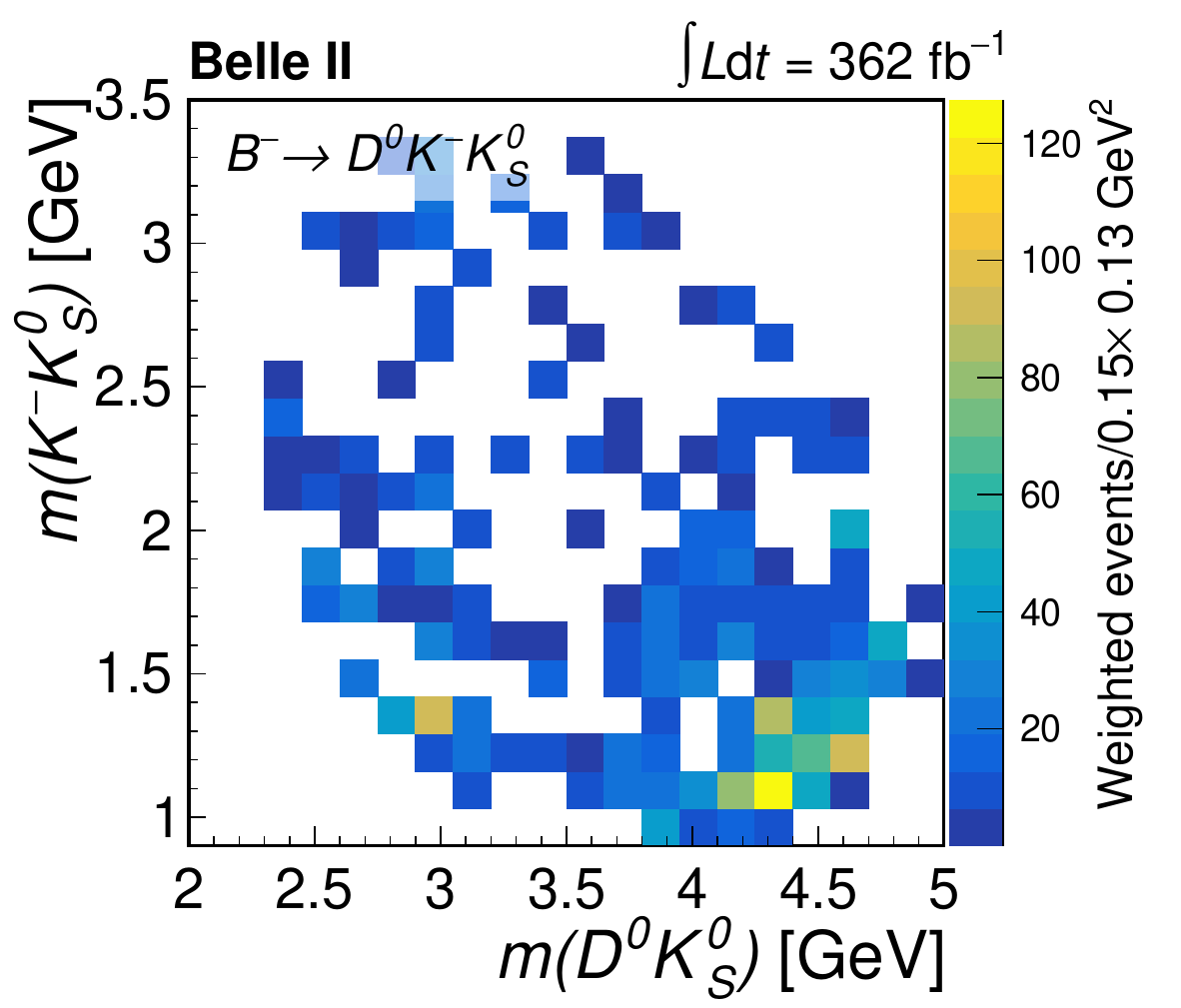}}
\subfigure{\includegraphics[width=0.37\columnwidth]{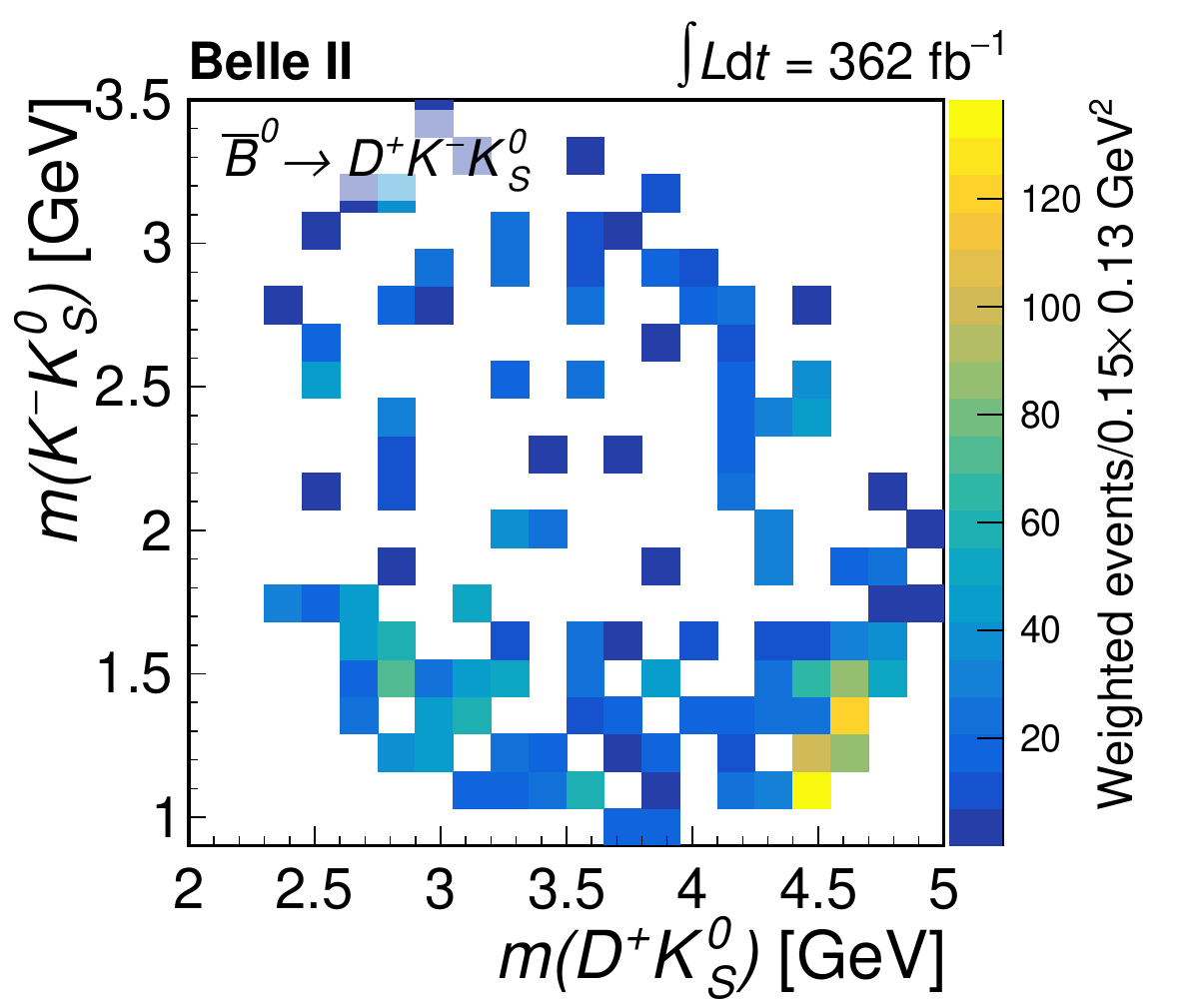}}
\subfigure{\includegraphics[width=0.37\columnwidth]{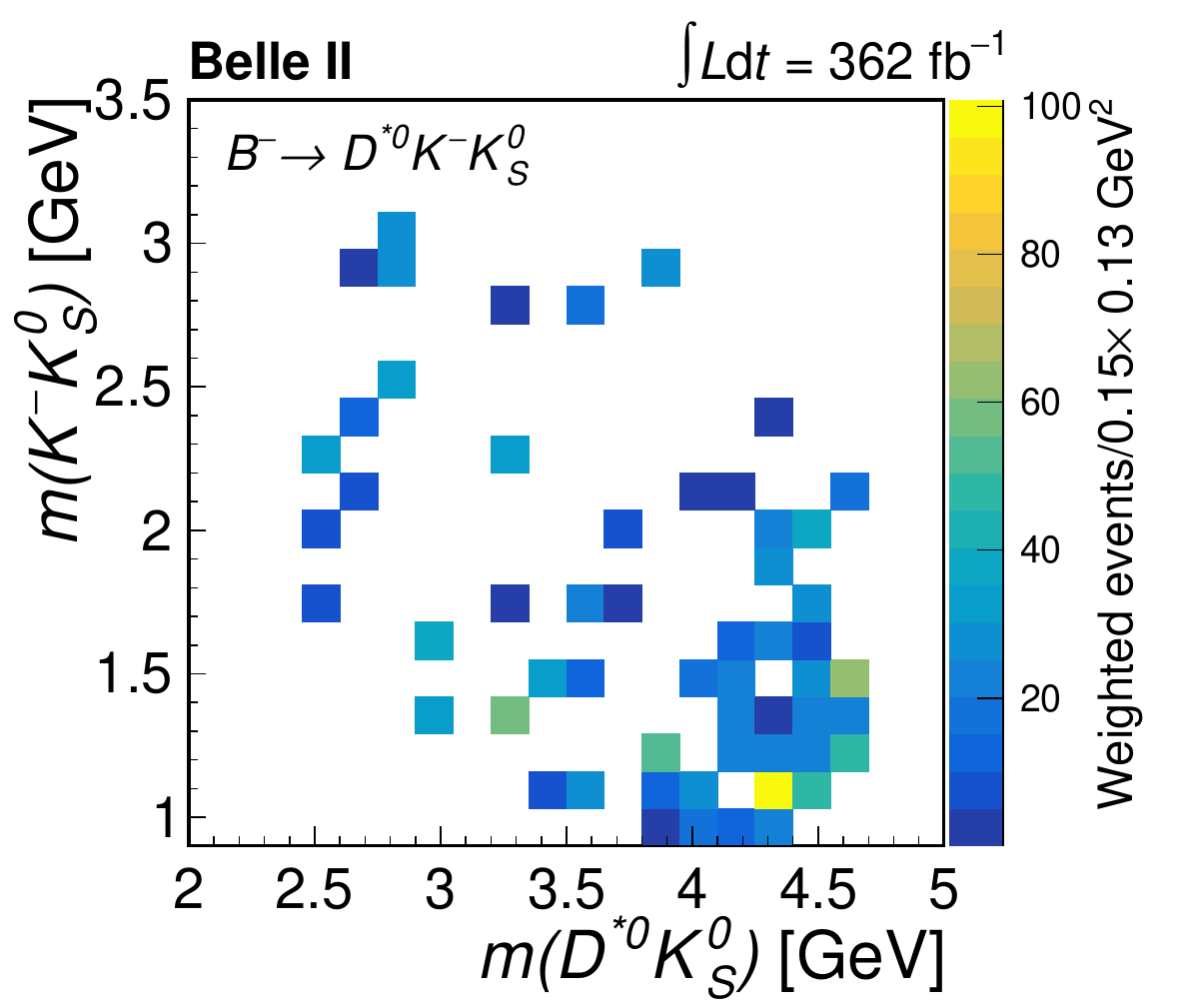}}
\subfigure{\includegraphics[width=0.37\columnwidth]{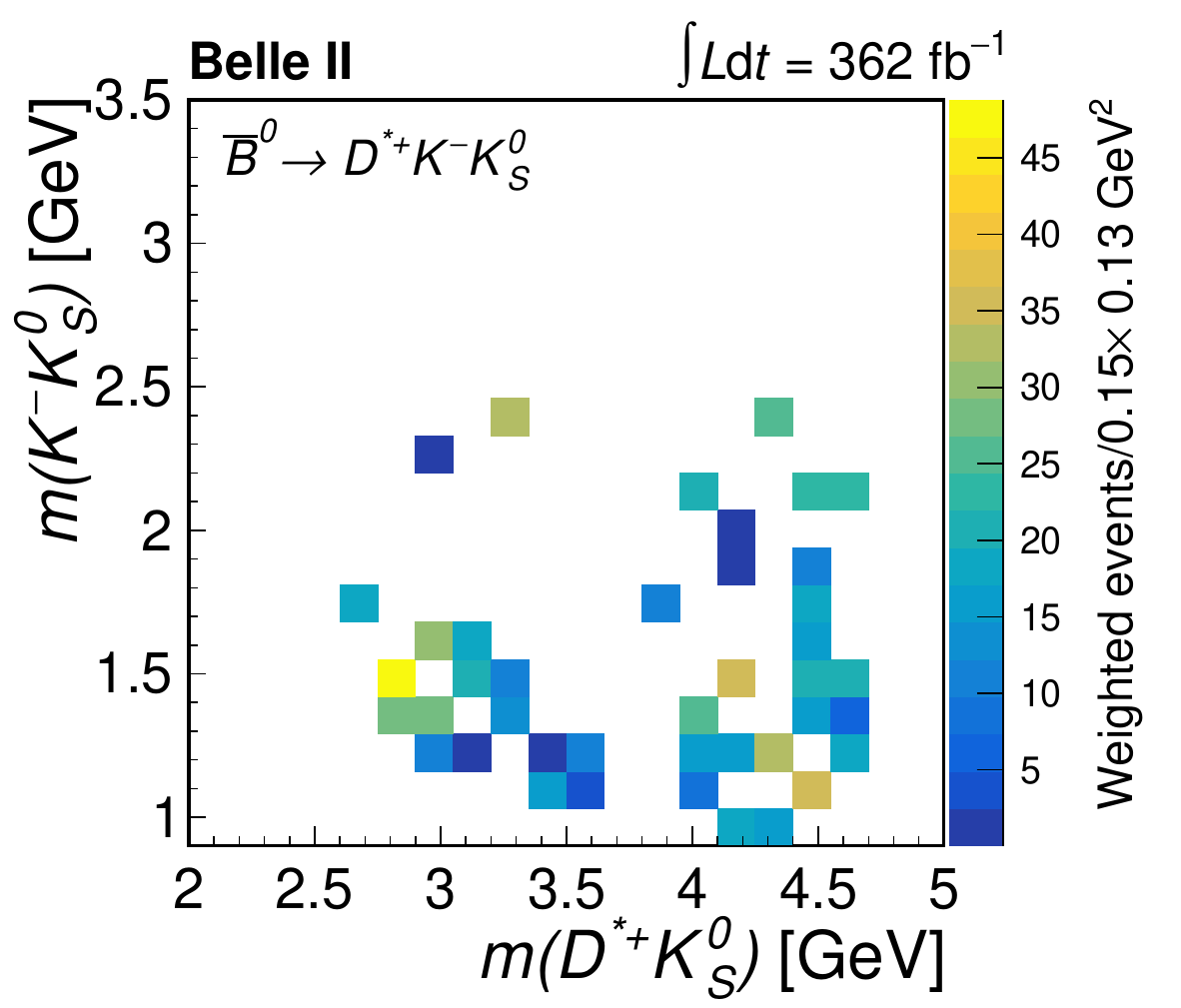}}
\subfigure{\includegraphics[width=0.37\columnwidth]{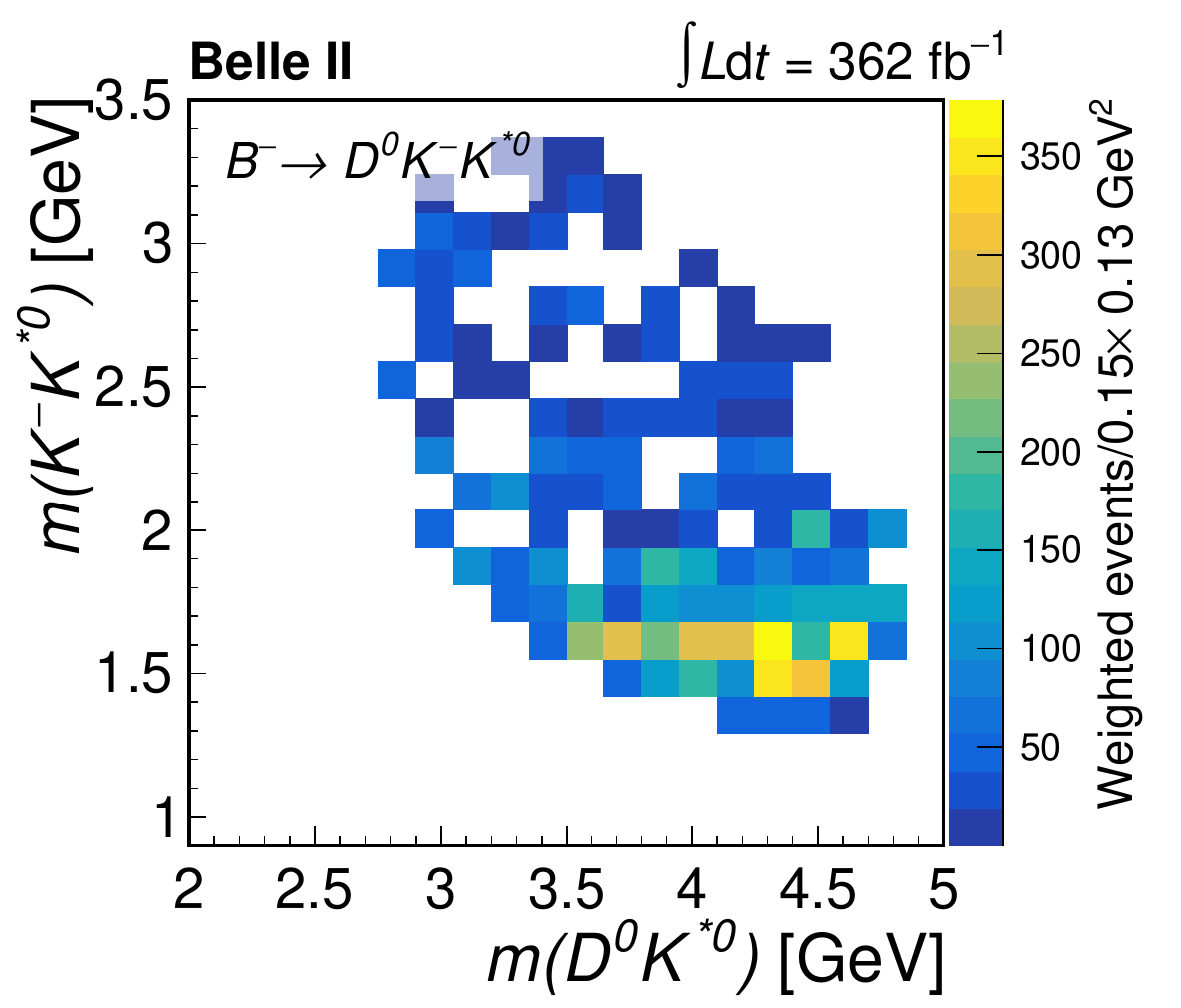}}
\subfigure{\includegraphics[width=0.37\columnwidth]{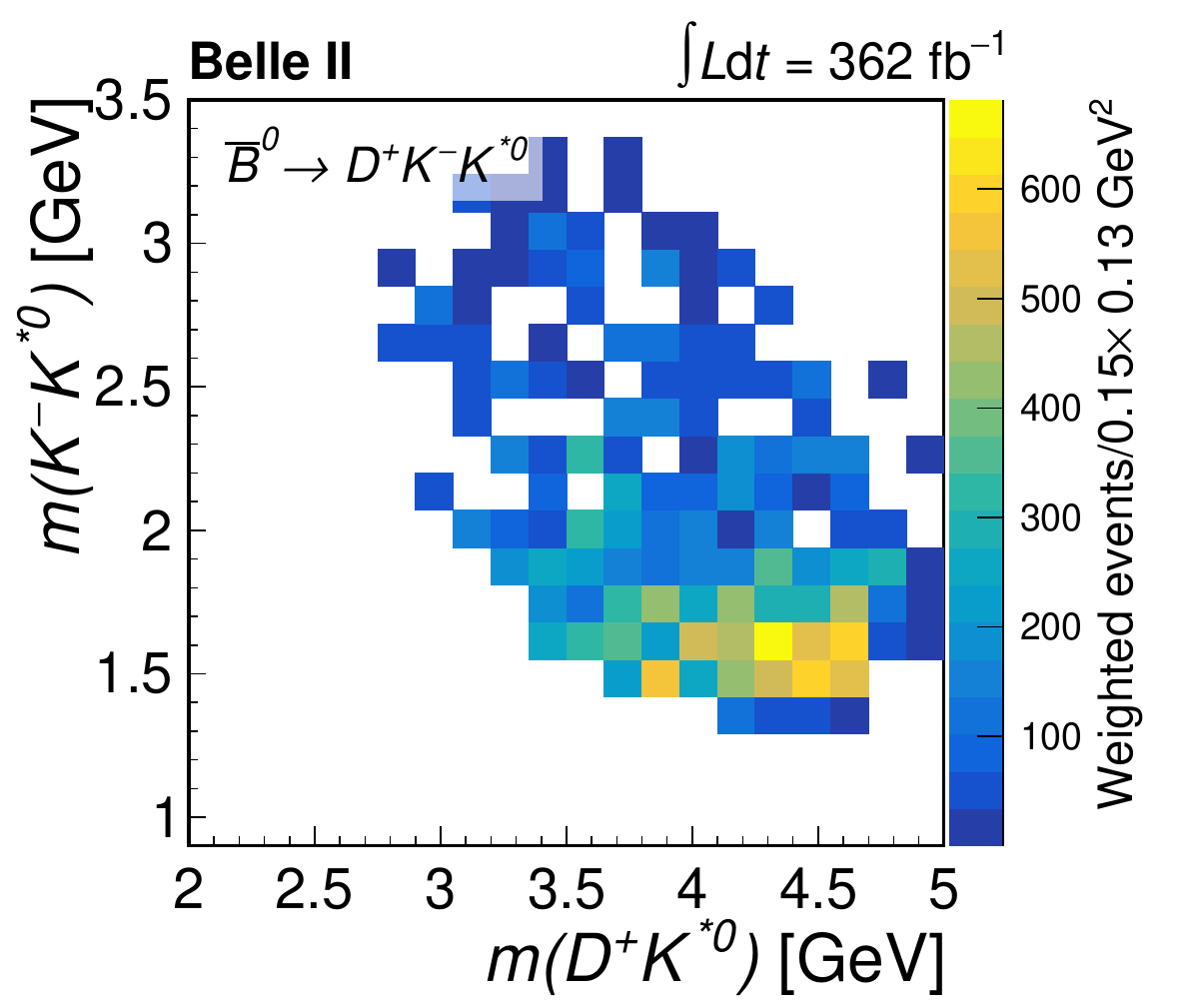}}
\subfigure{\includegraphics[width=0.37\columnwidth]{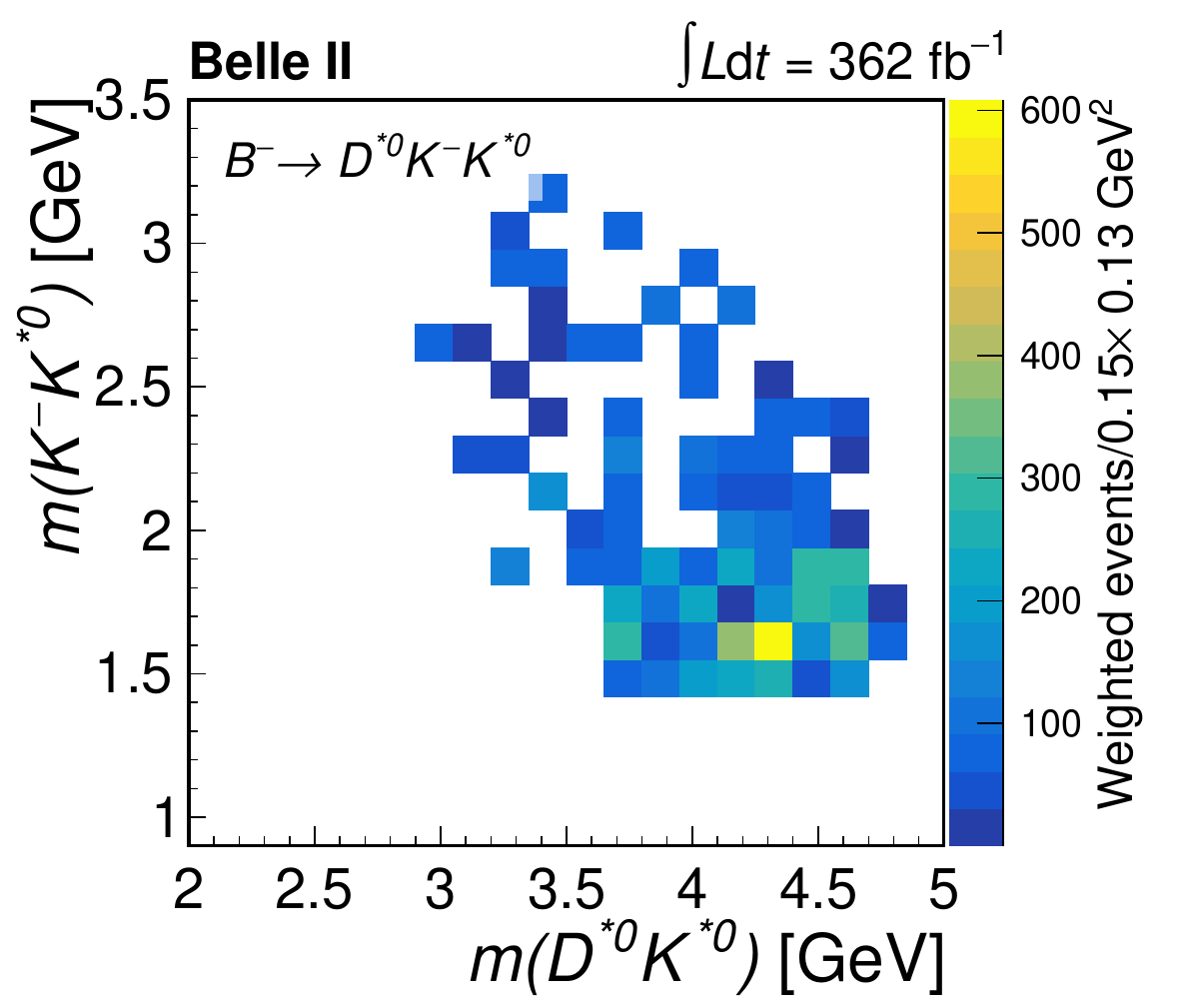}}
\subfigure{\includegraphics[width=0.37\columnwidth]{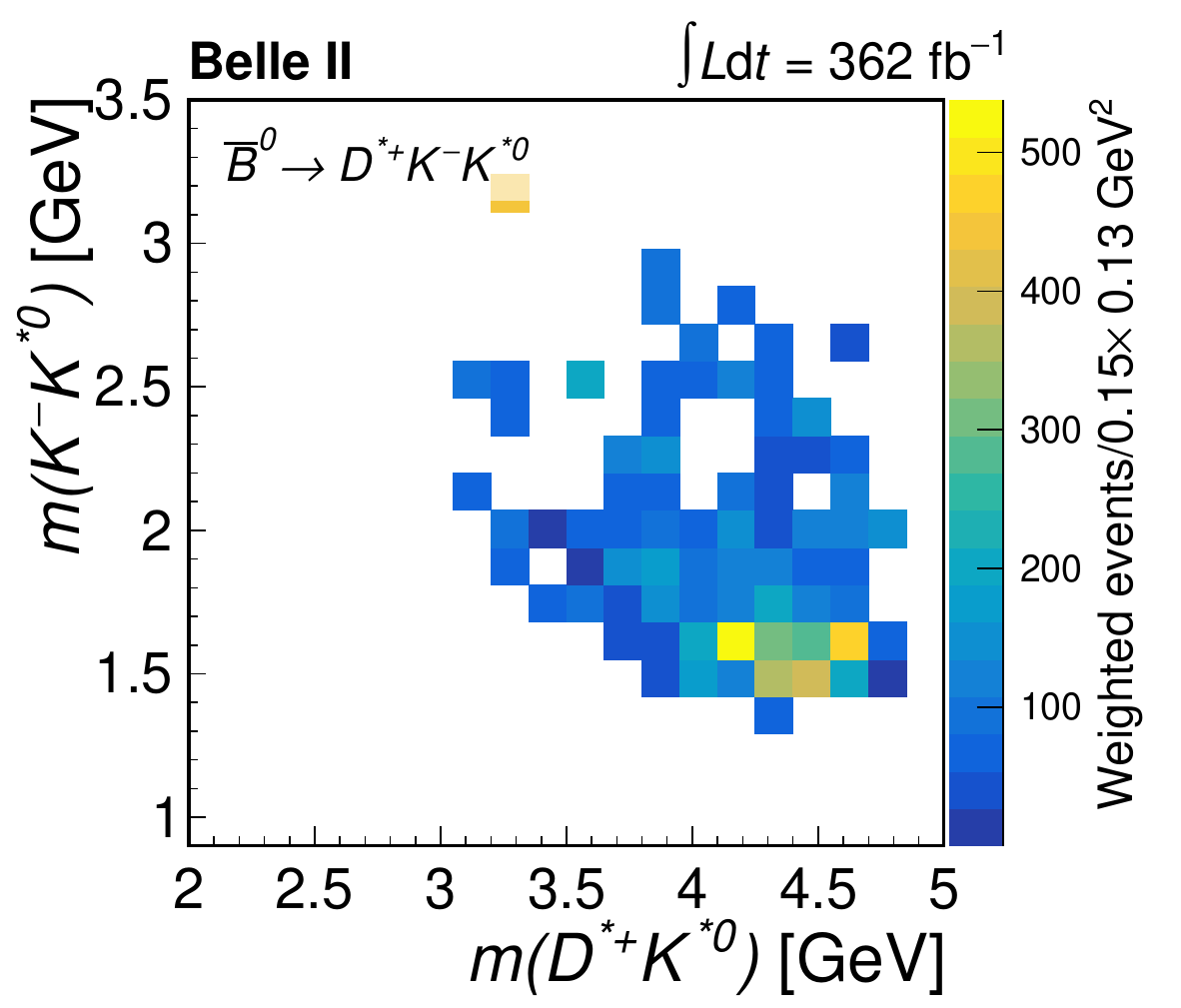}}
\caption{Background-subtracted and efficiency-corrected distribution of $\bigl(m(K^-K),m(DK)\bigr)$ for the $B^-\to D^0K^-K_S^0$ (first line, left),  $\overline B{}^0\to D^+K^-K_S^0$ (first line, right), $B^-\to D^{*0}K^-K_S^0$ (second line, left), $\overline B{}^0\to D^{*+}K^-K_S^0$ (second line, right), $B^-\to D^0K^-K^{*0}$ (third line, left), $\overline B{}^0\to D^+K^-K^{*0}$ (third line, right), $B^-\to D^{*0}K^-K^{*0}$ (fourth line, left), and $\overline B{}^0\to D^{*+}K^-K^{*0}$ (fourth line, right) channels.} \label{fig:Dalitz}
\end{figure}

\begin{figure}[!hbt]
\centering
\subfigure{\includegraphics[width=0.37\columnwidth]{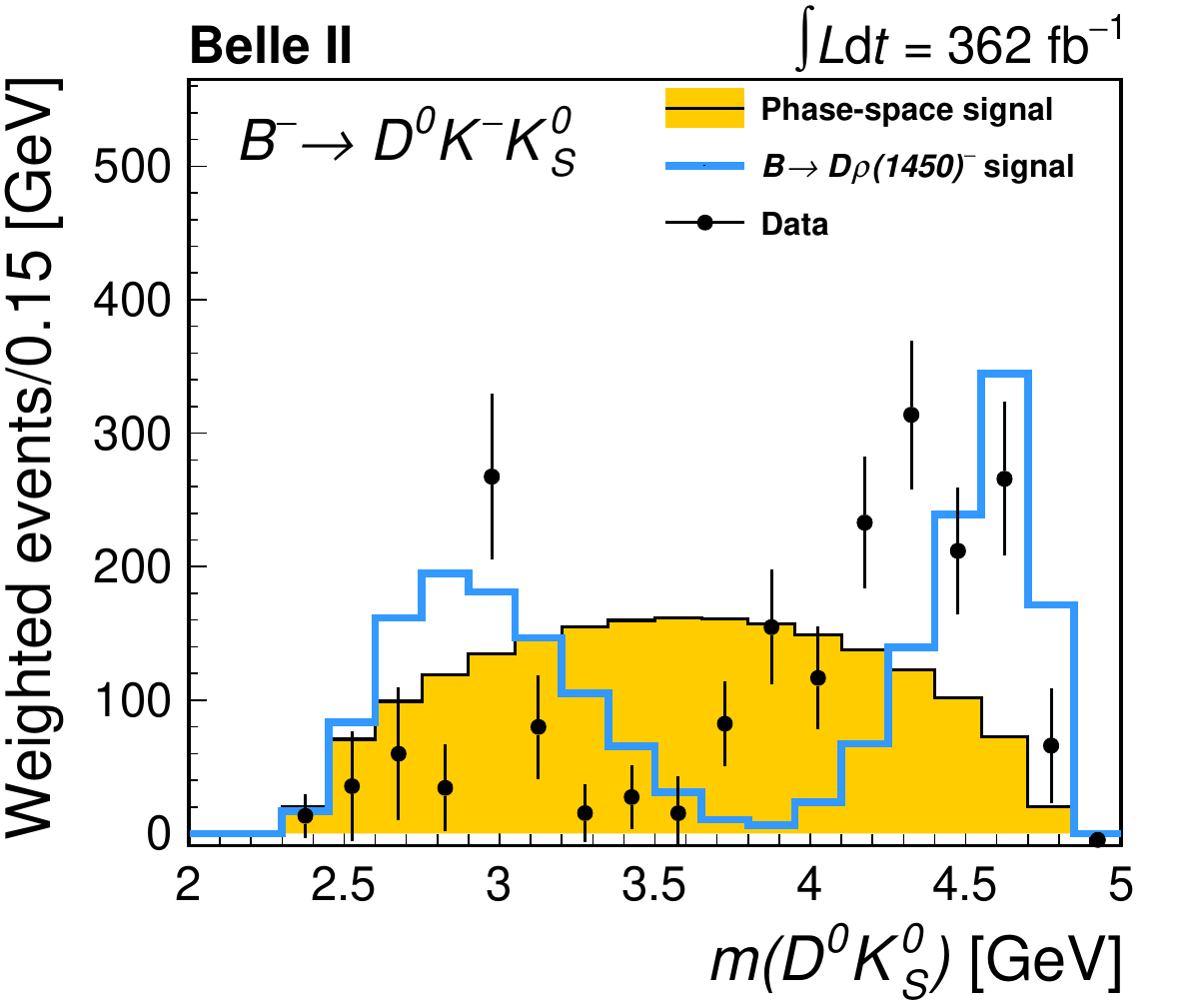}}
\subfigure{\includegraphics[width=0.37\columnwidth]{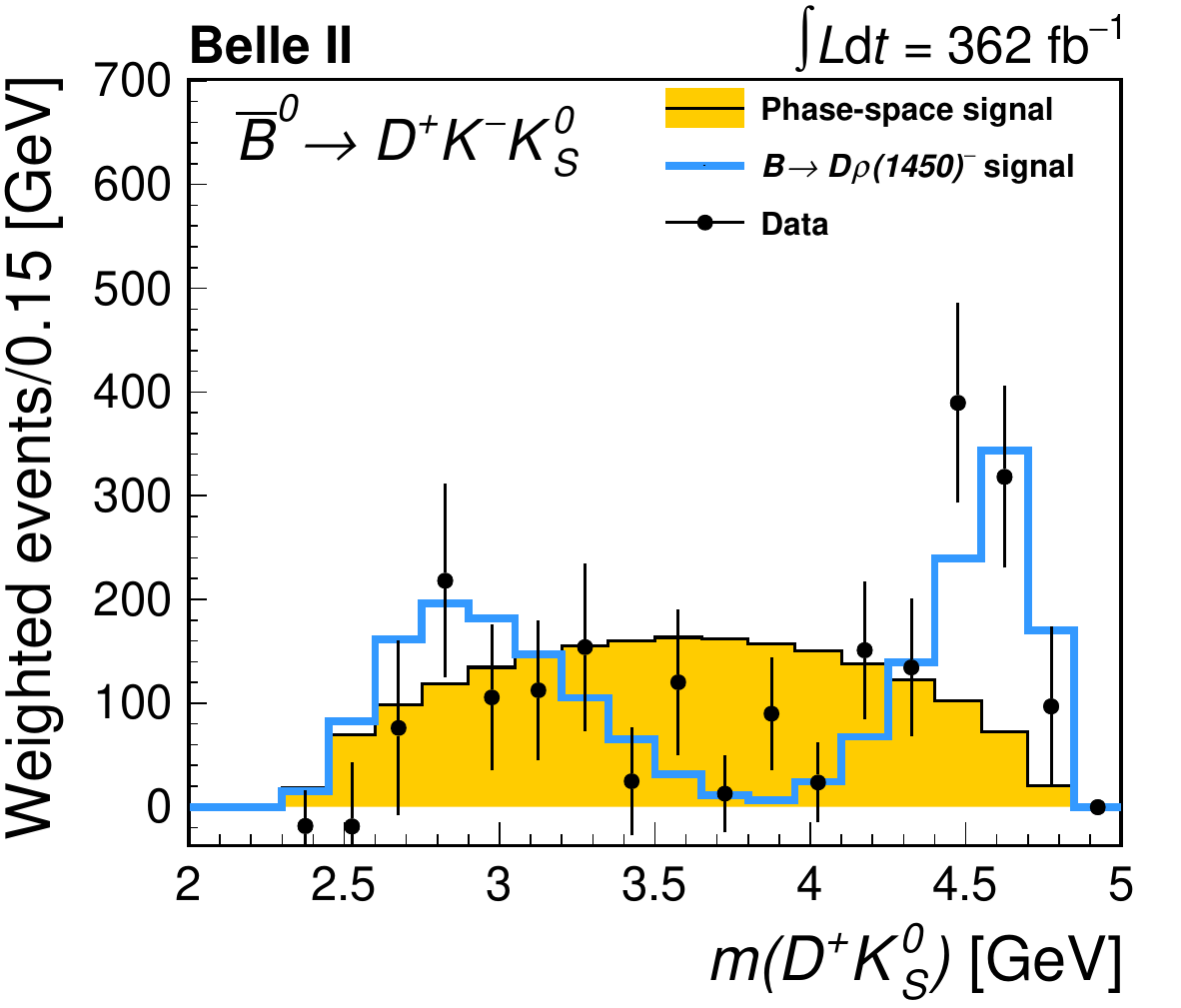}}
\subfigure{\includegraphics[width=0.37\columnwidth]{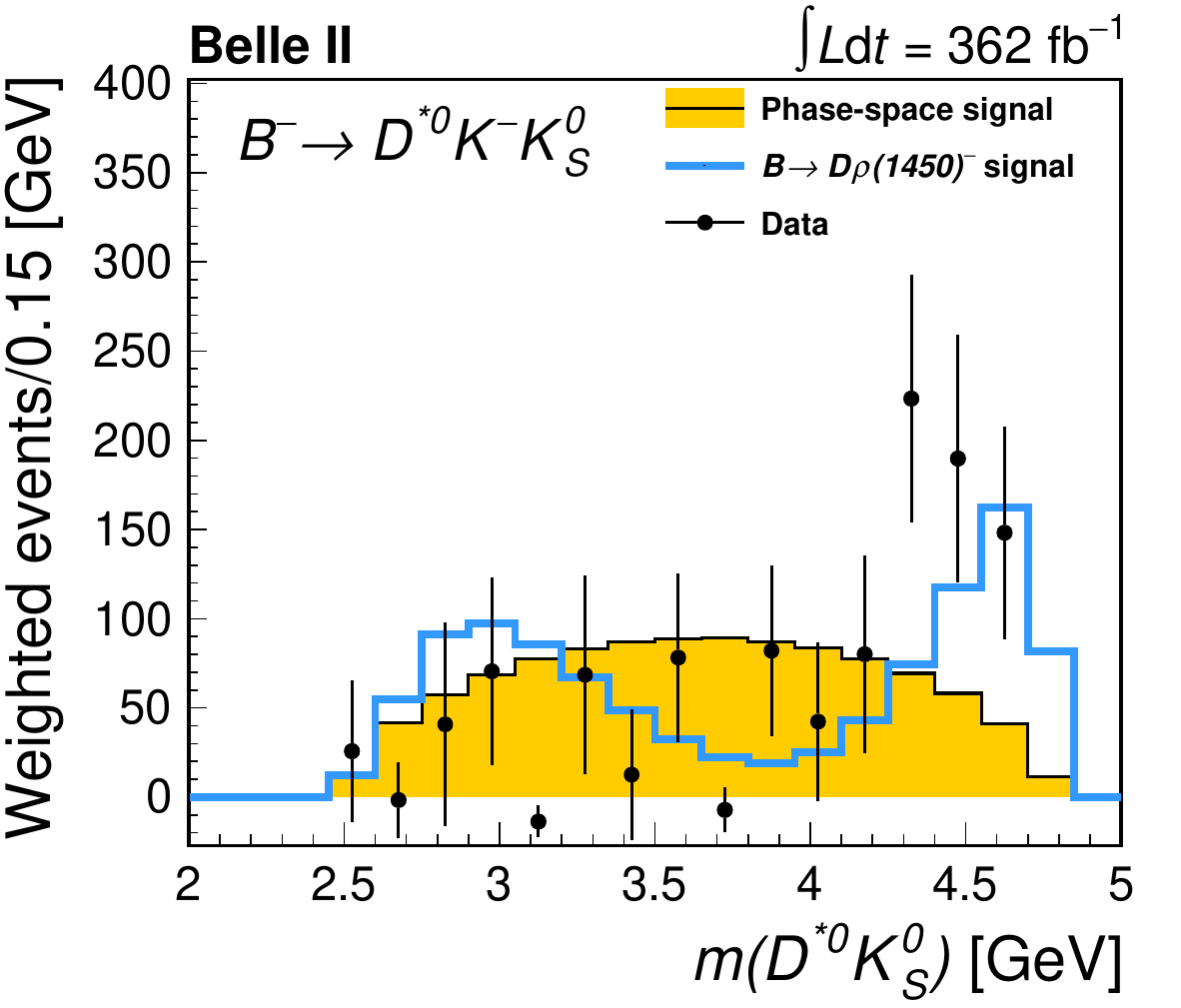}}
\subfigure{\includegraphics[width=0.37\columnwidth]{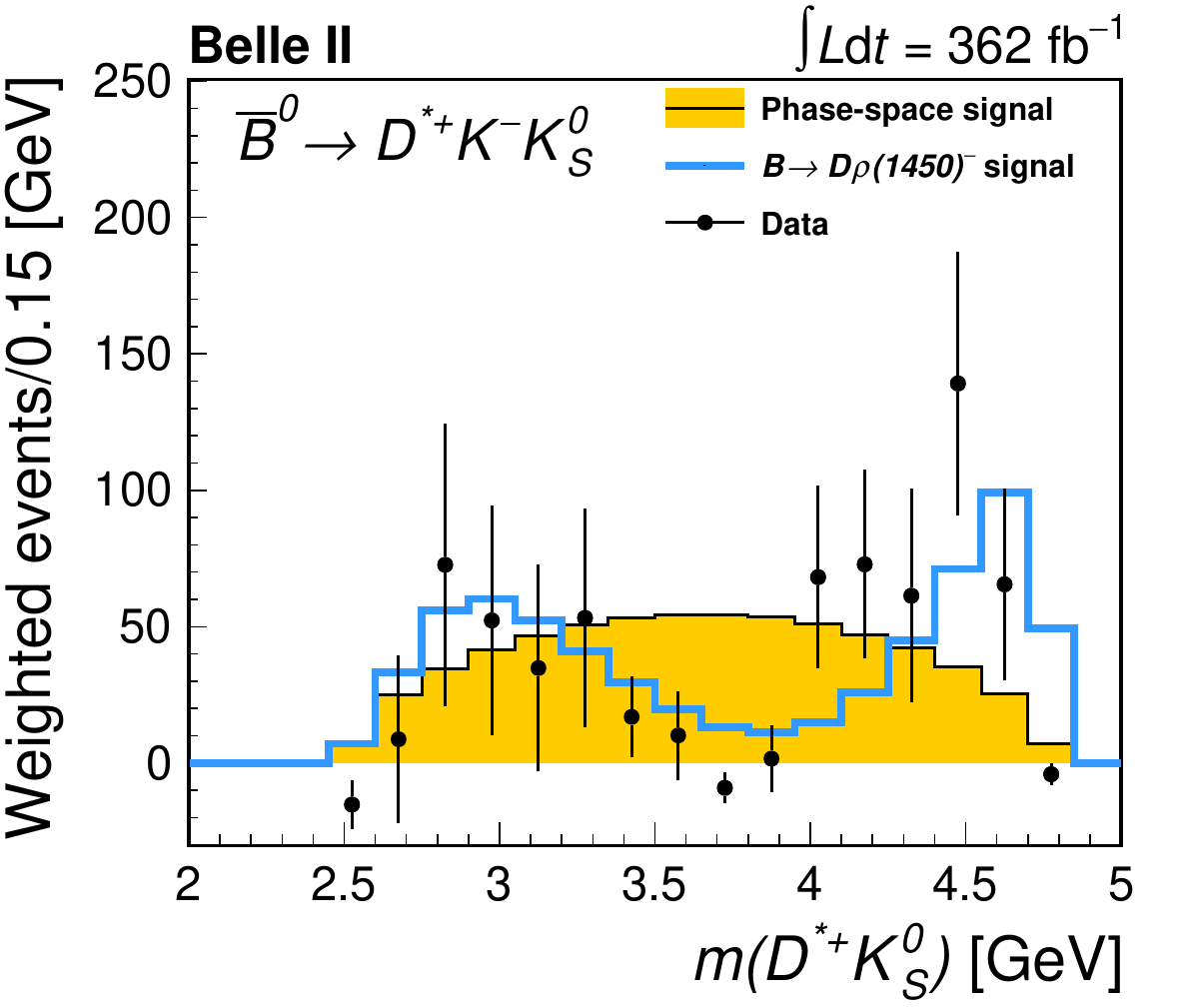}}
\subfigure{\includegraphics[width=0.37\columnwidth]{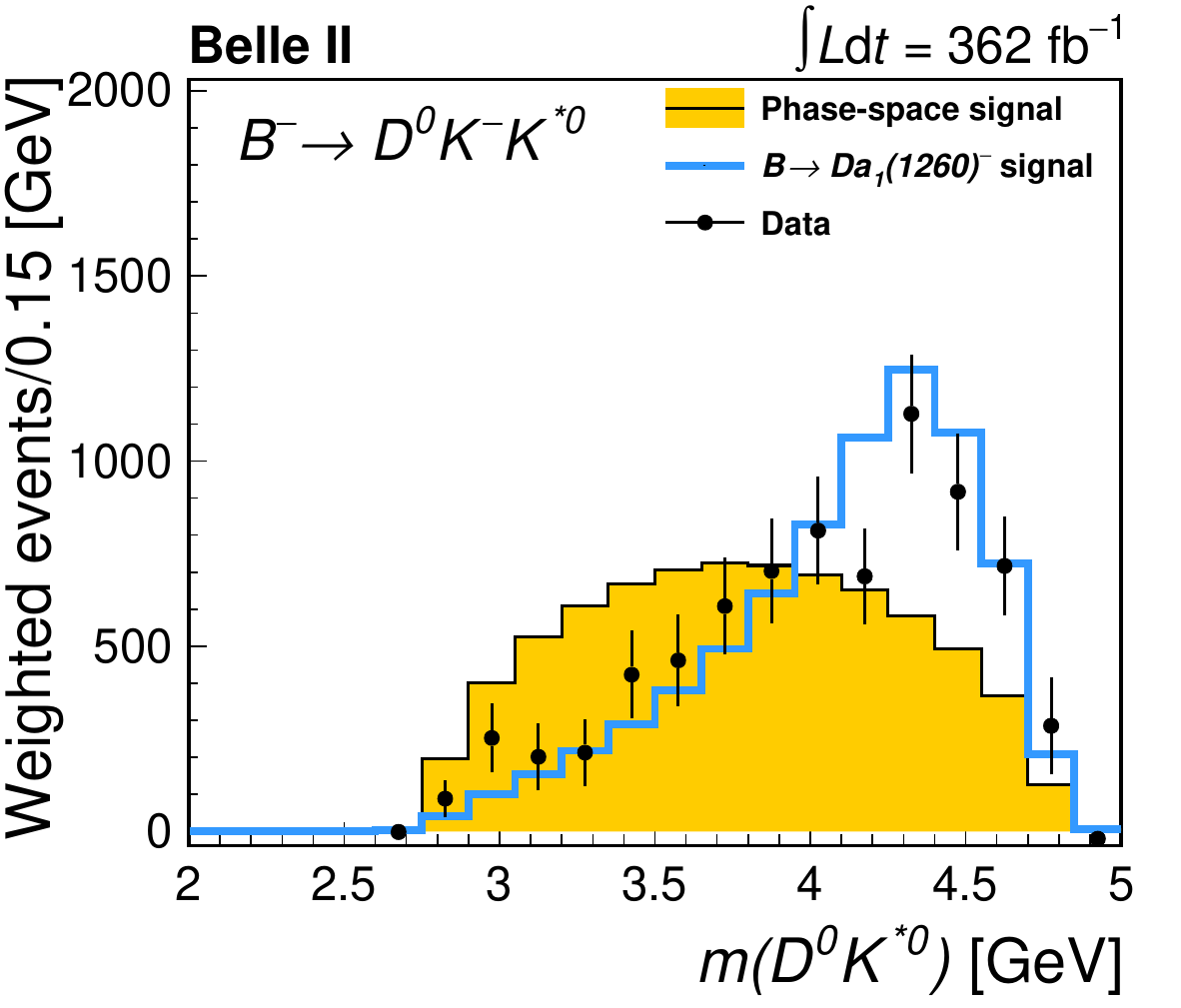}}
\subfigure{\includegraphics[width=0.37\columnwidth]{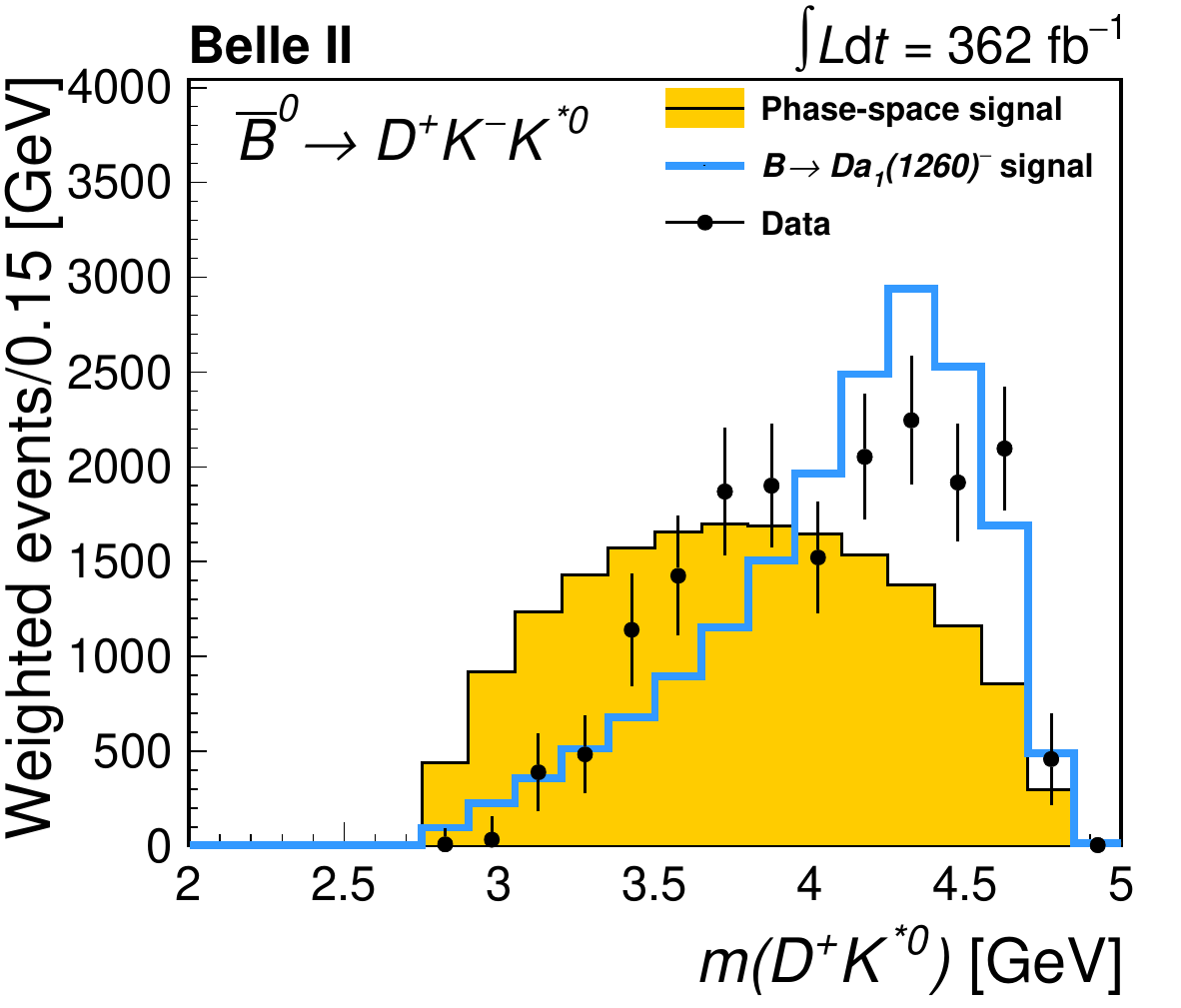}}
\subfigure{\includegraphics[width=0.37\columnwidth]{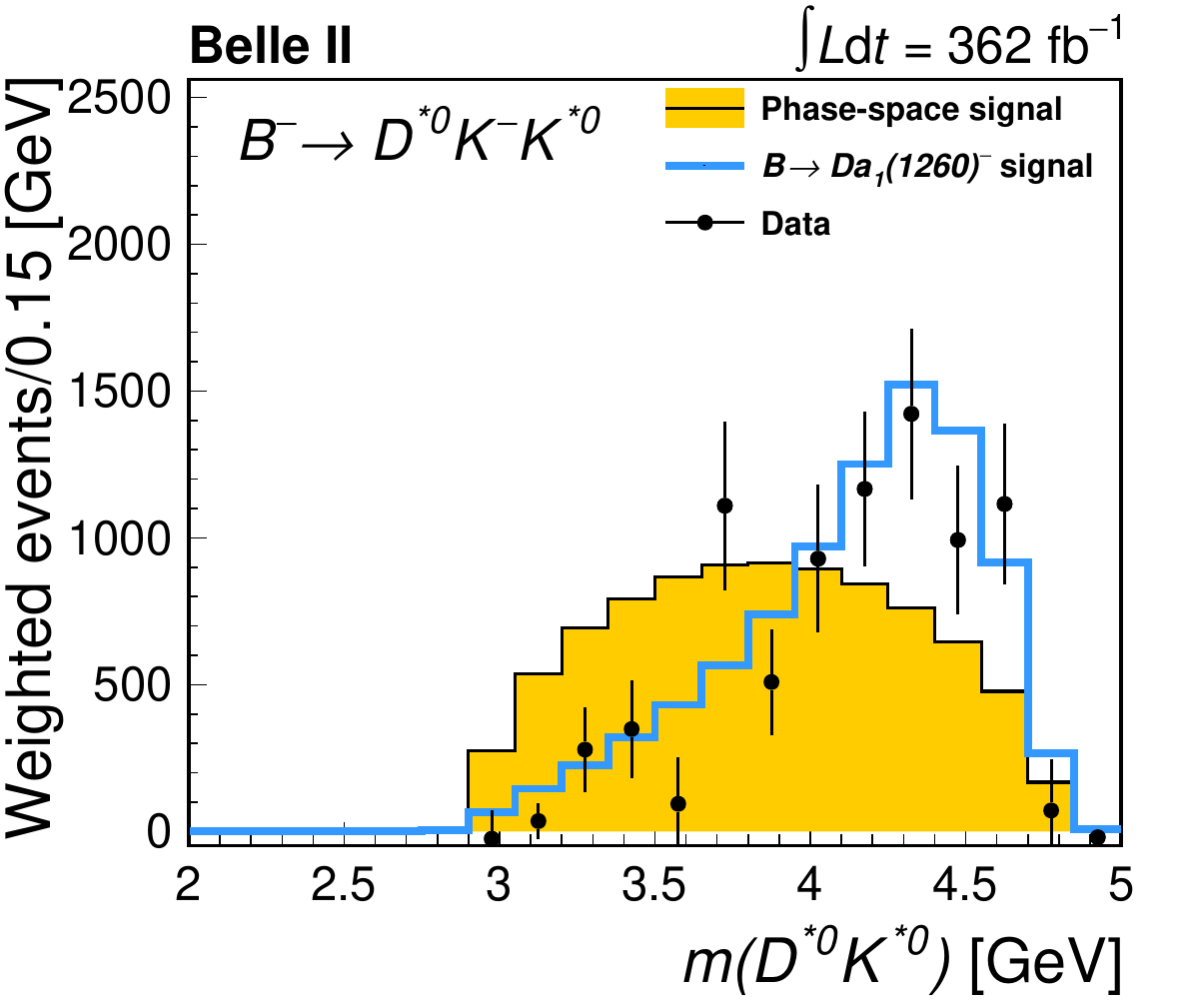}}
\subfigure{\includegraphics[width=0.37\columnwidth]{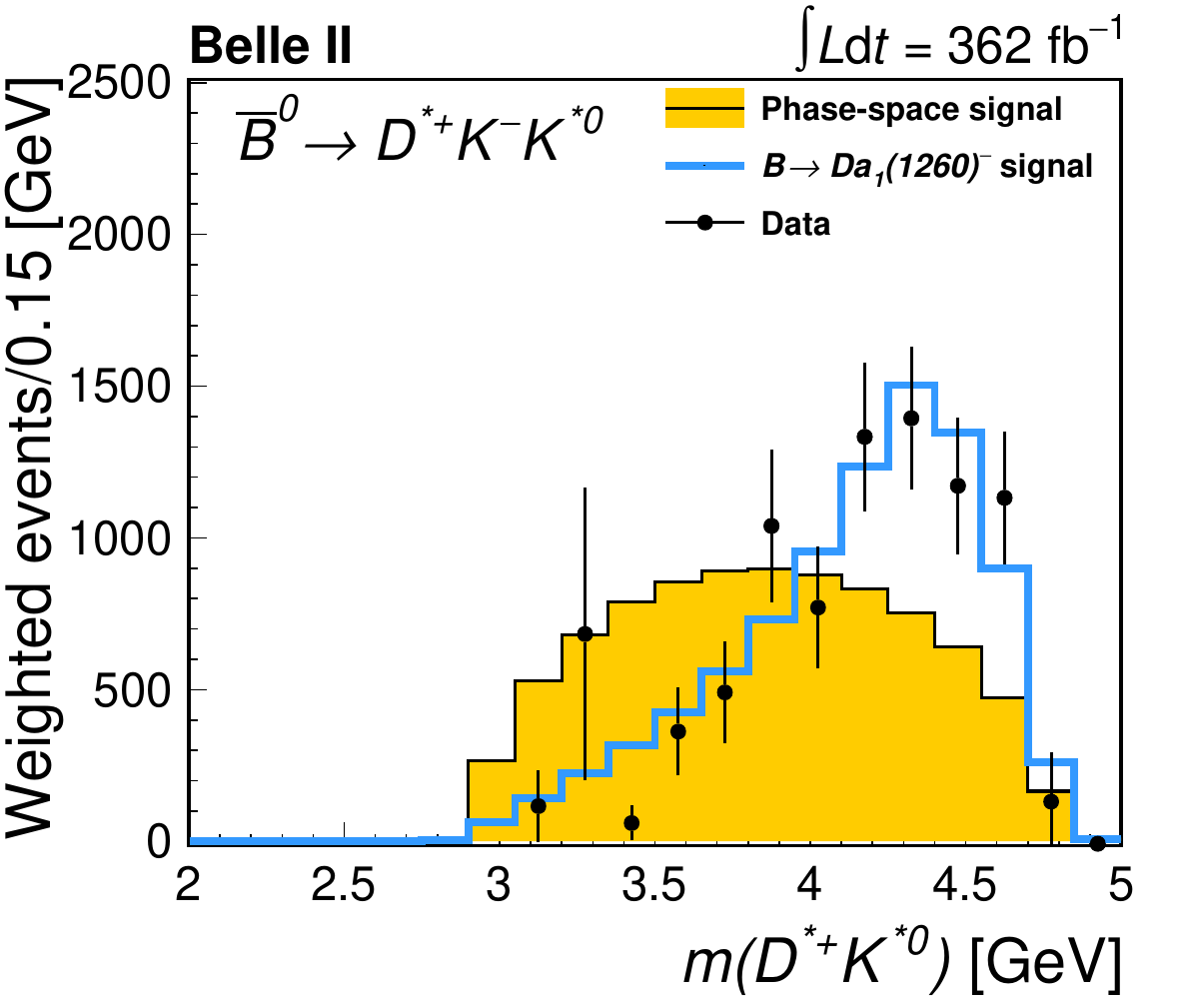}}
\caption{Background-subtracted and efficiency-corrected distribution of $m(DK)$ for the $B^-\to D^0K^-K_S^0$ (first line, left),  $\overline B{}^0\to D^+K^-K_S^0$ (first line, right), $B^-\to D^{*0}K^-K_S^0$ (second line, left), $\overline B{}^0\to D^{*+}K^-K_S^0$ (second line, right), $B^-\to D^0K^-K^{*0}$ (third line, left), $\overline B{}^0\to D^+K^-K^{*0}$ (third line, right), $B^-\to D^{*0}K^-K^{*0}$ (fourth line, left), and $\overline B{}^0\to D^{*+}K^-K^{*0}$ (fourth line, right) channels. The error bars represent the statistical uncertainty. A phase-space MC simulation and a resonant MC simulation at generator level, rescaled to the integral of the data distribution, are also shown for comparison. } \label{fig:effCorr_mDKst}
\end{figure}

\begin{figure}[!hbt]
\centering
\subfigure{\includegraphics[width=0.37\columnwidth]{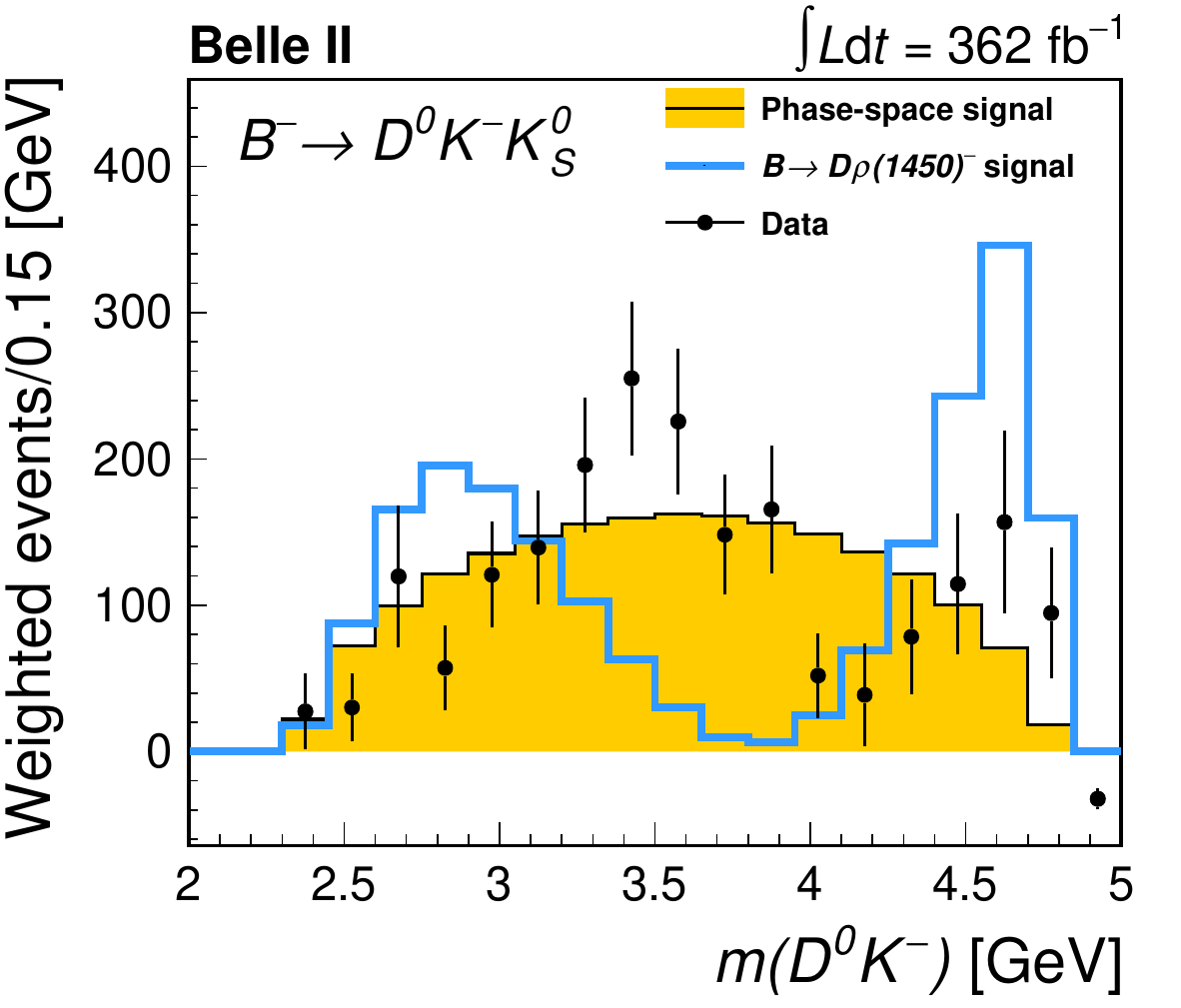}}
\subfigure{\includegraphics[width=0.37\columnwidth]{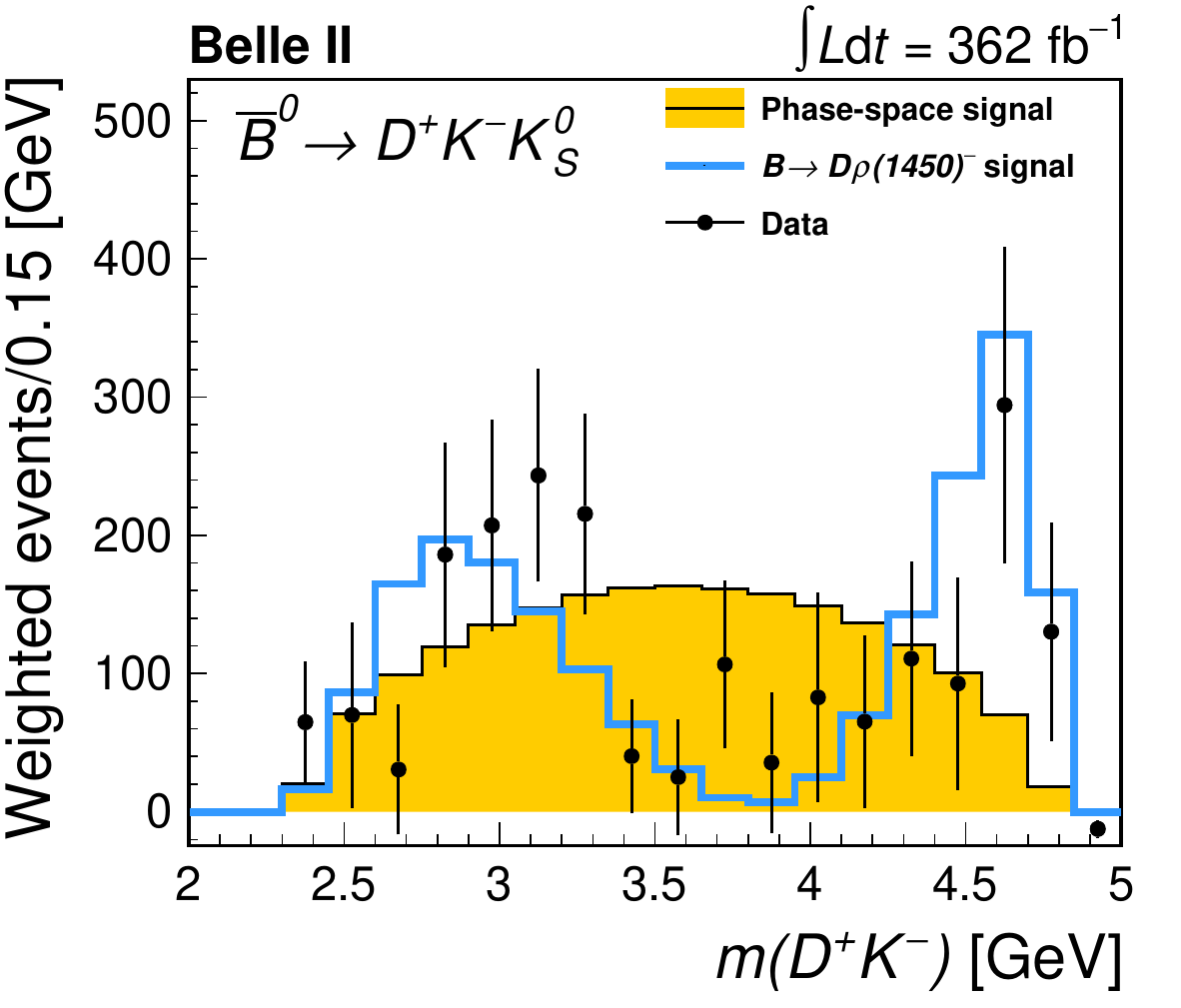}}
\subfigure{\includegraphics[width=0.37\columnwidth]{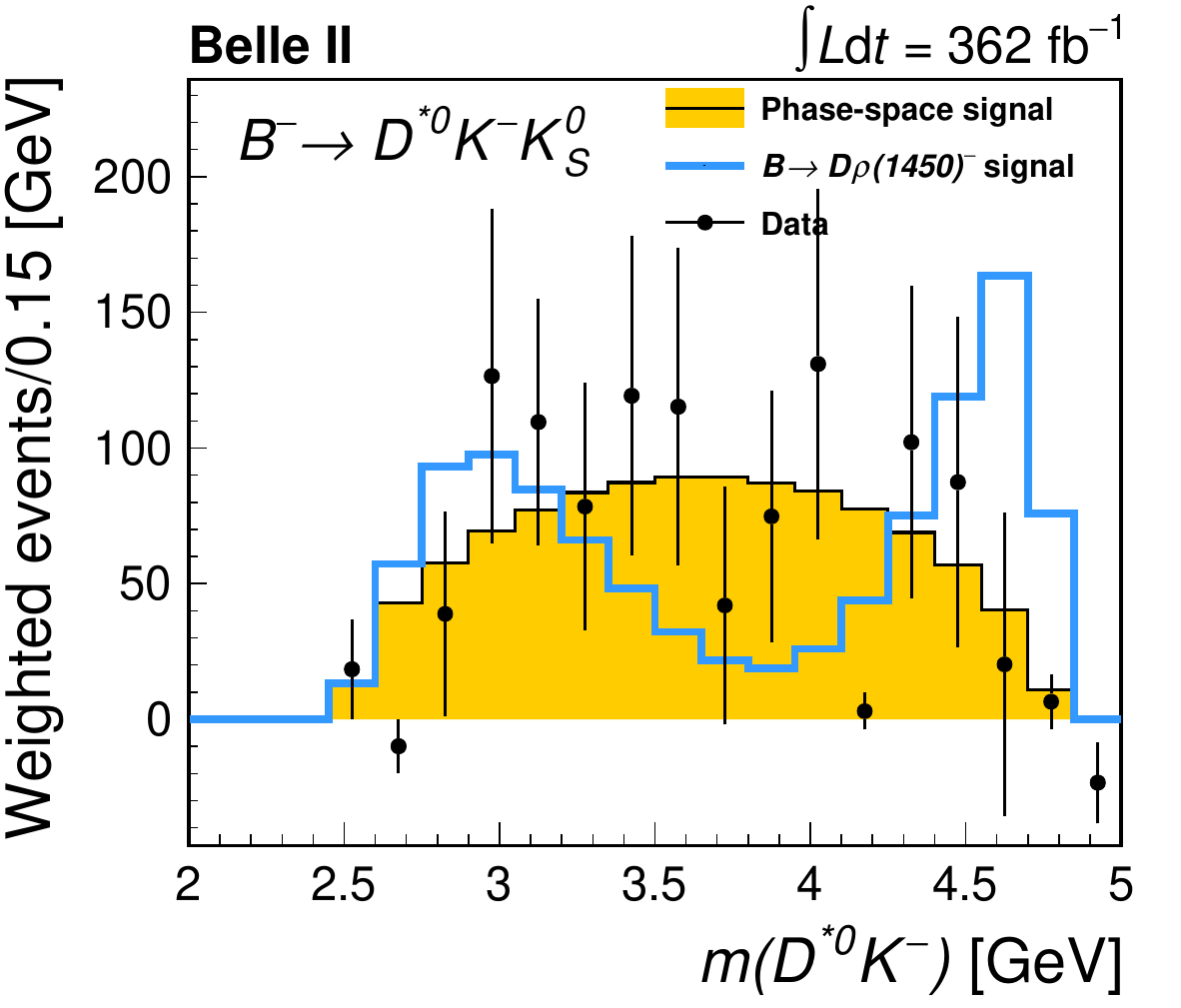}}
\subfigure{\includegraphics[width=0.37\columnwidth]{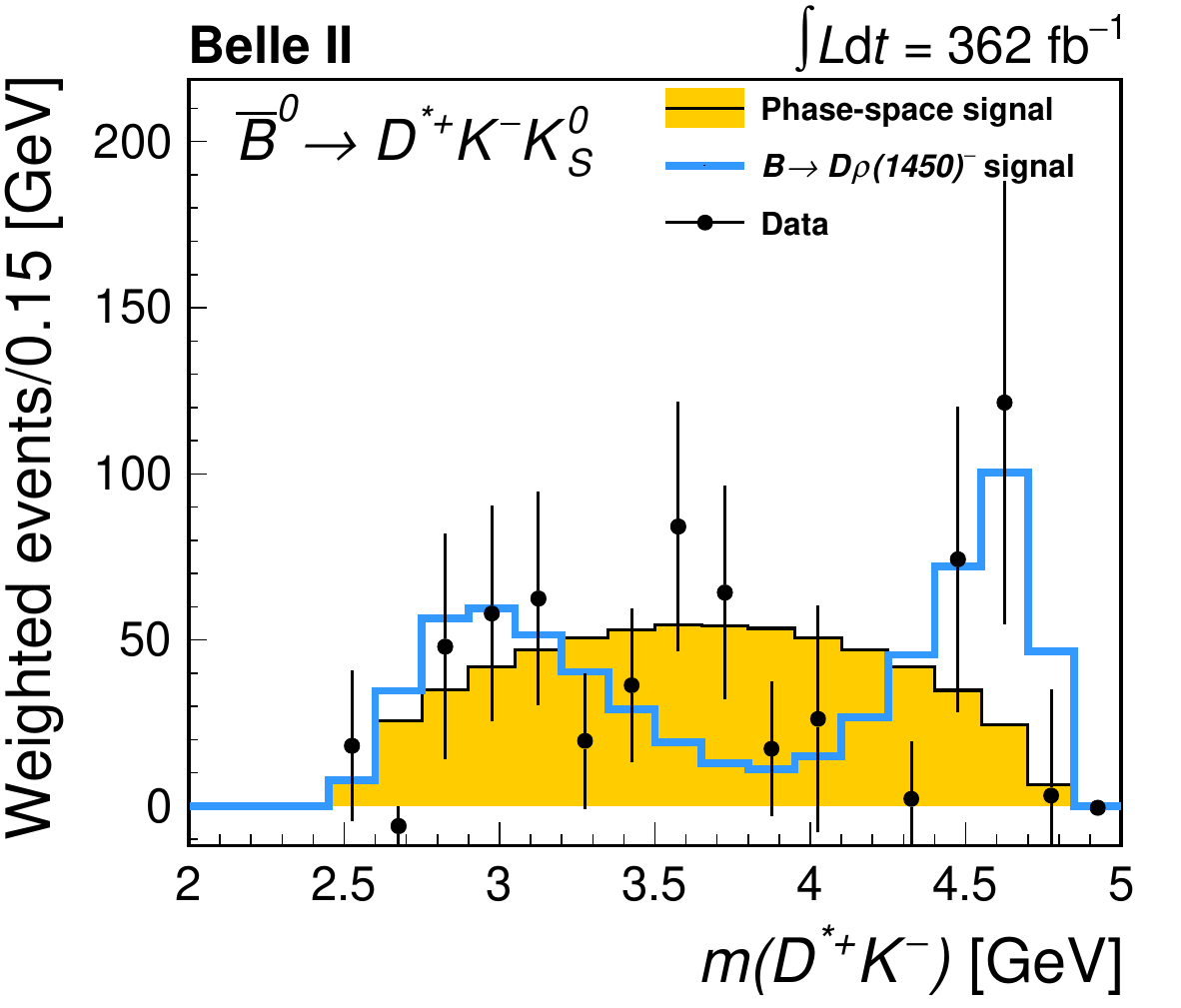}}
\subfigure{\includegraphics[width=0.37\columnwidth]{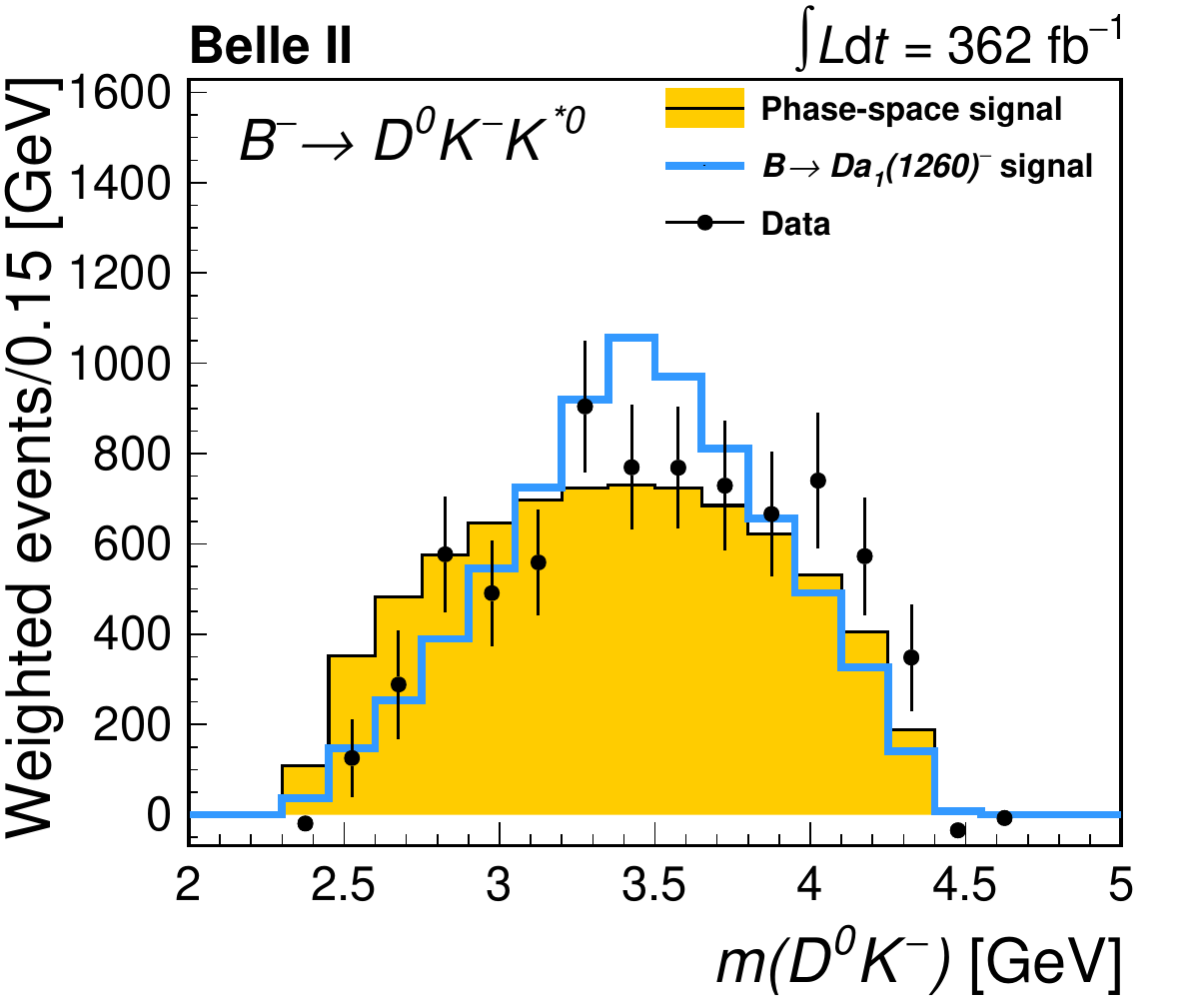}}
\subfigure{\includegraphics[width=0.37\columnwidth]{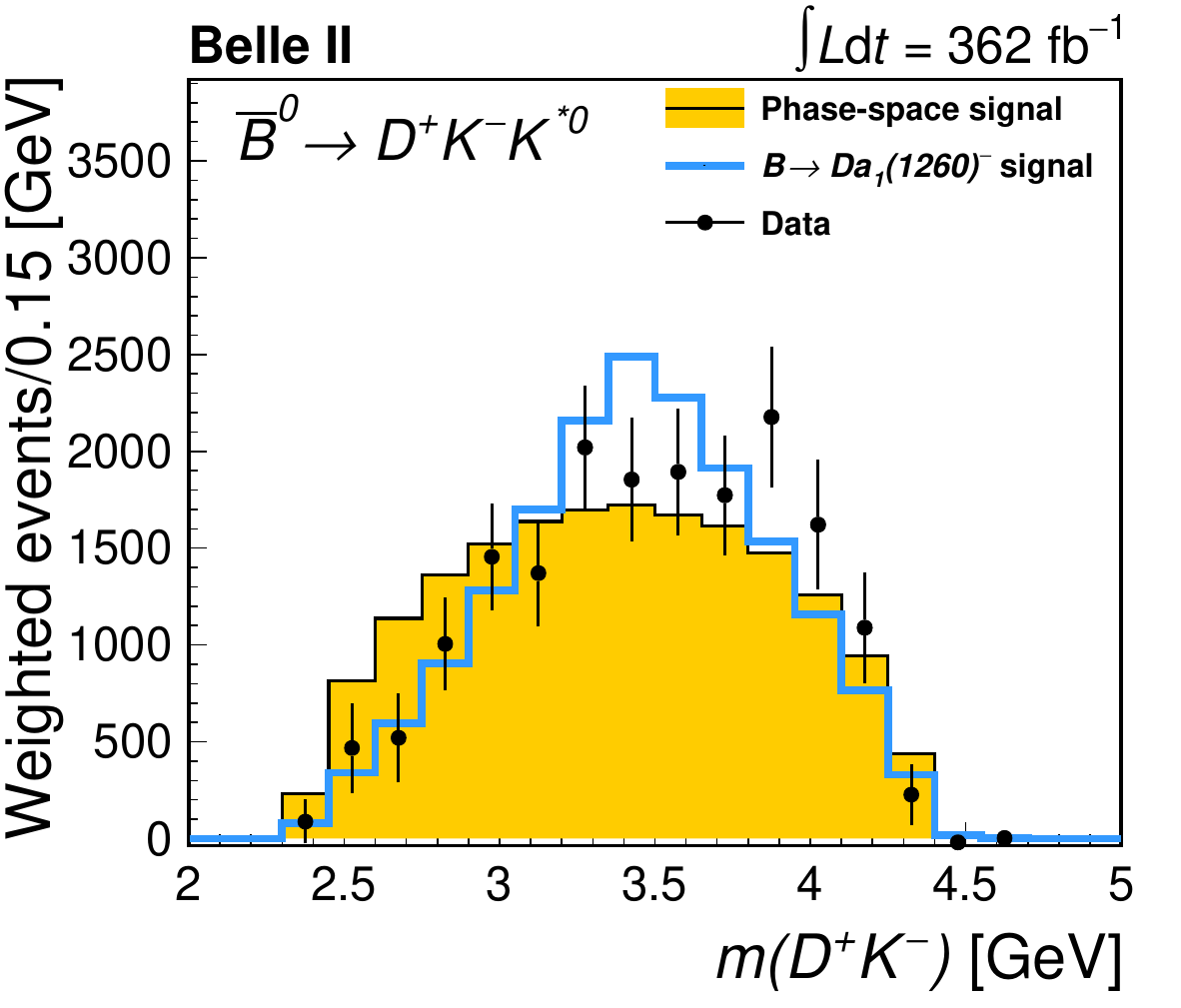}}
\subfigure{\includegraphics[width=0.37\columnwidth]{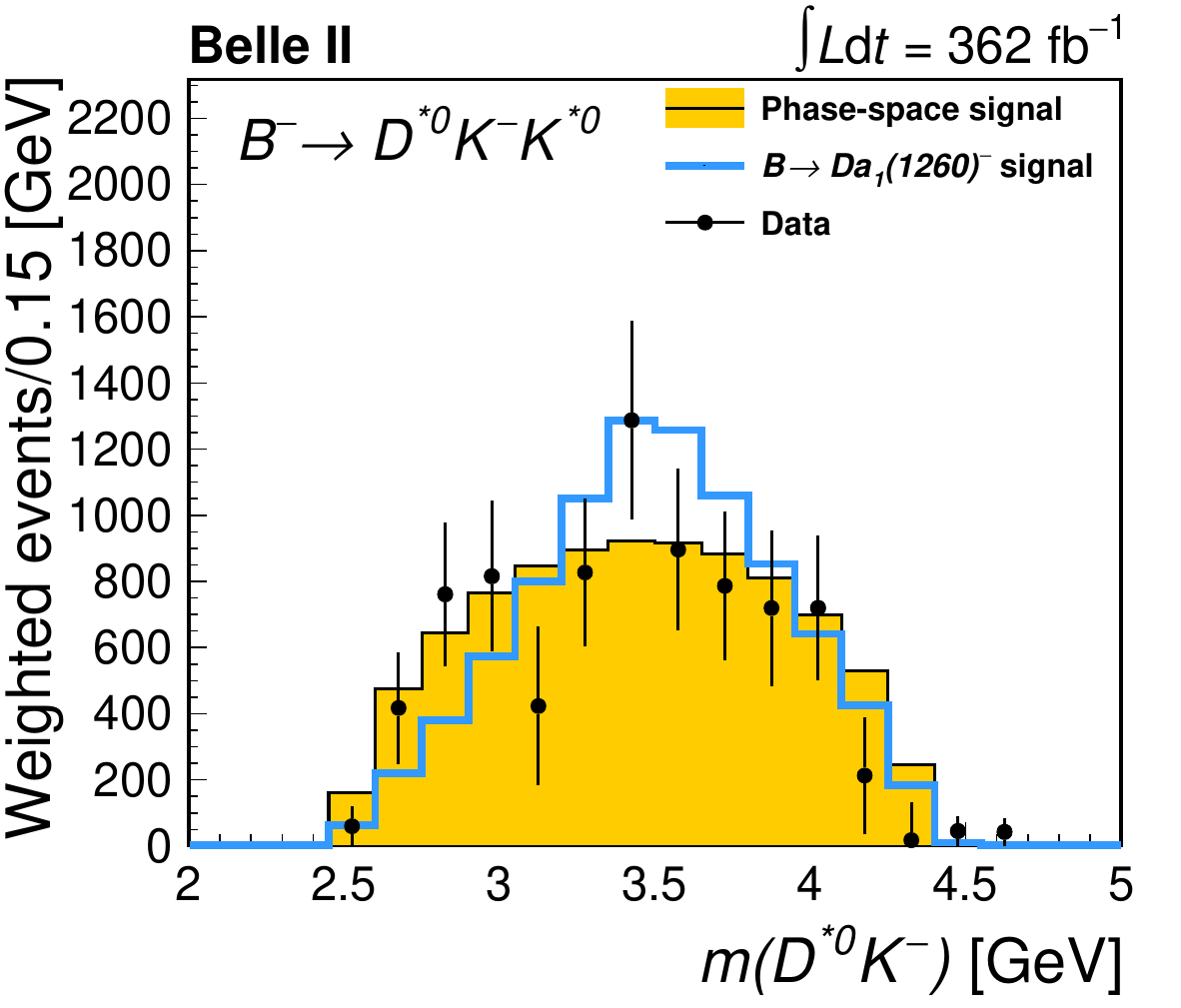}}
\subfigure{\includegraphics[width=0.37\columnwidth]{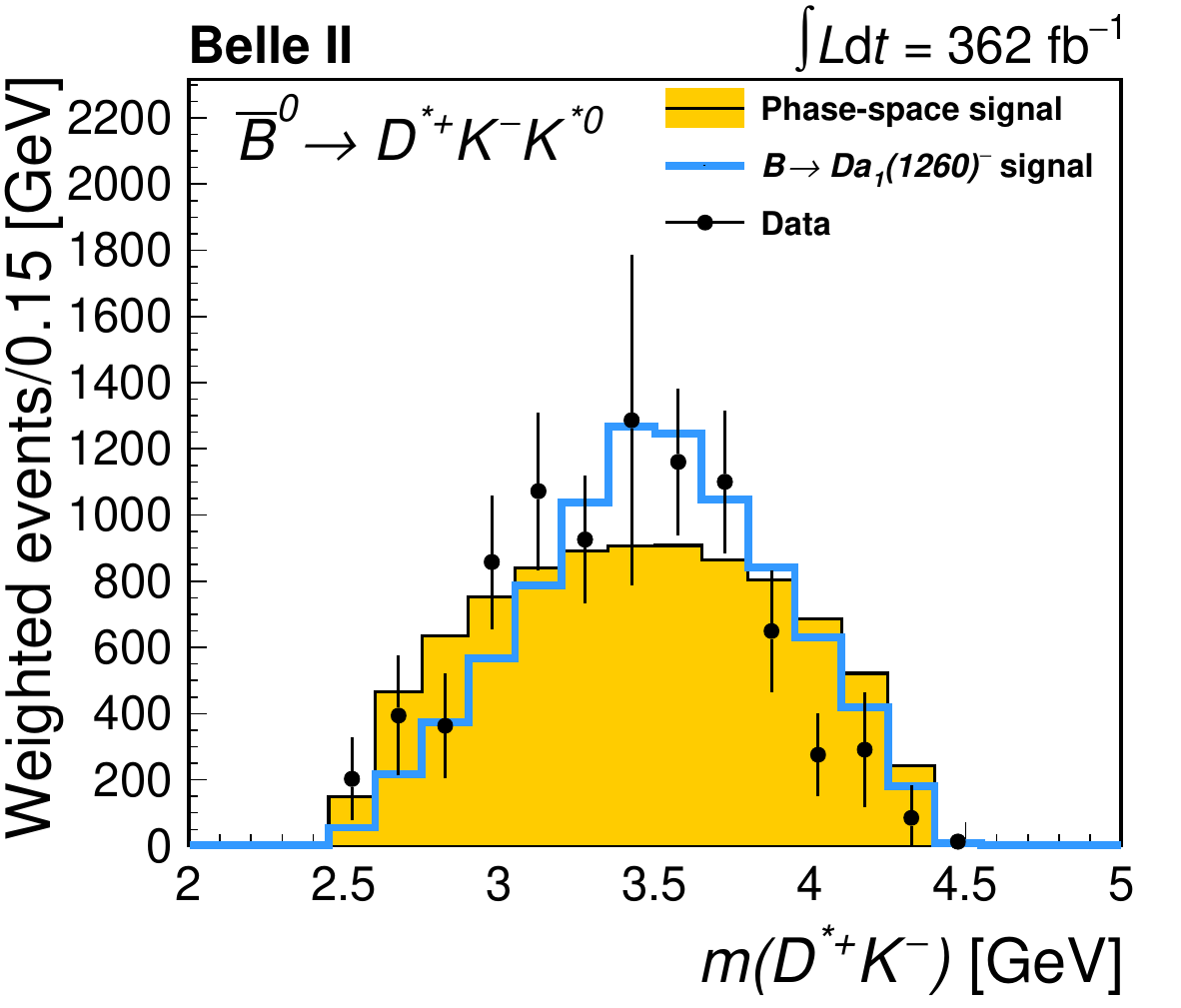}}
\caption{Background-subtracted and efficiency-corrected distribution of $m(DK^-)$ for the $B^-\to D^0K^-K_S^0$ (first line, left),  $\overline B{}^0\to D^+K^-K_S^0$ (first line, right), $B^-\to D^{*0}K^-K_S^0$ (second line, left), $\overline B{}^0\to D^{*+}K^-K_S^0$ (second line, right), $B^-\to D^0K^-K^{*0}$ (third line, left), $\overline B{}^0\to D^+K^-K^{*0}$ (third line, right), $B^-\to D^{*0}K^-K^{*0}$ (fourth line, left), and $\overline B{}^0\to D^{*+}K^-K^{*0}$ (fourth line, right) channels. The error bars represent the statistical uncertainty. A phase-space MC simulation and a resonant MC simulation at generator level, rescaled to the integral of the data distribution, are also shown for comparison.} \label{fig:effCorr_mDK}
\end{figure}

The $\bigl(m(K^-K),m(DK)\bigr)$ distributions are shown in Fig.~\ref{fig:Dalitz} for the eight $B\to DK^-K$ channels. The selection is applied together with the efficiency correction; $s$Weights are applied to subtract the background component. The distributions also show some structures, which are compatible with the resonant transition discussed in Sec.~\ref{sec:mKK}. 
The $m(DK)$ and $m(DK^-)$ projections show that the reflection of the $m(K^-K)$ resonant lineshape may produce such structures. These distributions are shown in Fig.~\ref{fig:effCorr_mDKst} and Fig.~\ref{fig:effCorr_mDK}.

\clearpage
\begin{table}[!htb]
\centering
\caption{Non-$K^{*0}$-resonant fraction, in percent of the signal, estimated for each of the four $K^{*0}$ channels. The first uncertainty is statistical, the second systematic.}\label{tab:NRfraction}
\begin{tabular}{cc}
\toprule
                     Channel &                  non-$K^{*0}$-resonant fraction [\%]  \\
                     \midrule
$ B^{-}\rightarrow D^{0}K^{-}K^{*0}$ &               $3.1\pm0.5 \pm 1.2$\\
$ \overline B{}^0\rightarrow D^{+}K^{-}K^{*0}$ &          $0.7\pm0.5 \pm 0.3$\\
$ B^{-}\rightarrow D^{*0}K^{-}K^{*0}$ &              $1.4\pm0.6 \pm 0.3$\\
 $ \overline B{}^0\rightarrow D^{*+}K^{-}K^{*0}$ &        $1.0\pm0.5 \pm 0.5$\\
\bottomrule
\end{tabular}
\end{table}

\clearpage

This work, based on data collected using the Belle II detector, which was built and commissioned prior to March 2019,
was supported by
Higher Education and Science Committee of the Republic of Armenia Grant No.~23LCG-1C011;
Australian Research Council and Research Grants
No.~DP200101792, 
No.~DP210101900, 
No.~DP210102831, 
No.~DE220100462, 
No.~LE210100098, 
and
No.~LE230100085; 
Austrian Federal Ministry of Education, Science and Research,
Austrian Science Fund
No.~P~34529,
No.~J~4731,
No.~J~4625,
and
No.~M~3153,
and
Horizon 2020 ERC Starting Grant No.~947006 ``InterLeptons'';
Natural Sciences and Engineering Research Council of Canada, Compute Canada and CANARIE;
National Key R\&D Program of China under Contract No.~2022YFA1601903,
National Natural Science Foundation of China and Research Grants
No.~11575017,
No.~11761141009,
No.~11705209,
No.~11975076,
No.~12135005,
No.~12150004,
No.~12161141008,
and
No.~12175041,
and Shandong Provincial Natural Science Foundation Project~ZR2022JQ02;
the Czech Science Foundation Grant No.~22-18469S 
and
Charles University Grant Agency project No.~246122;
European Research Council, Seventh Framework PIEF-GA-2013-622527,
Horizon 2020 ERC-Advanced Grants No.~267104 and No.~884719,
Horizon 2020 ERC-Consolidator Grant No.~819127,
Horizon 2020 Marie Sklodowska-Curie Grant Agreement No.~700525 ``NIOBE''
and
No.~101026516,
and
Horizon 2020 Marie Sklodowska-Curie RISE project JENNIFER2 Grant Agreement No.~822070 (European grants);
L'Institut National de Physique Nucl\'{e}aire et de Physique des Particules (IN2P3) du CNRS
and
L'Agence Nationale de la Recherche (ANR) under grant ANR-21-CE31-0009 (France);
BMBF, DFG, HGF, MPG, and AvH Foundation (Germany);
Department of Atomic Energy under Project Identification No.~RTI 4002,
Department of Science and Technology,
and
UPES SEED funding programs
No.~UPES/R\&D-SEED-INFRA/17052023/01 and
No.~UPES/R\&D-SOE/20062022/06 (India);
Israel Science Foundation Grant No.~2476/17,
U.S.-Israel Binational Science Foundation Grant No.~2016113, and
Israel Ministry of Science Grant No.~3-16543;
Istituto Nazionale di Fisica Nucleare and the Research Grants BELLE2;
Japan Society for the Promotion of Science, Grant-in-Aid for Scientific Research Grants
No.~16H03968,
No.~16H03993,
No.~16H06492,
No.~16K05323,
No.~17H01133,
No.~17H05405,
No.~18K03621,
No.~18H03710,
No.~18H05226,
No.~19H00682, 
No.~20H05850,
No.~20H05858,
No.~22H00144,
No.~22K14056,
No.~22K21347,
No.~23H05433,
No.~26220706,
and
No.~26400255,
and
the Ministry of Education, Culture, Sports, Science, and Technology (MEXT) of Japan;  
National Research Foundation (NRF) of Korea Grants
No.~2016R1\-D1A1B\-02012900,
No.~2018R1\-A2B\-3003643,
No.~2018R1\-A6A1A\-06024970,
No.~2019R1\-I1A3A\-01058933,
No.~2021R1\-A6A1A\-03043957,
No.~2021R1\-F1A\-1060423,
No.~2021R1\-F1A\-1064008,
No.~2022R1\-A2C\-1003993,
and
No.~RS-2022-00197659,
Radiation Science Research Institute,
Foreign Large-Size Research Facility Application Supporting project,
the Global Science Experimental Data Hub Center of the Korea Institute of Science and Technology Information
and
KREONET/GLORIAD;
Universiti Malaya RU grant, Akademi Sains Malaysia, and Ministry of Education Malaysia;
Frontiers of Science Program Contracts
No.~FOINS-296,
No.~CB-221329,
No.~CB-236394,
No.~CB-254409,
and
No.~CB-180023, and SEP-CINVESTAV Research Grant No.~237 (Mexico);
the Polish Ministry of Science and Higher Education and the National Science Center;
the Ministry of Science and Higher Education of the Russian Federation
and
the HSE University Basic Research Program, Moscow;
University of Tabuk Research Grants
No.~S-0256-1438 and No.~S-0280-1439 (Saudi Arabia);
Slovenian Research Agency and Research Grants
No.~J1-9124
and
No.~P1-0135;
Agencia Estatal de Investigacion, Spain
Grant No.~RYC2020-029875-I
and
Generalitat Valenciana, Spain
Grant No.~CIDEGENT/2018/020;
The Knut and Alice Wallenberg Foundation (Sweden), Contracts No.~2021.0174 and No.~2021.0299;
National Science and Technology Council,
and
Ministry of Education (Taiwan);
Thailand Center of Excellence in Physics;
TUBITAK ULAKBIM (Turkey);
National Research Foundation of Ukraine, Project No.~2020.02/0257,
and
Ministry of Education and Science of Ukraine;
the U.S. National Science Foundation and Research Grants
No.~PHY-1913789 
and
No.~PHY-2111604, 
and the U.S. Department of Energy and Research Awards
No.~DE-AC06-76RLO1830, 
No.~DE-SC0007983, 
No.~DE-SC0009824, 
No.~DE-SC0009973, 
No.~DE-SC0010007, 
No.~DE-SC0010073, 
No.~DE-SC0010118, 
No.~DE-SC0010504, 
No.~DE-SC0011784, 
No.~DE-SC0012704, 
No.~DE-SC0019230, 
No.~DE-SC0021274, 
No.~DE-SC0021616, 
No.~DE-SC0022350, 
No.~DE-SC0023470; 
and
the Vietnam Academy of Science and Technology (VAST) under Grants
No.~NVCC.05.12/22-23
and
No.~DL0000.02/24-25.

These acknowledgements are not to be interpreted as an endorsement of any statement made
by any of our institutes, funding agencies, governments, or their representatives.

We thank the SuperKEKB team for delivering high-luminosity collisions;
the KEK cryogenics group for the efficient operation of the detector solenoid magnet and IBBelle on site;
the KEK Computer Research Center for on-site computing support; the NII for SINET6 network support;
and the raw-data centers hosted by BNL, DESY, GridKa, IN2P3, INFN, 
and the University of Victoria.

\bibliographystyle{JHEP}
\bibliography{references}

\end{document}